\global\def\draftcontrol{0}

   \def\versionno{ viscoelastic}

\catcode`\@=11

\expandafter\ifx\csname draftcontrol\endcsname\relax\global\def\draftcontrol{0}
\fi

{\count255=\time\divide\count255 by 60
\xdef\hourmin{\number\count255}
\multiply\count255 by-60\advance\count255 by\time
\xdef\hourmin{\hourmin:\ifnum\count255<10 0\fi\the\count255}}
\def\draftdate{\number\month/\number\day/\number\year\ \ \ \hourmin }

\newcommand\makepapertitle{\par
  \begingroup
    \renewcommand\thefootnote{\@fnsymbol\c@footnote}%
    \def\@makefnmark{\rlap{\@textsuperscript{\normalfont\@thefnmark}}}%
    \long\def\@makefntext##1{\parindent 1em\noindent
            \hb@xt@1.8em{%
                \hss\@textsuperscript{\normalfont\@thefnmark}}##1}%
     \newpage
     \global\@topnum\z@   
     \@makepapertitle
     \thispagestyle{empty}\@thanks
  \endgroup
  \setcounter{footnote}{0}%
  \global\let\thanks\relax
  \global\let\makepapertitle\relax
  \global\let\@makepapertitle\relax
  \global\let\@thanks\@empty
  \global\let\@author\@empty
  \global\let\@date\@empty
  \global\let\@title\@empty
  \global\let\title\relax
  \global\let\author\relax
  \global\let\date\relax
  \global\let\and\relax
  \def\version{\let\version\@version\@gobble}
}
\def\@makepapertitle{
  \newpage
   \ifnum\draftcontrol=1 {}
   \version\versionno
   \vskip 3em%
   \else
   \hfill\hbox to 3cm {\parbox{4cm}{\@pubnum}\hss}%
   \vskip 3em%
   \fi
   \begin{center}%
   \let\footnoterule\relax%
   \let \footnote \thanks
     {\LARGE {\@title}}%
     \vskip 1.5em%
     {\normalsize
       \lineskip .5em%
       \begin{tabular}[t]{c}%
         \@author
       \end{tabular}\par}%
     \vskip 1.5em%
     {\@bstract}%
     \end{center}%
     \vskip 1.5em
     \@date%
   \par
}

\gdef\@pubnum{}
\def\pubnum#1{%
  \gdef\@pubnum{#1}}

\gdef\@bstract{}
\def\Abstract#1{%
  \gdef\@bstract{%
   \parbox{\textwidth-0pc}{%
   \centerline{\bf Abstract}\penalty1000%
\kern.2cm%
\noindent
\renewcommand\baselinestretch{1.0}%
{#1}}}
}

\def\ps@paper{\let\@mkboth\@gobbletwo%
     \ifnum\draftcontrol=1
    \def\@oddfoot{\hbox to \textwidth{\tiny \versionno \hfil\tiny\draftdate}%
    \hskip -\textwidth \hbox to \textwidth{\hfil\rm\thepage\hfil}}%
     \else\def\@oddfoot{\hbox to \textwidth{\hfil\rm\thepage\hfil}}
     \fi
     \let\@evenfoot\@oddfoot
}

\def\body{\clearpage
          \pagestyle{paper}
    }

\def\@version#1{\ifnum\draftcontrol=1
\typeout{}\typeout{#1}\typeout{}
\vskip3mm\centerline{\hbox{\fbox{\normalsize{\tt DRAFT -- #1 -- }
                   {\draftdate}}}}\vskip3mm
\fi}
\let\version\@version
\long\def\eqlabel#1{\ifnum\draftcontrol=1
                    \tag@false  
                    \tag*{(\theequation) \hbox to -0.2cm{\hspace{0cm}\small{#1}\hss}}
                    \refstepcounter{equation}
                    \edef\@currentlabel{\theequation}
                    \ltx@label{#1}          
                    \else
                    \label{#1}
                    \fi
                    }
\let\st@bibitem\@bibitem
\let\st@lbibitem\@lbibitem
\ifnum\draftcontrol=1
  \def\@bibitem#1{%
    \st@bibitem{#1}\a@@label{#1}\ignorespaces}
  \def\@lbibitem[#1]#2{%
    \st@lbibitem[#1]{#2}\a@@label{#2}\ignorespaces}
  \def\a@@label#1{%
    \gdef\a@lab{\smash{\normalfont\small#1}}
    \ifvmode
      \if@inlabel
        \global\setbox\@labels\hbox{%
          \llap{\a@lab\let\a@lab\relax
                \kern\@totalleftmargin\kern\marginparsep}%
          \box\@labels}%
      \fi
    \fi}
\fi

\documentclass[12pt,letterpaper]{article}

\usepackage{amsmath,amssymb,array,calc,epsfig,rotating,bm}
\usepackage[sort]{cite}
\usepackage{graphicx}
\usepackage{psfrag,verbatim}
\usepackage{xcolor}
\usepackage[utf8]{inputenc}

\ifnum\draftcontrol=1
\fi

\renewcommand\baselinestretch{1.25}
\renewcommand\section{\@startsection {section}{1}{\z@}%
                                   {-3.5ex \@plus -1ex \@minus -.2ex}%
                                   {2.3ex \@plus.2ex}%
                                   {\normalfont\large\bfseries}}
\renewcommand\subsection{\@startsection{subsection}{2}{\z@}%
                                   {-3.25ex\@plus -1ex \@minus -.2ex}%
                                   {1.5ex \@plus .2ex}%
                                   {\normalfont\normalsize\bfseries}}
\renewcommand\subsubsection{\@startsection{subsubsection}{3}{\z@}%
                                   {-3.25ex\@plus -1ex \@minus -.2ex}%
                                   {1.5ex \@plus .2ex}%
                                   {\normalfont\normalsize\it}}
\renewcommand\paragraph{\@startsection{paragraph}{4}{\z@}%
                                   {-3.25ex\@plus -1ex \@minus -.2ex}%
                                   {1.5ex \@plus .2ex}%
                                   {\normalfont\normalsize\bf}}

\numberwithin{equation}{section}

\usepackage{hyperref}

\def\revise#1       {\raisebox{-0em}{\rule{3pt}{1em}}%
                     \marginpar{\raisebox{.5em}{\vrule width3pt\
                     \vrule width0pt height 0pt depth0.5em
                     \hbox to 0cm{\hspace{0cm}{%
                     \parbox[t]{4em}{\raggedright\footnotesize{#1}}}\hss}}}}

\newcommand\nxt[1]  {\\\fnxt#1}
\newcommand{\ie}{{\it i.e.,}\ }
\newcommand{\eg}{{\it e.g.,}\ }

\def\cale         {{\cal E}}
\def\calf         {{\cal F}}
\def\calg         {{\cal G}}

\def\calk         {{\cal K}}

\def\calm         {{\cal M}}
\def\caln         {{\cal N}}
\def\calo         {{\cal O}}
\def\calp         {{\cal P}}
\def\calq         {{\cal Q}}

\def\cals         {{\cal S}}
\def\calt         {{\cal T}}

\usepackage{etoolbox}
\usepackage{epsfig}
\usepackage{rotating}

\newcommand\testletter[1]{
{\rotatebox[origin=c]{90}{#1}}
}

\def\zet          {{\mathbb Z}}

\def\del          {\partial}

\def\Re           {{\rm Re\hskip0.1em}}
\def\Im           {{\rm Im\hskip0.1em}}

\def\sqr#1#2{{\vcenter{\vbox{\hrule height.#2pt
 \hbox{\vrule width.#2pt height#1pt \kern#1pt
 \vrule width.#2pt}\hrule height.#2pt}}}}

\newcommand{\cool}{\ensuremath{%
  \mathchoice{\includegraphics[height=2ex]{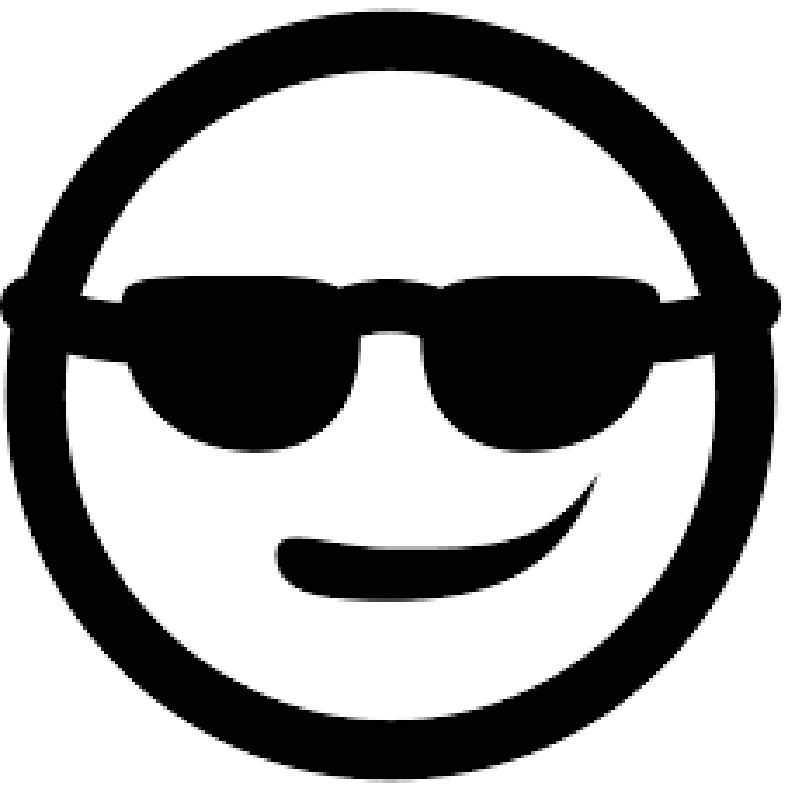}}
    {\includegraphics[height=2ex]{cool}}
    {\includegraphics[height=1.5ex]{cool}}
    {\includegraphics[height=1ex]{cool}}
}}

\def\AB#1{{\color{magenta}{AB[#1]}}}

\def\a{\alpha}
\def\b{\beta}
\def\w{\omega}

\def\dd{\delta}

\def\e{\epsilon}

\def\g{\gamma}

\def\aa1{\phi}
\def\cc1{\psi}

\def\arctanh{{\rm arctanh}}

\def\k{\kappa}

\def\l{\lambda}

\def\k{\kappa}

\def\t{\tau}

\catcode`\@=12

\begin{document}

\title{\bf Holographic Viscoelastic Hydrodynamics}

\date\today

\author{
Alex Buchel$^{1,2,3}$ and Matteo Baggioli$^{\,\cool}$\\[0.4cm]
\it $^1$Department of Applied Mathematics\\
\it $^2$Department of Physics and Astronomy\\ 
\it University of Western Ontario\\
\it London, Ontario N6A 5B7, Canada\\
\it $^3$Perimeter Institute for Theoretical Physics\\
\it Waterloo, Ontario N2J 2W9, Canada\\[0.4cm]
\it $\cool$ Crete Center for Theoretical Physics\\ \it Institute for Theoretical and Computational Physics\\
\it Department of Physics, University of Crete\\
\it 71003 Heraklion, Greece\\[0.1cm]
abuchel$@$perimeterinstitute.ca,  mbaggioli$@$physics.uoc.gr \thanks{CCTP-2018-5 , ITCP-IPP 2018/24}
}

\Abstract{
Relativistic fluid hydrodynamics, organized as an effective field
theory in the velocity gradients, has zero radius of convergence due
to the presence of non-hydrodynamic excitations.  Likewise, the theory of
elasticity of brittle solids, organized as an effective field theory in
the strain gradients, has zero radius of convergence due to the
process of the thermal nucleation of cracks.  Viscoelastic materials
share properties of both fluids and solids. We use holographic gauge
theory/gravity correspondence to study all order hydrodynamics of
relativistic viscoelastic media.
}

\makepapertitle

\body

\version\versionno
\tableofcontents

\section{Introduction}\label{intro}
The applications of the gauge theory/string theory correspondence \cite{Maldacena:1997re,Aharony:1999ti} towards non-equilibrium dynamics
of relativistic plasma led to a modern perspective on hydrodynamics \cite{Baier:2007ix,Bhattacharyya:2008jc,Romatschke:2009kr}
as an effective field theory (EFT) where the role of 
higher-dimensional operators and their coupling constants is being played by higher-order gradient combinations of the
local fluid velocity field $u^\mu$, in irreducible representations of the symmetry group of the theory,
and the transport coefficients correspondingly. For example, in the case of uncharged conformal fluids
in background metric $g^{\mu\nu}$, the local stress-energy
tensor $T^{\mu\nu}$ takes the form \cite{Baier:2007ix}
\begin{equation}
T^{\mu\nu}=\e\ u^\mu u^\nu+ P(\e)\ \Delta^{\mu\nu}+\sum_{i=1}^\infty \Pi^{\mu\nu}_{(i)}\,,\qquad \Delta^{\mu\nu}=g^{\mu\nu}+u^{\mu} u^{\nu},
\eqlabel{tmunuder}
\end{equation}
where $\e$ and $P(\e)$ are the local energy density and the pressure, and the $\Pi^{\mu\nu}_{(i)}$ collects all operators involving
a total of $i$ derivatives of the local velocity and the metric. Explicitly,
\begin{itemize}
\item for $i=1$:
\begin{equation}
\begin{split}
&\Pi_{(1)}=\eta(\e)\ \calo_{1,1}\\
&\calo_{1,1}^{\mu\nu}=-\Delta^{\mu\a}\Delta^{\nu\b}\left(\nabla^\a u^\b+\nabla^\b u^\a\right)+\frac 23\Delta^{\mu\nu}\ (\nabla\cdot u)\,,\qquad
\end{split}
\eqlabel{pi1}
\end{equation}
where $\eta$ is the shear viscosity;
\item for $i=2$:
\begin{equation}
\begin{split}
&\Pi_{(2)}=\eta \tau_\Pi\ \calo_{2,1}+\kappa\ \calo_{2,2}+\l_1\ \calo_{2,3}+\l_2\ \calo_{2,4}+\l_3\ \calo_{2,5}\\
&\calo_{2,1}=\biggl[\calo_{1}-\calo_2-\frac 12\calo_3-2\calo_5\biggr]\,,\qquad \calo_{2,2\cdots 5}=\calo_{2\cdots 5}
\end{split}
\eqlabel{pi2}
\end{equation}
where $\tau_\Pi$ is the shear relaxation time, and $\kappa$ and $\l_{1\cdots 3}$ are the four additional second-order transport coefficients.
See eq.~(3.7) in \cite{Baier:2007ix}  for the definition of $\calo_{1\cdots 5}^{\mu\nu}$.
\end{itemize}
For non-conformal relativistic fluids there is an additional first-order in the gradients operator, $\calo_{1,2}$,
\begin{equation}
\Pi_{(1)}=\eta(\e)\ \calo_{1,1}+\zeta(\e)\ \calo_{1,2}\,,\qquad \calo_{1,2}^{\mu\nu}=-\Delta^{\mu\nu}\ (\nabla\cdot u)
\eqlabel{o12}
\end{equation}
with the coupling constant being the bulk viscosity $\zeta$, and $15$ second-order in the gradients operators with $10$ independent
transport coefficients \cite{Romatschke:2009kr,Bhattacharyya:2012nq}. A substantial effort of the community\footnote{We will not review this progress here.}
was dedicated to the computation of the transport coefficients from the first principles, using the framework of the holographic correspondence. 

While it was clear that from the gauge theory/gravity correspondence it is possible in
principle to extend the program of
\cite{Baier:2007ix,Romatschke:2009kr},
to arbitrary order in the velocity gradients, it was not until the work of \cite{Heller:2013fn}
when this was practically implemented for the boost-invariant flow of $\caln=4$ supersymmetric Yang-Mills (SYM)
theory. Using the first 240 terms of the gradient expansion, the authors observed
a factorial growth of gradient contributions at large orders, which indicated a zero radius of convergence of
the hydrodynamic series. Furthermore, they identified the leading singularity in the Borel transform of the hydrodynamic energy density
with the lowest nonhydrodynamic excitation corresponding to a `nonhydrodynamic' quasinormal mode on the gravity side. 

A related conclusion was reached in the analysis of the nonlinear elastic theory of
brittle materials \cite{1996PhRvL..77.1520B,1997PhRvE..55.7669B}. Consider a 2D brittle material with a Young's modulus $Y$, a Poisson ratio $\sigma$, and a
'brittle crack surface tension' $\a$. At a finite temperature $T$, due to thermal nucleation of cracks,
the free energy  $\calf$ of the material subject to the external uniform tension $P$ (negative compression)  
develops an essential singularity,
\begin{equation}
\Im \calf(P)=\frac{2 T}{|P|}\ \left(\frac{T Y}{\l^2(1-\sigma^2)}\right)^{1/2} \left(2\pi \frac{A}{\l^2}\right)\
\exp\left\{-\frac{4Y\a^2}{\pi T P^2(1-\sigma^2)}\right\}\,,\qquad |P|\ll Y
\eqlabel{fcrack}
\end{equation}
where $A$ is the elastic material area and $\l$ is the ultraviolet cutoff in the theory (roughly, the interatomic distance). 
Assuming that the free energy is an analytic function in the complex $P$ plane except for a branch cut along $P\in (-\infty,0]$,
a Cauchy representation for \eqref{fcrack} implies that the power series for the inverse bulk modulus
\begin{equation}
\frac{1}{K(P)}\equiv -\frac 1V \left(\frac{\del V}{\del P}\right)_T=-\frac{1}{PA}\left(\frac{\del F}{\del P}\right)_T=\sum_{n=0}^\infty c_n P^n
\eqlabel{defk}
\end{equation}
is an asymptotic expansion with
\begin{equation}
\frac{c_{n+1}}{c_n}\to -n^{1/2}\ \left(\frac{\pi T(1-\sigma^2)}{8Y\a^2}\right)^{1/2}\,,\qquad {\rm as}\qquad n\to\infty
\eqlabel{defcnratio}
\end{equation}
\ie, the high-order terms $c_n$ in the perturbative expansion for the inverse bulk modulus roughly grow as $(n/2)!$. For brittle
3D elastic materials the scaling is $c_{n+1}/c_n\sim n^{1/4}$ as $n\to \infty$. 
We should stress that unlike the boost-invariant $\caln=4$ SYM conformal relativistic hydrodynamics,
where the higher-order terms were explicitly calculated and the essential
singularity in the Borel-resumed expression was identified with the non-perturbative effects (the nonhydrodynamic modes in plasma),
in the nonlinear elastic theory,  the brittle non-perturbative effects (cracks) were conjectured to imply the
zero radius of convergence of the perturbative in strain expansion.

Fluids and solids are different. Most notably, fluids lack the shear elastic modulus, and as a result
the transverse sound modes in fluids are non-propagating (purely dissipative). Viscoelastic materials are rather interesting
as they share both the properties of fluids and solids: they flow (and thus one can formulate for them the hydrodynamic theory) and they
also exhibit  elastic characteristics when undergoing deformation. Recent advances in engineering holographic viscoelastic materials
\cite{Alberte:2017cch,Alberte:2017oqx,Esposito:2017qpj,Amoretti:2017frz,Grozdanov:2018ewh} opened a possibility to explore large-order in the velocity gradients
hydrodynamics of materials with a control parameter that smoothly interpolates between the fluids and solids.
This is precisely the focus of the paper: following the set-up of \cite{Buchel:2016cbj,Buchel:2018ttd} we
study the asymptotic properties of viscoelastic hydrodynamics for the holographic models inspired by
\cite{Amoretti:2017frz} and the earlier work \cite{Donos:2013eha}. \textit{En passant} we analyze, within linear response, the viscoelastic properties of the model and in particular the elastic moduli and the viscosities and we make contact with the results from the QNMs spectrum.

In the next section \ref{model} we introduce the holographic model describing the viscoelastic media. Translational invariance is explicitly broken as in \cite{Donos:2013eha}. Typical holographic observables,
the one- and two-point correlation functions, 'feel' this explicit symmetry breaking through a messenger sector
experiencing explicit or spontaneous flavor symmetry breaking. Next, we discuss the thermodynamics of the model in section \ref{thermodynamics}. We follow in section \ref{elastic} highlighting the elastic features of
our holographic media. In section \ref{hydro} we  discuss all-order hydrodynamics of the
model undergoing homogeneous and isotropic expansion. We conclude in section \ref{conclude}. We provide more details about the model and the computations in the appendices \ref{App1}, \ref{App3}, \ref{App4}.

\section{The model}\label{model}

Our model realizes the idea of holographic $Q$-lattices introduced in \cite{Donos:2013eha}.
It shares the viscoelastic properties of the related models \cite{Amoretti:2017frz,Baggioli:2014roa,Alberte:2017cch,
Alberte:2017oqx,Alberte:2015isw,Alberte:2016xja,Andrade:2017cnc,Baggioli:2015gsa,
Esposito:2017qpj,Grozdanov:2018ewh}.

Our effective action is defined by\footnote{We do not have a string theory embedding of the model. As a result, the boundary interpretation
of a higher-derivative bulk coupling $\l_2$ is unclear.} 
\begin{equation}
\begin{split}
S=\frac{1}{16\pi G_N}\int_{\calm_{5}}d^5 x \sqrt{-g} \biggl[&R+12-\frac12 (\del\phi)^2-\frac 14 (1+\g \phi^2)\
F^2+\frac{\Delta(4-\Delta)}{2}\ \phi^2\\
&-\frac {1}{2}\phi^2\ \sum_{I=1}^3 \biggl\{\l_1 (\del\psi_I)^2+\l_2  \left((\del\psi_I)^2\right)^2\biggr\}\biggr]
\end{split}
\eqlabel{action}
\end{equation}
where $\Delta$ is the scaling dimension of the boundary operator dual to the bulk field $\phi$ and the bulk coupling constants satisfy $\l_i>0, \gamma>0$. The action \eqref{action} can be thought as a generalization of an effective Maxwell-Einstein-Hilbert action with three complex scalar fields with $SU(3)$ symmetry (see App.\ref{App1} for more details about it). 

In this paper we discuss states of the holographic viscoelastic media \eqref{action}
with explicit breaking of the translational invariance, \ie we turn on the non-normalizable components of the
'axions'\footnote{The fields $\psi^I$ are genuine axions only when $\lambda_1=1$ and $\lambda_2=0$. Thinking about these
fields as axions allows for an intuitive understanding why the holographic media shares viscoelastic properties. As we demonstrate explicitly later, the media enjoys nonzero shear elastic modulus for all $\{\lambda_1,\lambda_2\}$.} $\psi_I$:
\begin{equation}
\psi_I=k \,\dd_I^J\, x_J\,,\qquad k > 0
\eqlabel{defaxion}
\end{equation}
where $x_{J=1\cdots 3}$ are the spatial directions of the boundary gauge theory. We retain the $SO(3)$ invariance in the spatial coordinates. As a definition of the model, we assume the periodicity of
$\psi_I$ in the field space as in eq.\eqref{defpsi}\\ 
The ansatz \eqref{defaxion} implies that all the spatial coordinates are
periodically identified:
\begin{equation}
x_J \sim x_J+\frac{2\pi}{k\sqrt{3}} 
\eqlabel{defaxion2}
\end{equation}
thus, we have a cubic spatial lattice. The remaining gravitational fields are taken to depend only
on the $AdS$ radial coordinate $x$ and the time $t$:
\begin{equation}
\begin{split}
&ds^2=-2dt \left(\frac{dx}{x^2}+A(t,x)\ dt\right)+\Sigma(t,x)^2\ d\boldsymbol{x}^2\,,\qquad
d\boldsymbol{x}^2=\sum_{J=1}^3 (dx_J)^2\\
& A_\mu=\dd_\mu^t\ a_0(t,x)\,,\qquad \phi=\phi(t,x)
\end{split}
\eqlabel{ansatz}
\end{equation}
The AdS boundary is at $x\to 0_+$, and the background metric of the dual gauge theory is taken
as
\begin{equation}
ds_4^2=-dt^2+a(t)^2\ d\boldsymbol{x}^2\,,\qquad \Longrightarrow\qquad  \lim_{x\to 0_+} x\, \Sigma(t,x) = a(t)
\eqlabel{flrw}
\end{equation}
The equations of motions obtained from \eqref{action} are shown in Appendix \ref{App1}.

Generically, the geometry \eqref{ansatz} will have an apparent horizon $x_{AH}$, located at 
\begin{equation}
d_+\Sigma\bigg|_{x=x_{AH}}=0
\eqlabel{defahg}
\end{equation}
 see \cite{Chesler:2013lia}. Following \cite{Booth:2005qc,Figueras:2009iu} we associate the non-equilibrium entropy density 
$\cals$ with the Bekenstein-Hawking entropy of the apparent horizon (AH) in the
geometry\footnote{The location of the AH, and hence the definition of the non-equilibrium entropy, is not unique.
Here, the choice of the slicing --- fixed radial coordinate $x$ in \eqref{ansatz} --- is motivated by the
spatial symmetry of the boundary non-equilibrium states considered. Such a definition correctly reproduces the
hydrodynamic limit (see \cite{Buchel:2016cbj}), and guarantees that the comoving entropy production 
rate \eqref{erate} is non-negative, see \cite{Buchel:2017pto}.},  
\begin{equation}
a(t)^3 \cals= \frac{\Sigma^3}{4G_N} \bigg|_{x=x_{AH}}
\eqlabel{noneqs}
\end{equation}
Taking the derivative of the entropy density and using the holographic equations of motion we find
\begin{equation}
\frac{d(a^3 \cals)}{dt}=\frac{2 x^2 (\Sigma^3)' (d_+\phi)^2}{
(a_0')^2 (1+\gamma \phi^2) x^4
+\phi^2 \left(\Delta (\Delta-4)+\frac{3\l_1 k^2}{\Sigma^2}+\frac{3 \l_2 k^4}{\Sigma^4}\right)-24}
\eqlabel{erate}
\end{equation}

We conclude this section with the following observations.
\begin{itemize}
\item Our system admits a general solution describing the homogeneous and isotropic expansion of a plasma in
the generic background metric \eqref{flrw}:
\begin{equation}
\begin{split}
&\phi=0\,,\qquad \Sigma=\frac {a(t)}{x}\,,\qquad a_0=\frac{Q(x_h^2a(t)^2-x^2)}{a(t)^3}\,,\\
&A=\frac{1}{2x^2}-\frac{\dot a(t)}{xa(t)}+\frac{x^4Q^2}{6a(t)^6}-\frac{x^2}{a(t)^4}\left(\frac 16 x_h^2 Q^2+\frac{1}{2x_h^4}\right)
\end{split}
\eqlabel{rnflrw}
\end{equation}
From \eqref{defahg}, the dynamical (apparent) horizon $x_{AH}$ is located at
\begin{equation}
x_{AH}=x_h a(t)
\eqlabel{defah}
\end{equation}
The corresponding (dynamical) thermal characteristics of the state are
\begin{equation}
\begin{split}
&\mu=\frac{Qx_h^2}{a(t)}\,,\qquad \calq= \frac{Q}{a(t)^3}\,,\qquad T=\frac{1}{a(t)}\left(\frac{1}{\pi x_h }
-\frac{Q^2x_h^5}{6\pi }\right)\\
& \cals=\frac{2\pi}{x_h^3 a(t)^3}\,,\qquad
\cale=\frac{1}{a(t)^4}\left(\frac{3}{2x_h^4}+\frac{Q^2x_h^2}{2}\right)+\frac{3{\dot a}(t)^4}{a(t)^4}\\
&\calp=\frac 13 \cale  -\frac{4\dot a(t)^2\ddot a(t)}{a(t)^3}
\end{split}
\eqlabel{thermods}
\end{equation}
where $\mu$ is the chemical potential, $\calq$ the charge density, $T$ the temperature, $\cals$ the entropy density, $\cale$ the energy density and finally $\calp$ the pressure.\\
Notice that there is no comoving entropy production:
\begin{equation}
\frac {d}{dt}\left\{ a(t)^3 \cals \right\}=0
\eqlabel{comoving}
\end{equation}
This, along with the conformal anomaly contributions to the energy density and the pressure  in \eqref{thermods},
expresses the statement that the  state \eqref{rnflrw} is a Weyl transform of the thermal equilibrium state with $a(t)=1$,
see \cite{Buchel:2017pto}. To avoid unnecessary cluttering of the formulas we set $8\pi G_N=1$ henceforth. 
\item In the static limit $a(t)=1$ we obtain the familiar $AdS_5$ Reissner-Nordstr\"om black brane solution where $x_h$ is the position of the horizon and $Q$ the charge density of the dual theory.
\item Despite the fact that translational invariance is broken explicitly due to \eqref{defaxion},
the thermal equilibrium states of our viscoelastic media \eqref{rnflrw} do not feel this breaking
because the messenger field $\phi$ vanishes identically. In this case the discrete $\zet_2$ symmetry
$\phi \leftrightarrow -\phi$ is unbroken.
Asymptotically near the boundary\footnote{When $\Delta$ is an integer the asymptotic expansion includes the $\ln x$ terms.}, we have
\begin{equation}
\phi=\dd_\Delta\ x^{4-\Delta} \left[1+\cdots\right]+ f_\Delta(t)\ x^\Delta \left[1+\cdots\right]\,,\qquad 2\le\Delta < 4
\end{equation}
where a constant $\dd_\Delta$ is a source term (coupling constant of the dual operator), and $f_\Delta$ is its expectation
value. When $\dd_\Delta\ne 0$, the parity $\zet_2$ symmetry is explicitly broken; it can also be spontaneously broken when
$\dd_\Delta=0$ provided the energy density of the state with $\calq\ne 0$ is sufficiently small. Whenever this $\zet_2$ symmetry
is broken, the one-point correlation function of the stress-energy tensor is sensitive to the explicit
translational invariance breaking \eqref{defaxion}.
\end{itemize}

\section{Phase diagram of the holographic viscoelastic media}\label{thermodynamics}

While it is straightforward to analyze the phase diagram of the equilibrium thermal states of the
viscoelastic media holographically represented by \eqref{action}, the task is  daunting  given
the dimensionality of the parameter space of the model: the bulk couplings $\{\g,\l_1,\l_2\}$, the lattice spacing
$\Delta x\equiv 2\pi/(k\sqrt{3})$, the scaling dimension $\Delta$ of the relevant operator $\calo_\phi$, and the boundary
conformal symmetry breaking coupling constant $\dd_\Delta$. These parameters have to be supplemented
by the temperature $T$ and the chemical potential $\mu$ (for a grand canonical ensemble equilibrium states),
or the energy density $\cale$ and the charge density $\calq$ (for a microcanonical ensemble equilibrium states).
In what follows, we mostly restrict the bulk parameters of the holographic dual to the choice
\begin{equation}
\Delta=\{2,3\}\,,\qquad \{\g,\l_1,\l_2\}=\{1,0,1\} \label{benchmark}
\end{equation}
The lattice spacing $\Delta x$ is allowed to vary to interpolate from the more-fluid-like behavior $k\ll T$ to the more-solid-like behavior
$k\gg T$ of our viscoelastic media.

We begin in section \ref{bfology} discussing the linearized stability of the $\zet_2$-invariant RN state (\eqref{rnflrw} with $a(t)=1$) at extremality ($T=0$).
We continue with the linearized stability analysis in the microcanonical and grand canonical ensembles and determine
the critical energy density and temperature. We discuss the fully non-linear phase diagram of the $\Delta=2$ and $\dd_2=0$ model
in section \ref{nonlinear}.

\subsection{BFology and critical phenomena}\label{bfology}

The $\zet_2$ invariant thermal equilibrium states of the viscoelastic media
represented by the bulk geometry \eqref{rnflrw} in the static limit $a(t)=1$, do not feel the explicit translational
invariance breaking \eqref{defaxion}. Following \cite{Hartnoll:2008kx}, we expect that close to the extremality,
these states with $Q\ne 0$ will become unstable to spontaneous $\zet_2$ symmetry breaking due to the
condensation of $\phi$. Of course, the onset of the instability is sensitive to $\Delta x$ (or $k$),
as well as to the bulk couplings $\{\g,\l_1,\l_2\}$. Indeed, the linearized $\phi$-fluctuations about the RN background satisfy 
\begin{equation}
\begin{split}
0=&\del_{\t z}^2\phi-\frac{(z^2-1)(q^2z^4-3z^2-3)}{6}\ \del^2_{zz}\phi-
\frac{3q^2z^6-q^2z^4-3z^4-9}{6z}\ \del_z \phi\\
&-\frac {3}{2z}\ \del_\t \phi+\frac{m_{eff}^2}{2  z^2}\ \phi\,,\\
m_{eff}^2\equiv &\Delta(\Delta-4)-z^2(4\gamma q^2 z^4-3\lambda_2 s^4 z^2-3 \lambda_1 s^2)\,,\qquad \phi=\phi(\t,z)
\end{split}
\eqlabel{phieom}
\end{equation}
where we introduced 
\begin{equation}
\t\equiv\frac{t}{x_h}\,,\qquad z\equiv \frac{x}{x_h}\in [0,1]\,,\qquad Q=\frac{q}{x_h^3}\,,\qquad k=\frac{s}{x_h}
\eqlabel{defzqs}
\end{equation}

\begin{figure}[t]
\begin{center}
  \includegraphics[width=2.75in]{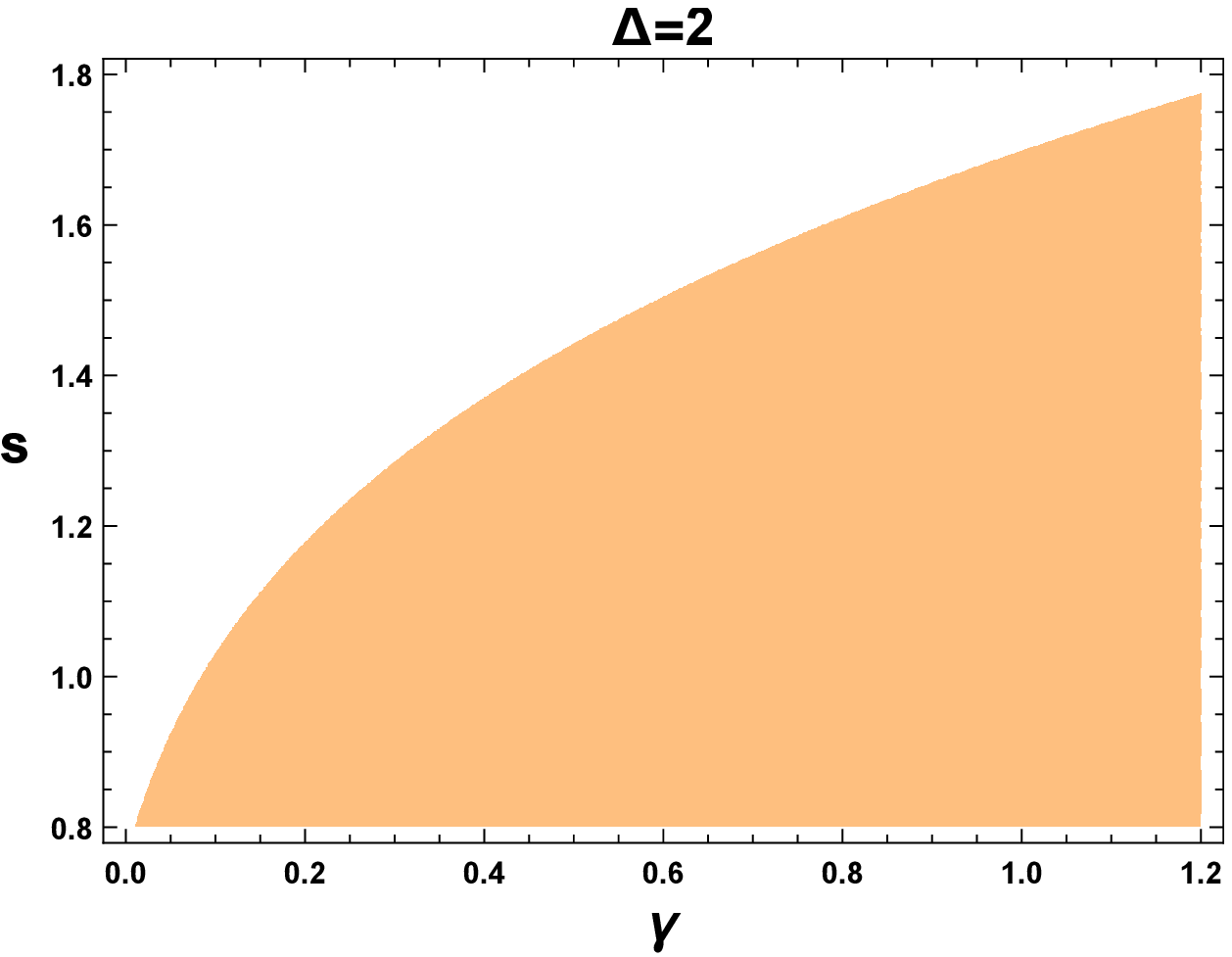}\quad
  \includegraphics[width=2.75in]{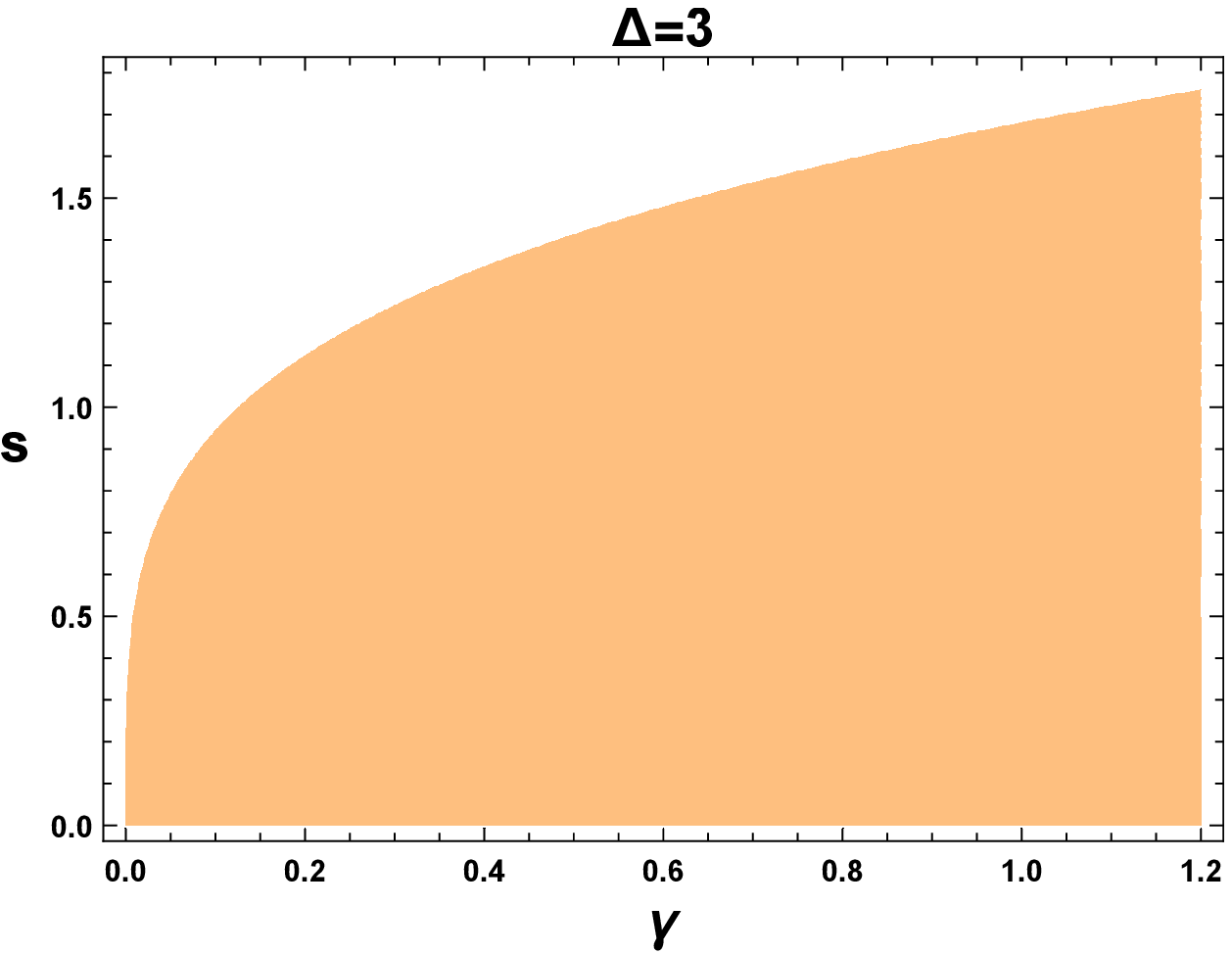}
\end{center}
 \caption{Instability of the $\zet_2$ symmetric phase (the green region)
 of the holographic viscoelastic media for the choice of parameters \eqref{benchmark}, see \eqref{bfbound}.
}\label{figure1}
\end{figure}

In the next section we determine the precise critical energy and temperature for the presence of the normalizable mode of
\eqref{defzqs}. For now, we determine the boundaries of the gravitational bulk parameter space region that would
guarantee the onset of the instability at the extremality of the $\zet_2$ symmetric solution. This ``BFology'' analysis
is done applying the Breitenlohner-Freedman (BF) bound to $\phi$ scalar in the infrared, \ie $z\to 1$, of the extremal RN
solution which can be achieved by setting $q\to \sqrt{6}_-$. In the latter case the
near horizon (IR) geometry is that of $AdS_2\times R^3$ with the radius of the AdS being
\begin{equation}
L_2^2=\frac{L^2}{12}=\frac{1}{12}
\eqlabel{ads2}
\end{equation}
and the effective IR mass of $\phi$:
\begin{equation}
m_{AdS_2}^2=m_{eff}^2\bigg|_{\{z=1,q=\sqrt{6}\}}=\Delta\,(\Delta-4)-24\g+3\,\lambda_2\, s^4+3\,\lambda_1 \,s^2
\eqlabel{mads2}
\end{equation}
The BF bound implying the onset of the instability reads
\begin{equation}
m_{AdS_2}^2 L_2^2< -\frac 14\qquad \Longleftrightarrow\qquad \biggl(\Delta\,(\Delta-4)-24\,\g+3\,\lambda_2 \,s^4+3\,\lambda_1 \,s^2+3\biggr)<0
\eqlabel{bfbound}
\end{equation}
Some examples of regions of the gravitational bulk parameter space leading to spontaneous breaking of $\zet_2$
symmetry are shown in fig.~\ref{figure1}. Since $\lambda_i>0$ increasing the (dimensionless) lattice spacing $s$ makes the instability more difficult to appear. On the contrary, the coupling $\g$ enhances the instability.

We move now to identify critical points for the onset of the $\zet_2$ symmetry breaking instability in microcanonical and grand canonical ensembles for our benchmark model \eqref{benchmark}.  We allow the lattice spacing $\Delta x$ to
vary. To this end, we need to find the normalizable solutions of \eqref{phieom} at the threshold of instability:
\begin{equation}
\phi\equiv \phi(z)\,,\qquad \phi=z^\Delta\left(1+\calo(z^4)\right)\,,\qquad \phi=f_{\Delta,0}^h+\calo((1-z))
\eqlabel{bcphi}
\end{equation}
where we exploited the linearity of \eqref{phieom} to fix the normalizable coefficient at the boundary to $1$.

\begin{figure}[t]
\begin{center}
  \includegraphics[width=2.875in]{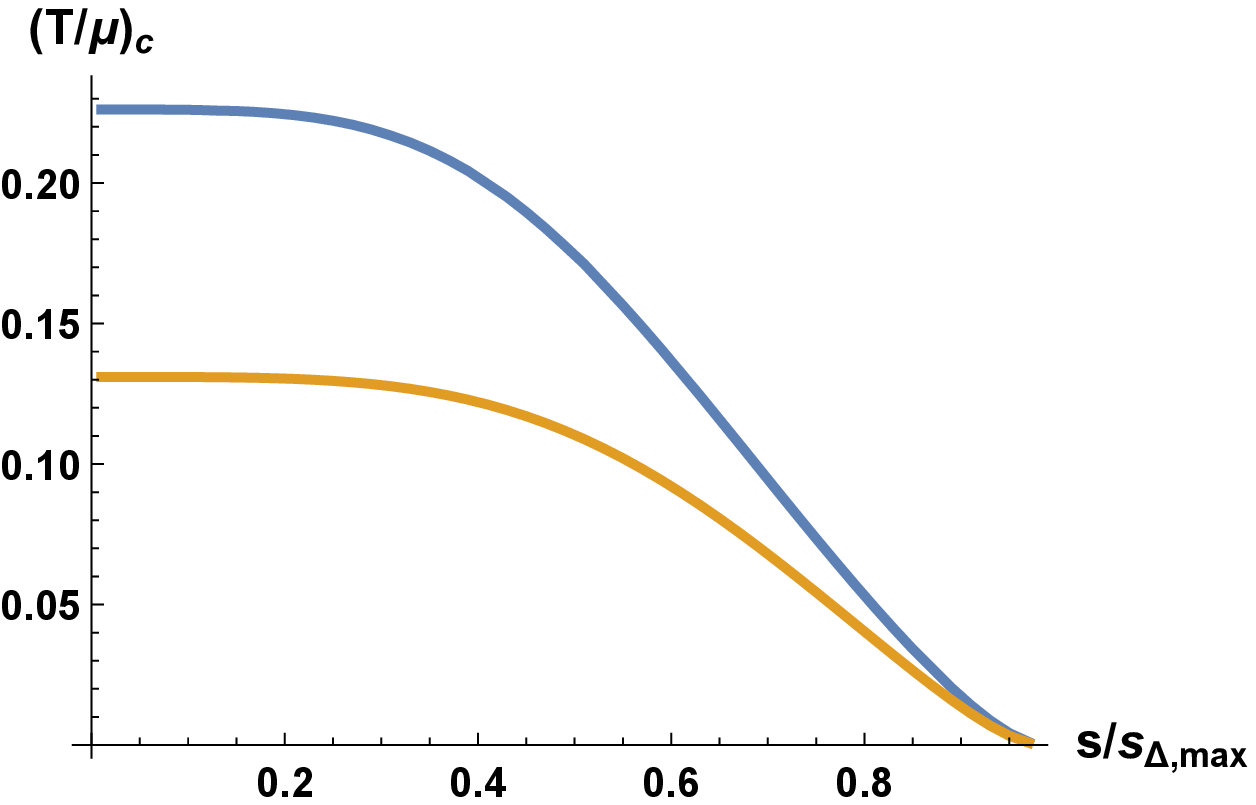}\quad
  \includegraphics[width=3.05in]{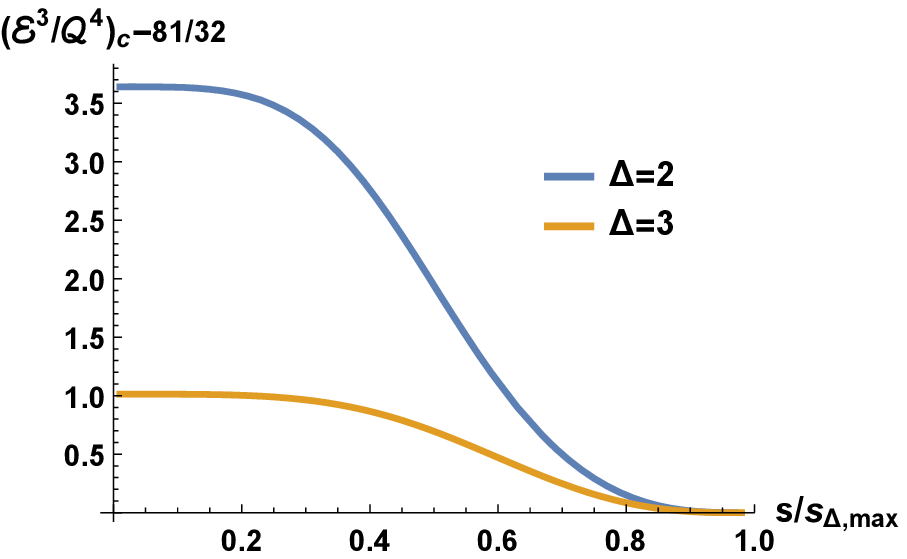}
\end{center}
 \caption{Critical value of $T/\mu$ and $\frac{\cale^3}{\calq^4}$ below which $\zet_2$ symmetry
 of the RN phase  is spontaneously broken. $\frac{\cale^3}{\calq^4}|_{extremal}=\frac{81}{32}$ and 
 $s_{\Delta,max}$ is the maximal value of the
 bulk parameter $s$  \eqref{defzqs}, determined by \eqref{maxs}.  The blue and orange curves refer to $\Delta=2,3$.   
 }\label{figure3}
\end{figure}

Numerical results for the onset of the instability are presented in fig.~\ref{figure3} (in both microcanonical and grand canonical ensembles). The onset is pushed to the extremality once $s$ reaches
the boundary of the instability region,  see \eqref{bfbound}:
\begin{equation}
s_{\Delta,max}=\left(7+\frac{\Delta(4-\Delta)}{3}\right)^{1/4}=\begin{cases}
&\frac{5^{1/2}}{3^{1/4}}\,,\qquad \Delta=2\\
&8^{1/4}\,,\qquad \Delta=3
\end{cases}
\eqlabel{maxs}
\end{equation}

\subsection{Spontaneous symmetry breaking and phase diagram}\label{nonlinear}

In the previous section we discussed the critical phenomena  in the $\zet_2$-invariant sector of the model 
\eqref{action} both in microcanonical and grand canonical ensembles, due to the presence of $\Delta=\{2,3\}$
operators. The onset of the instabilities identifies a bifurcation point on the thermal phase diagram of the 
system with a new $\zet_2$-broken phase branching off the $\zet_2$-symmetric phase. The symmetry broken
phase, when it exists: 
\nxt (a) dominates the symmetric phase both in microcanonical and grand canonical ensembles (as in \cite{Hartnoll:2008kx})
\nxt (b) does not dominate the symmetric phase in either ensembles (as in some models in \cite{Buchel:2009ge,Bosch:2017ccw})
\nxt (c) dominates the symmetric phase in microcanonical ensemble, but not in a canonical ensemble
(as in some models in \cite{Donos:2011ut,Buchel:2017map})

\begin{figure}[t]
\begin{center}
  \includegraphics[width=2.85in]{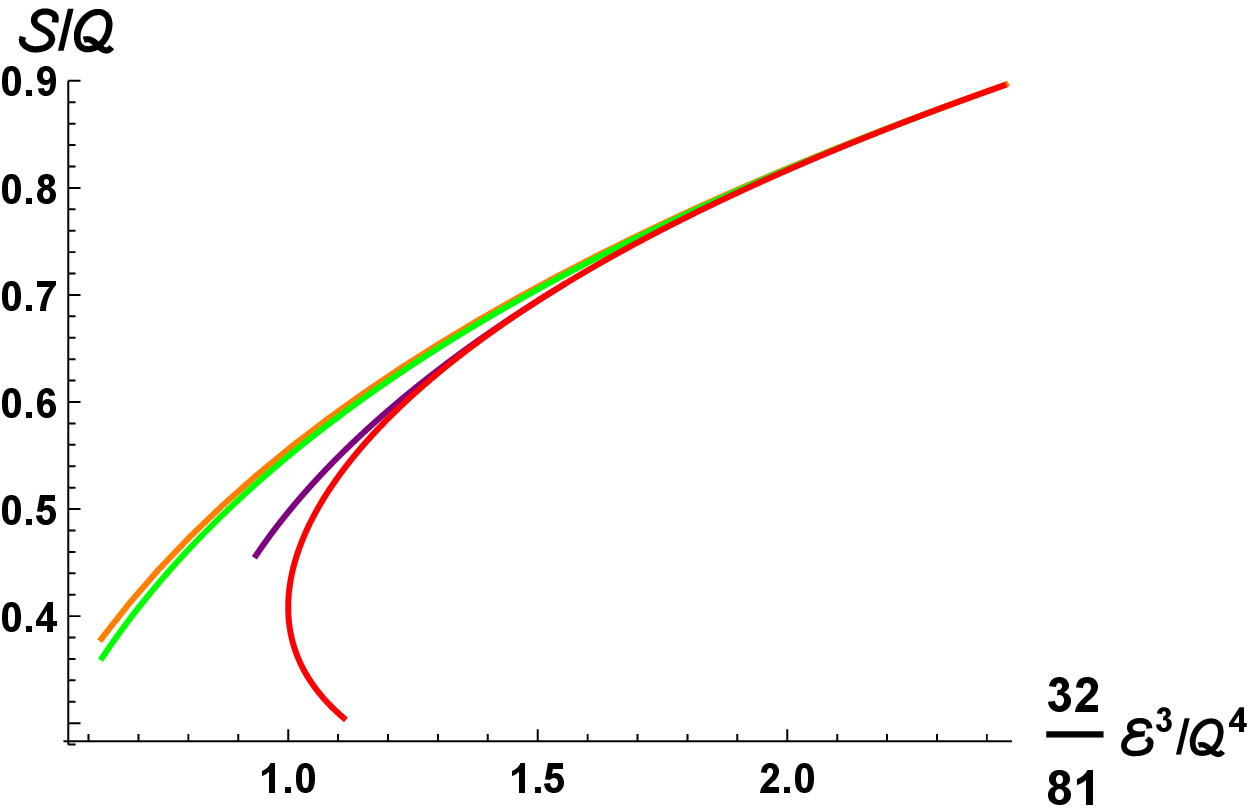}\quad
  \includegraphics[width=2.9in]{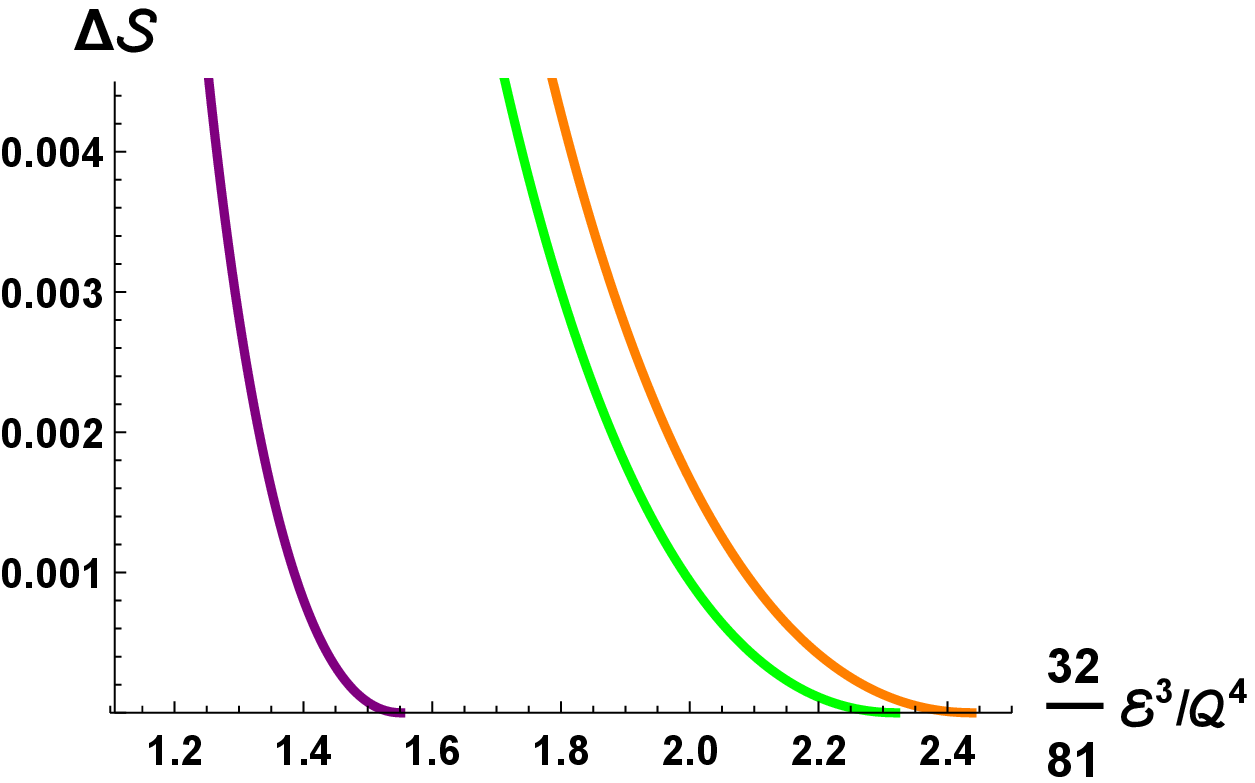}
\end{center}
 \caption{Phase diagram of the holographic viscoelastic media with spontaneous
 breaking of  $\zet_2$ symmetry due to a dimension $\Delta=2$ operator. The red curve
 is $\zet_2$ symmetric phase; the $\{$orange,green,purple$\}$ curves 
 represent $\zet_2$-broken phases with $s/s_{2,max}=\{0,\frac 13,\frac 12\}$. The right panel enhances the
 entropy differences between the symmetry broken and the symmetric phases at fixed energy density $\cale$
 and the charge density $\calq$.} \label{figure5} 
\end{figure}

In our model the case (a) is realized. In fig.~\ref{figure5} we present the phase diagram of the model
\eqref{action} with $\{\g,\l_1,\l_2\}=\{1,0,1\}$, $\Delta=2$ and $\dd_2=0$ in microcanonical
ensemble\footnote{The story in  grand canonical ensemble is equivalent.}.  The left panel
represent the dimensionless ratio $\frac{\cale^3}{\calq^4}$ of the energy density $\cale$ and the charge density $\calq$,
normalized to its extremal value $\frac{\cale^3}{\calq^4}|_{extremal}=\frac{81}{32}$, versus
the dimensionless ratio $\frac{\cals}{\calq}$ involving the entropy density $\cals$.
The red curve is the equilibrium $\zet_2$ symmetric phase, and the $\{$orange,green,purple$\}$ curves 
 represent $\zet_2$-broken phases with $s/s_{2,max}=\{0,\frac 13,\frac 12\}$.
The right panel show the entropy differences between the symmetry broken and the symmetric phases. 
The onset of the instabilities
 (the location of the bifurcation points) agree to better than
 $1.5\times 10^{-3}\%$ with computations presented in
 section \ref{bfology}. 
The analysis used to determined the phase diagram are standard and we will not detail them here. 
 
\section{Viscoelastic properties}\label{elastic}

It is convenient to study elastic properties of the model \eqref{action} in
Fefferman-Graham  (FG) coordinates system. Thus, we assume the following gravitational ansatz for the equilibrium states:
\begin{equation}
ds^2=-A\ dt^2 +B\ dx^2 + C\ d\boldsymbol{x}^2 \,,\qquad A_\mu=\dd_\mu^t\ a_0\,,\qquad \psi_I=k\dd_I^J x_J
\eqlabel{fgmetric}
 \end{equation}
where $\{A,B,C,a_0\}$ and $\phi$ are functions of the radial coordinate only.
The corresponding EOMs are given in Appendix \ref{App1}.

To determine the retarded correlation function $\calg_{x_1x_2,x_1x_2}^R(\w)$ that encodes the shear elastic modulus
and the shear viscosity of the media \cite{PhysRevA.6.2401} we study the following time-dependent metric fluctuation
\begin{equation}
ds^2\to ds^2+ 2 h(t,x) C(x)\ dx_1dx_2
\eqlabel{scalarceom}
\end{equation}
where $C$ is an unperturbed metric warp factor, see \eqref{fgmetric}.
It is straightforward to verify that fluctuations \eqref{scalarceom} decouple from all the
other fluctuations at the linearized level. Assuming the usual Fourier expansion $h=e^{-i \w t} H(x)$ we find
\begin{equation}
0=H''+\frac 12 H' \left( \ln \frac{AC^3}{B}\right)'
+H \left(\frac{B \w^2}{A}-\frac{B \phi^2 k^2}{C^2} \left(2 k^2 \lambda_2+C \lambda_1\right)\right)
\eqlabel{heom}
\end{equation}
We look for a solution of \eqref{heom} normalized to $1$ near the $AdS_5$ boundary, $x\to 0$,
and representing an incoming wave at the horizon, $x\to x_h$.  
From the near-boundary asymptotic expansion  of the solution,
\begin{equation}
H=1+h_4^b(\w)\ x^4+\calo\left(x^{5}\right)
\eqlabel{expaH}
\end{equation}
the corresponding  retarded correlation function reads
\begin{equation}
\calg_{x_1x_2,x_1x_2}^R(\w)=- \frac{h_4^b(\w)}{4\pi G_N}
\eqlabel{corrfunc}
\end{equation}
and it determines the shear elastic modulus $G$ and the shear
viscosity\footnote{In presence of the spontaneous/explicit breaking of translational invariance the definition of the viscosity, its Kubo formula and its relation with the momentum diffusion constant become subtle \cite{Alberte:2016xja,Burikham:2016roo,Ciobanu:2017fef}. Nevertheless the usual expression in terms of Kubo formula still has a clear physical interpretation as the rate of entropy production due to a strain \cite{Hartnoll:2016tri}. For completeness let us just summarize the main results of \cite{Hartnoll:2016tri}. Given an external strain source $\delta g_{xy}= t \Delta$, where $\Delta$ is a constant, the corresponding entropy production induced by the work of such external source is obtained as:
\begin{equation}
\dot{s}\,=\,\frac{\eta}{T}\,\Delta^2
\end{equation}
which very suggestively can be rewritten as:
\begin{equation}
\frac{1}{T}\,\frac{d \log s}{dt}\,=\,\left(\frac{\eta}{s}\right)\,\left(\frac{\Delta}{T}\right)^2
\end{equation}
Whenever temperature is the only scale in the system it is natural to expect $\Delta/T$ being order unit. Most importantly the previous equation can be used to give a more universal definition of the $\eta/s$ ratio which does not rely on hydrodynamics. Moreover we can rewrite the KSS bound as a bound on the entropy production of the system:
\begin{equation}
t_{pl}\,\frac{d \log s}{d t}\,\geq 1
\end{equation}
where $t_{pl}$ is the Planckian time $t_{pl}\equiv \hbar/k_B T$.
} $\eta$ as 
\begin{equation}
G=\lim_{\w\to 0}\Re\left[  \calg_{x_1x_2,x_1x_2}^R(\w)\right]\,,
\qquad \eta=-\lim_{\w\to 0}\ \frac 1\w\ \Im  \left[\calg_{x_1x_2,x_1x_2}^R(\w)\right]
\eqlabel{viscotransport}
\end{equation}

\subsection{Viscoelastic properties to leading order in $\zet_2$
symmetry breaking}

The $\zet_2$ symmetry breaking  of the viscoelastic media state  (either explicit or spontaneous)
is mediated to the gravitational bulk scalar $\phi$. 
In general, to find the shear elastic modulus $G$ and the shear viscosity $\eta$ \eqref{viscotransport}
one has to solve \eqref{heom}
numerically\footnote{We will present an example of this computation in section \ref{fullg}.}. Semi-analytic
computations are possible to leading order in the symmetry breaking, \ie to order $\calo(\phi^2)$. We outline the details of these computations in Appendix \ref{App3}.
In the main text we just collect the final expressions for $\{G,\eta\}$
to leading order in the $\zet_2$ symmetry breaking:
\begin{equation}
\begin{split}
&16\pi G_N\ G=\int_{0}^{x_h}
dx\ \sqrt{\frac{A_{\zet_2}B_{\zet_2}}{C_{\zet_2}}}(\l_1 k^2 C_{\zet_2}+2k^4\l_2)\phi^2\\
&\frac{\eta}{\cals}=\frac{1}{4\pi}\biggl(1-  \int_{0}^{x_h}
dx\ \sqrt{\frac{A_{\zet_2}B_{\zet_2}}{C_{\zet_2}}}(\l_1 k^2 C_{\zet_2}+2k^4\l_2)\phi^2
\biggl\{\int_0^x  dy\ \sqrt{\frac{A_{\zet_2}B_{\zet_2}}{C_{\zet_2}^3}}\biggr\}\biggr) 
\end{split}
\eqlabel{getalead}
\end{equation}
where we used the fact that to this order it is consistent to replace the metric
warp factors with those of the $\zet_2$-symmetric RN solution (see \eqref{rnfg} for details). Notice that with an explicit $\zet_2$ symmetry breaking
$G$ is divergent when $\l_1\ne 0$
and $\Delta\ge 3$; it is divergent as well with $\l_1=0$ and $\l_2\ne 0$ provided $\Delta=4$. 
When $k=0$, the shear elastic modulus vanishes and the shear viscosity attains its universal ratio \cite{Buchel:2003tz}. In other words for $k=0$ the system is a perfect strongly coupled fluid with no elastic properties and the viscosity saturating the KSS bound \cite{Kovtun:2004de}.

\begin{figure}[t]
\begin{center}
  \includegraphics[width=2.8in]{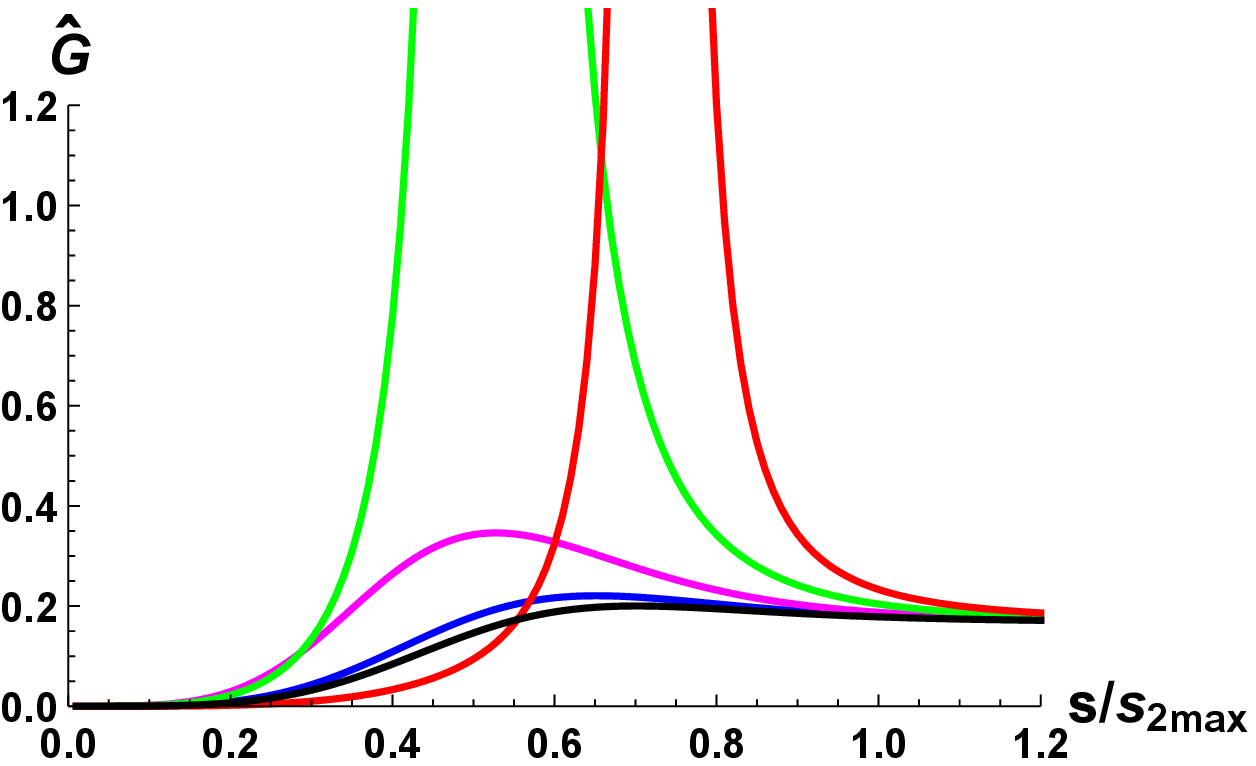}\quad 
  \includegraphics[width=3in]{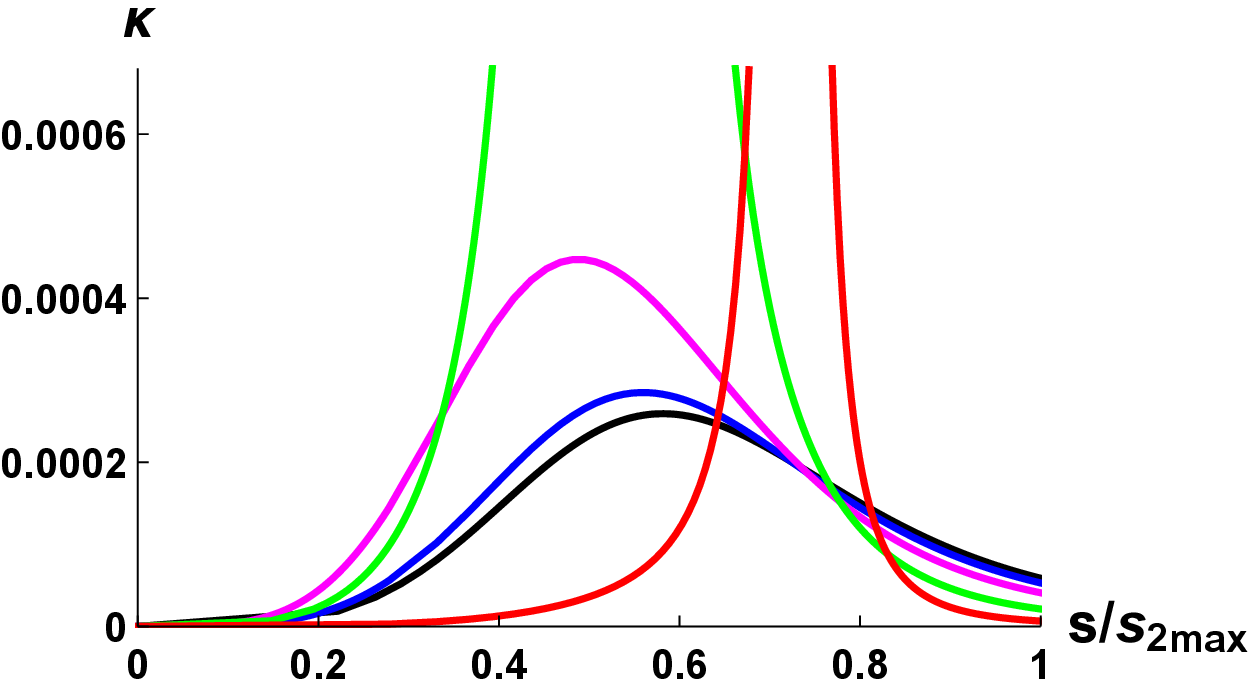}
\end{center}
 \caption{Shear elastic modulus (left panel) and the shear viscosity (right panel)
 of the holographic viscoelastic media to leading order in the explicit $\zet_2$ symmetry breaking
 (see \eqref{getaphi2} for the parametrization). The color coding corresponds to
 different values of the ratio of temperature $T$ to the chemical potential $\mu$:
 $\frac T\mu=\frac {1}{12}$ (red),  $\frac T\mu=\frac {1}{6}$ (green),
  $\frac T\mu=\frac {1}{3}$ (magenta) and  $\frac T\mu=\frac {2}{3}$ (blue).  The former two values
  are below the critical values of ${T}/{\mu}|_{crit}$ at $s=0$, and the latter two are above.
  Black curves correspond to neutral viscoelastic media, \ie $\mu=0$.
 } \label{figure6} 
\end{figure}

\begin{figure}[t]
\begin{center}
  \includegraphics[width=4in]{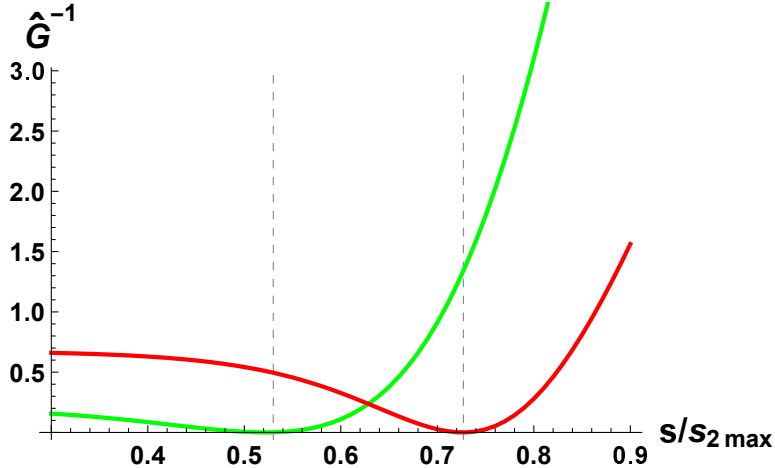}
\end{center}
 \caption{Shear elastic modulus  of the holographic viscoelastic media
 becomes very large (diverges in $\calo(\dd_2^2)$ approximation)
at the critical point.
 The color coding corresponds to
 different values of the ratio of temperature $T$ to the chemical potential $\mu$:
 $\frac T\mu=\frac {1}{12}$ (red),  $\frac T\mu=\frac {1}{6}$ (green) --- both
 below the critical values of ${T}/{\mu}|_{crit}$ at $s=0$. The dashed lines indicate the critical points, in perfect agreement with fig.\ref{figure3}.
 } \label{figure7} 
\end{figure}

 We conclude this section with a sample of results obtained using\footnote{Other cases can be analyzed in
a similar fashion.} \eqref{getalead}. We focus on the
case of an explicit $\zet_2$ symmetry breaking of the model with $\{g,\l_1,\l_2\}=\{1,0,1\}$ and $\Delta=2$. 
It is important to remember that because \eqref{getalead} was obtained  to order $\calo(\dd_\Delta^2)$,
large/divergent results are outside the approximation, and the full nonlinear in the bulk scalar $\phi$
analysis must be performed. We define for simplicity
\begin{equation}
\frac{16\pi G_N\ G }{\dd_2^2}=\hat{G}\,,\qquad \frac \eta \cals= \frac{1}{4\pi}\left(1-\frac{\dd_2^2}{T^4}\ 
\k\right)
\eqlabel{getaphi2}
\end{equation}
The results of the computations (whose details are given in Appendix \ref{App3}) are presented in figs.~\ref{figure6}-\ref{figure7}.
We use $\frac{T}{\mu}=\{\frac {1}{12},\frac16\}$ (red,green) which are below the critical temperature
for the $\zet_2$ spontaneous symmetry breaking (see fig.~\ref{figure1}), and $\frac{T}{\mu}=\{\frac {1}{3},\frac23\}$ (magenta,blue) which are above the critical temperature.
 Black curves correspond to neutral viscoelastic media, \ie $\mu=0$.
Both the shear elastic modulus
and the correction to the shear viscosity become large (divergent in $\calo(\dd_2^2)$ approximation)
at the critical point. This is shown in \ref{figure7} for  $\frac{T}{\mu}=\{\frac {1}{12},\frac16\}$ ---
the zeros in $\hat{G}^{-1}$ agree to better than $1\%$ with the results
reported in fig.~\ref{figure3}. Note that $\kappa>0$ implying the violation of the 
KSS bound\footnote{The violation of the KSS bound in anisotropic setting 
has been observed earlier \cite{Jain:2015txa}.}.

\subsection{Viscoelastic transport of the spontaneously broken $\zet_2$ phase}\label{fullg}

We discuss here the viscoelastic transport (elastic shear and bulk moduli) and the shear viscosity
of the spontaneously broken $\zet_2$ phase in the model \eqref{action}.
We focus on the parameter set $\{\g,\l_1,\l_2\}=\{1,0,1\}$ and $\Delta=2$ ($\dd_2=0$, \textit{i.e.} no explicit source, since
we consider the spontaneous symmetry breaking).

The technical details about the computations are given in Appendix \ref{App4}.

\begin{figure}[t]
\begin{center}
  \includegraphics[width=2.6in]{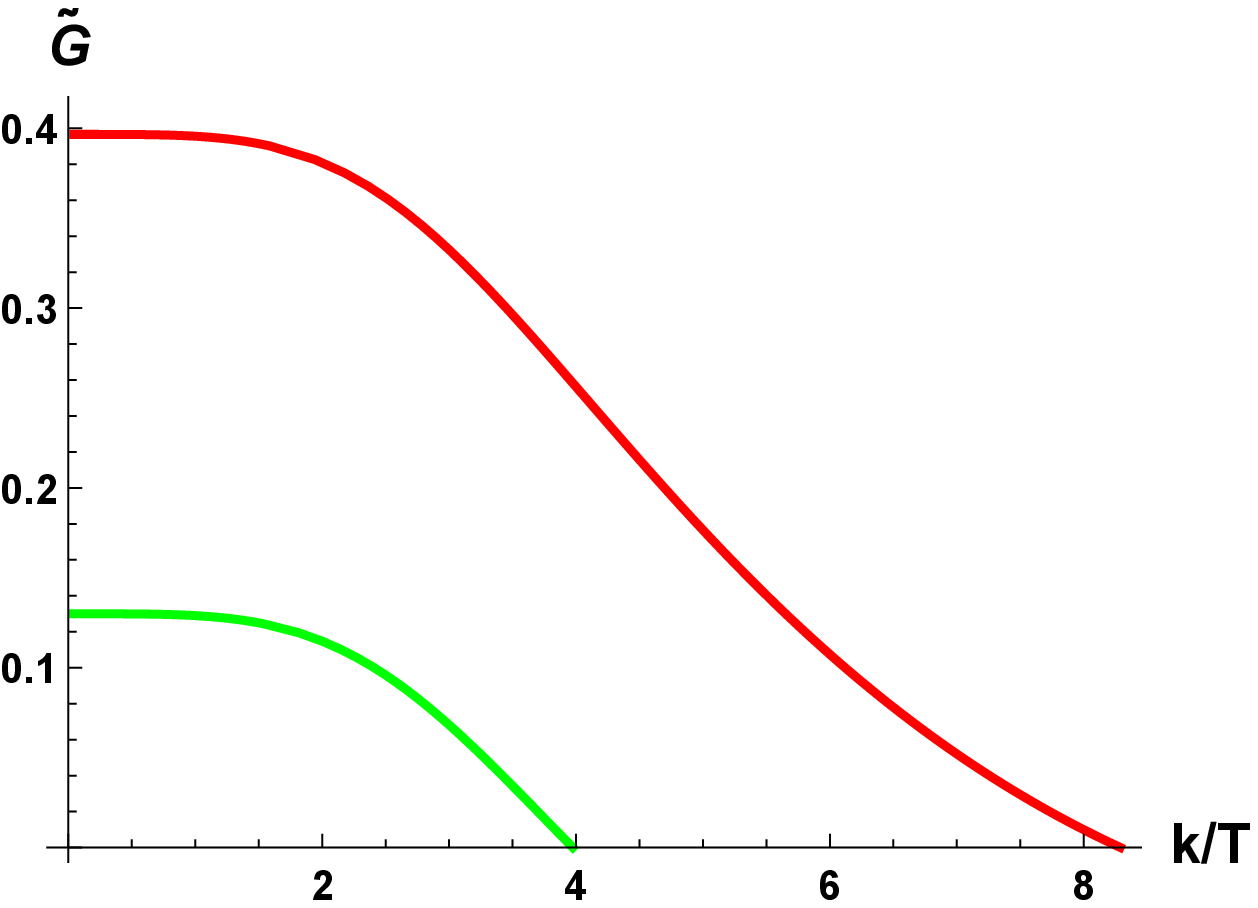}\quad
  \includegraphics[width=2.8in]{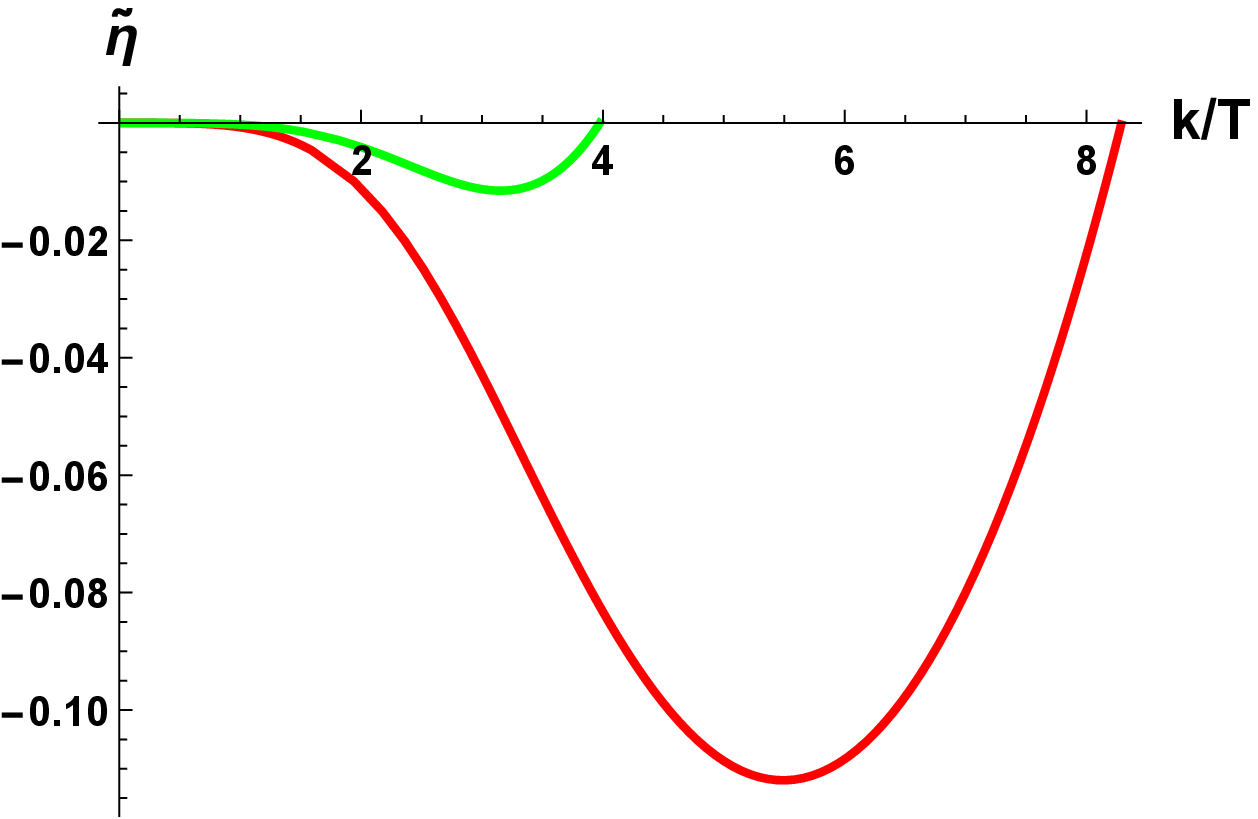}
\end{center}
 \caption{The reduced shear elastic modulus $\tilde{G}=16\pi G_N G/k^4$ (left panel)
 and the reduced shear 
viscosity   $\tilde{\eta}= (4\pi \eta/\cals-1)$  (right panel) as a functions of $k/T$
for select values of $\frac{T}{\mu}=\{\frac {1}{12},\frac 16\}$, $\{$red,green$\}$ curves, at the
criticality.} \label{figure8} 
\end{figure}

In the $\zet_2$ symmetric phase the shear modulus and the viscosity can be simply obtained 
\begin{equation}
\left\{\frac{G}{k^4}=0\,,\ \frac{\eta}{\cals}=\frac{1}{4\pi}\right\}
\end{equation}
and they correspond to a fluid state which saturates the KSS bound and has no shear elastic properties.\\
In fig.~\ref{figure8} we collect the numerical results for
$\tilde G\equiv 16\pi G_N G/k^4$ and $\tilde{\eta}\equiv (4\pi \eta/\cals-1)$
for select values of $\frac T\mu$ as a function of $\frac kT$. Notice that both
the shear elastic modulus and the deviation from the universal shear viscosity vanish in the limit
$\frac{k}{T}\to 0$; they also vanish for large enough $\frac{k}{T}$, when the $\zet_2$ symmetry is
restored (see fig.~\ref{figure3}). Red curves correspond to $\frac T\mu=\frac{1}{12}$ at the critical point, while
the green curves correspond to $\frac T\mu=\frac{1}{6}$ at the critical point.

\begin{figure}[t]
\begin{center}
  \includegraphics[width=2.75in]{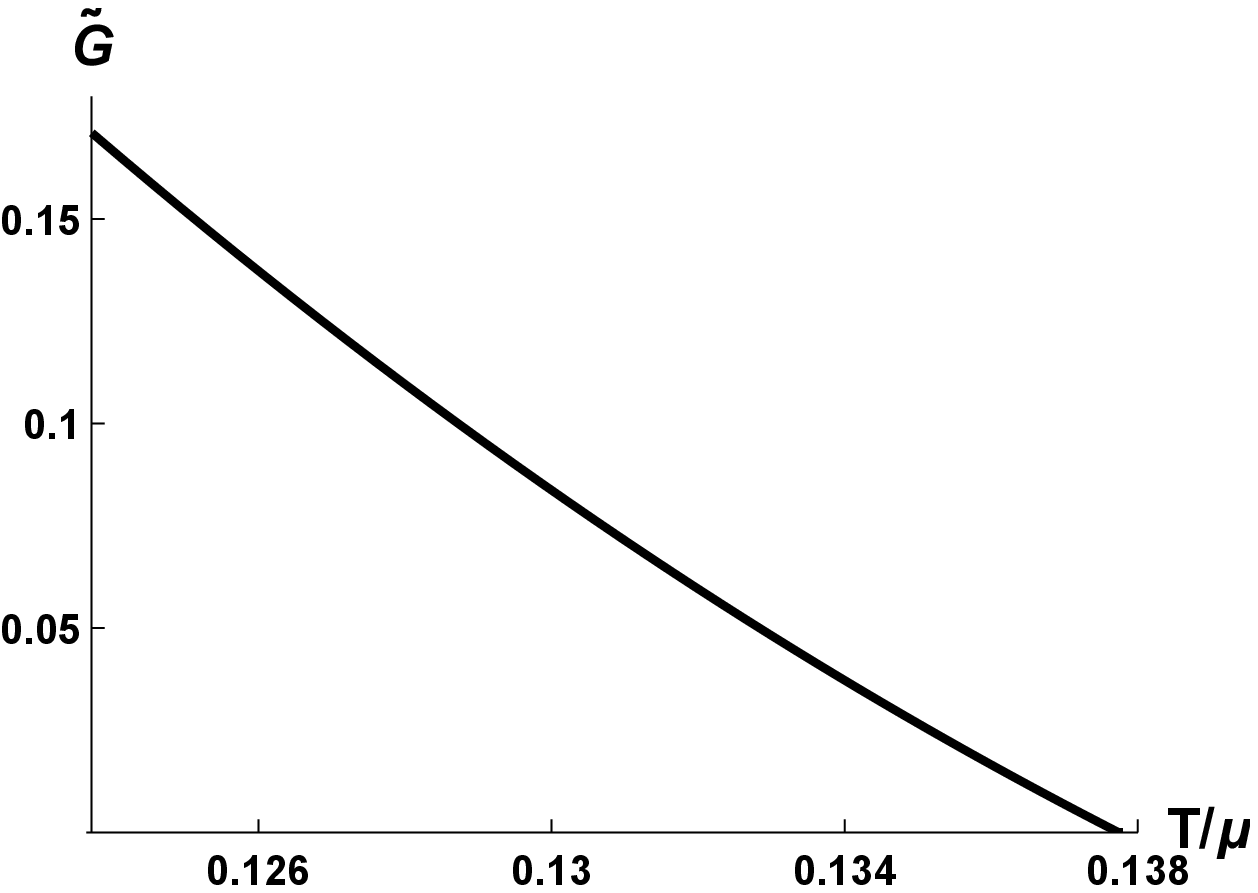}\quad
  \includegraphics[width=2.9in]{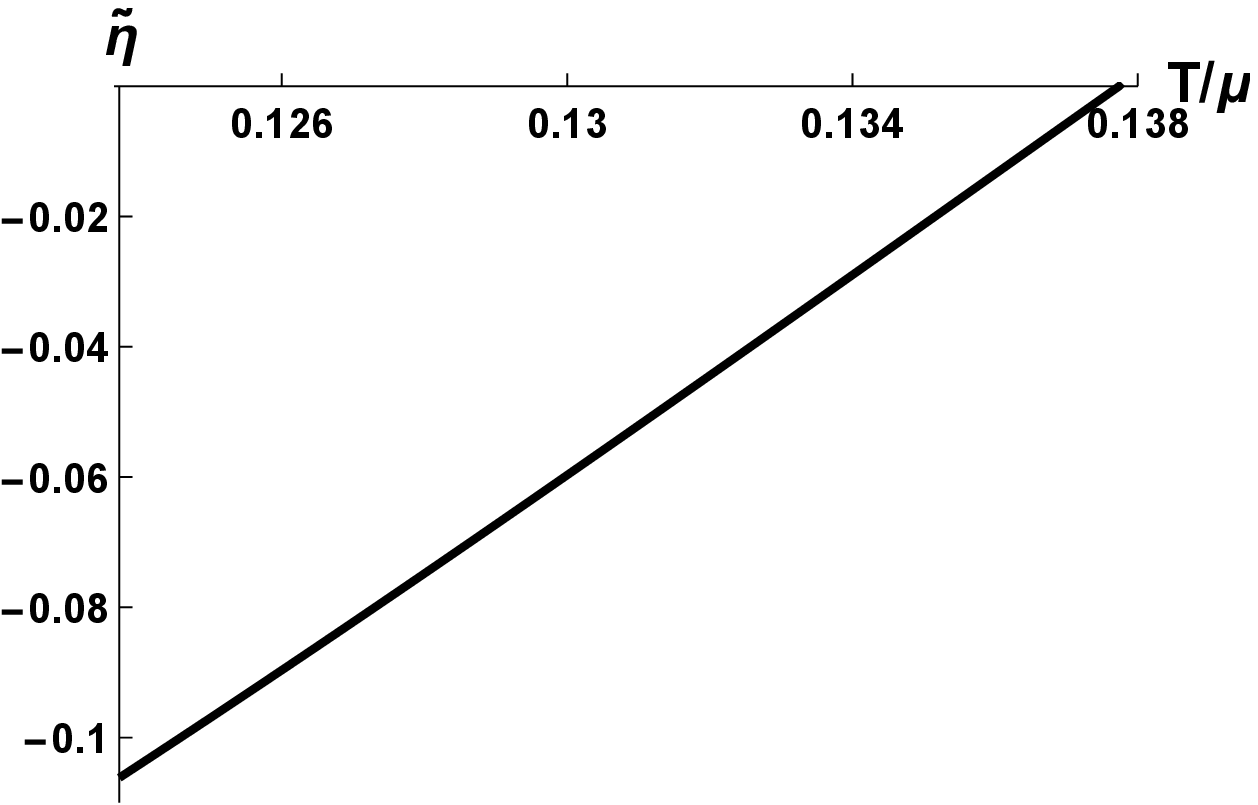}
\end{center}
 \caption{The reduced shear elastic modulus $\tilde{G}=16\pi G_N G/k^4$ (left panel)
 and the reduced shear 
viscosity   $\tilde{\eta}= (4\pi \eta/\cals-1)$  (right panel) as a functions of $T/\mu$
close to criticality at fixed value $\frac kT=5$.} \label{figure9} 
\end{figure}

In fig.~\ref{figure9} we study behavior of  $\tilde{G}$ and $\tilde{\eta}$ in the vicinity of the critical
point. Both quantities vanish at criticality with the same critical exponent:
\begin{equation}
\tilde{G}\ \propto \left(1-\frac{T}{T_{crit}}\right)^{\a_G}\,,\qquad
\tilde{\eta}\ \propto \left(1-\frac{T}{T_{crit}}\right)^{\a_\eta}\,,\quad \a_G=\a_\eta=1\,.
\eqlabel{critgeta}
\end{equation}

\begin{figure}[t]
\begin{center}
  \includegraphics[width=2.75in]{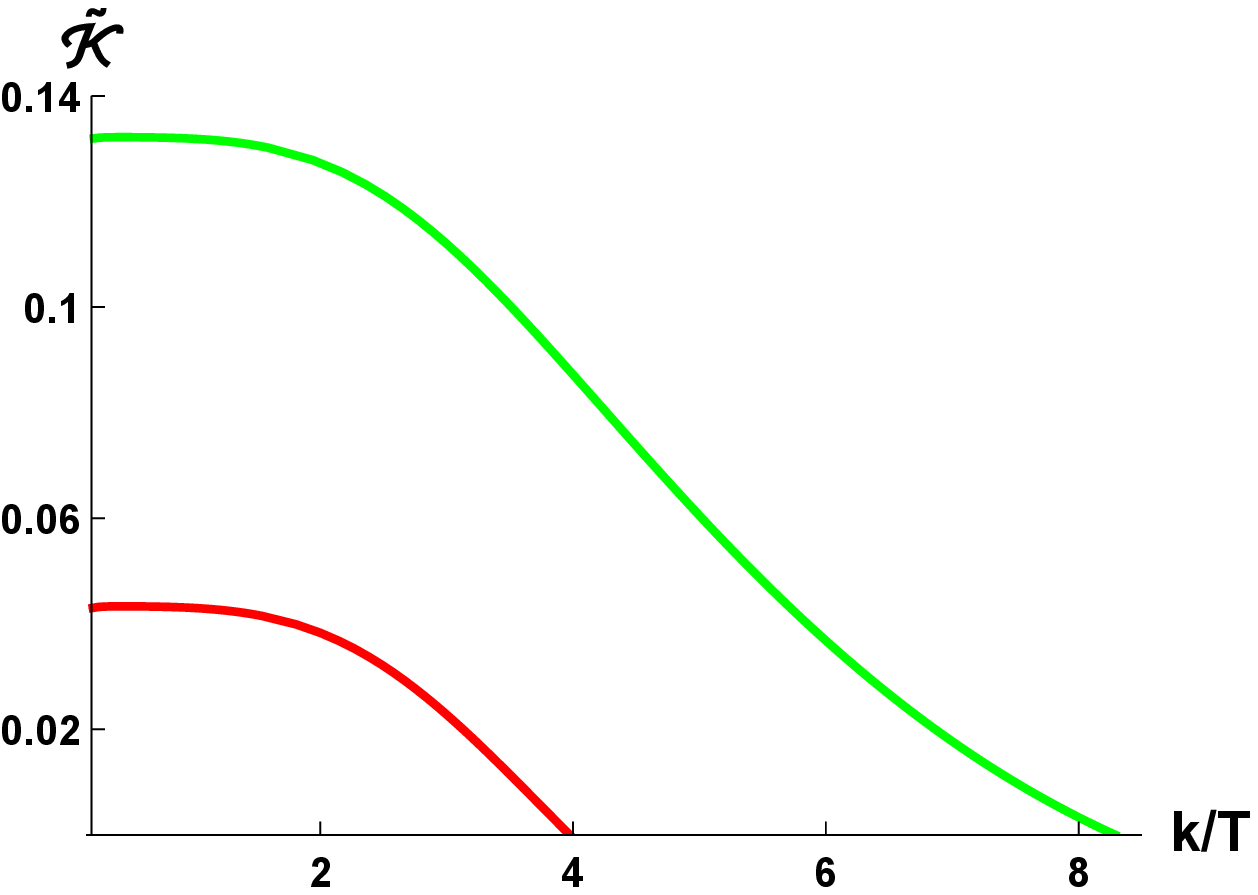}\quad
  \includegraphics[width=2.75in]{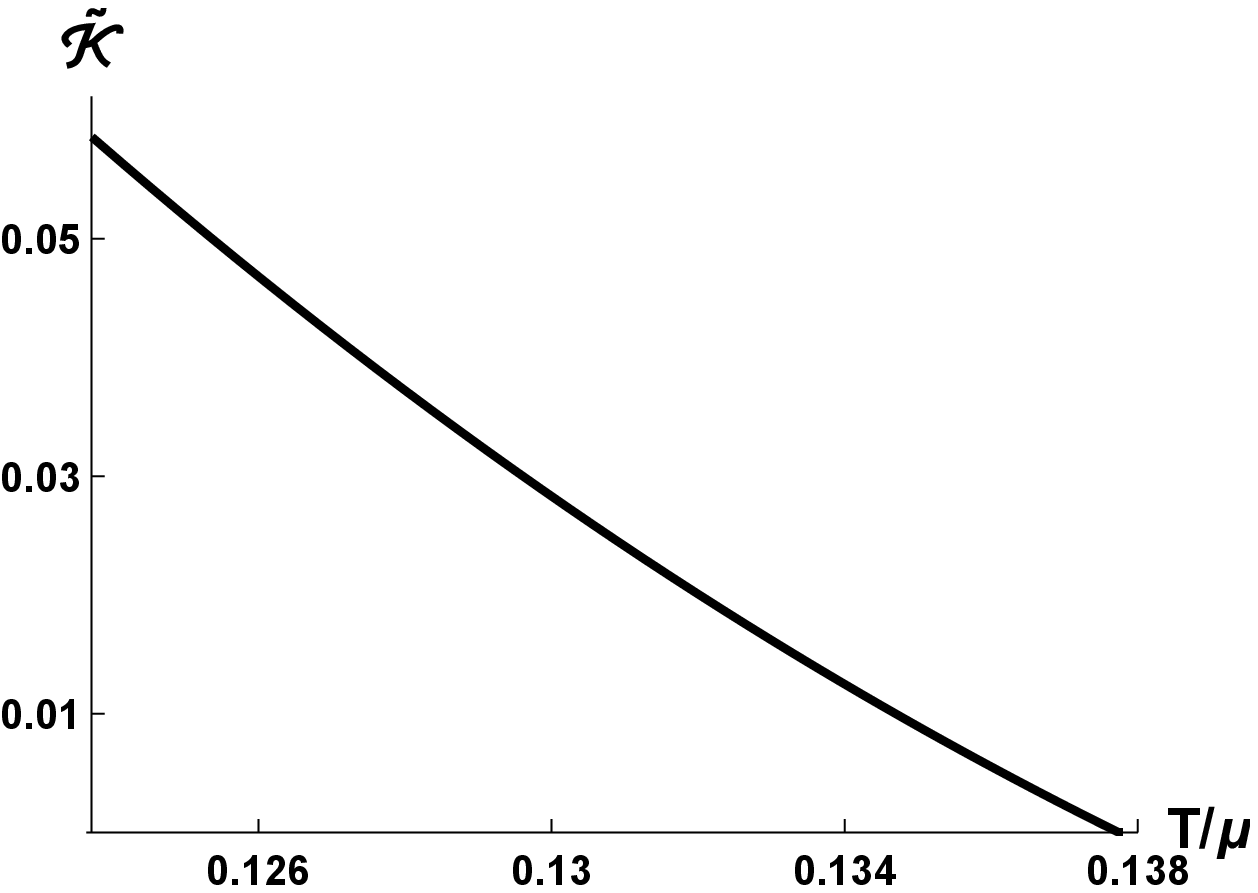}
\end{center}
 \caption{The reduced bulk elastic modulus $\tilde{\calk}=16\pi G_N \calk/k^4$ (left panel) as a functions of $k/T$
for select values of $\frac{T}{\mu}=\{\frac {1}{12},\frac 16\}$, $\{$red,green$\}$ curves, at the
criticality. The right panel presents the same quantity as a function of
$T/\mu$ close to criticality at fixed value $\frac kT=5$.
} \label{figure9a} 
\end{figure}

So far we focused on the shear elastic modulus. There is a simple way to compute the bulk elastic modulus
of the media, provided that the conformal symmetry is unbroken --- of course, this is the case
for the phase of the model with the spontaneous $\zet_2$ symmetry breaking discussed in this section.
To proceed with the computations, notice that the equilibrium
stress tensor of our media is compatible with the one of an isotropic crystal \cite{Lubensky,landau7}:
\begin{equation}
T_{crystal}^{IJ}=\left[\calp+\calk\ \partial \cdot \Theta\right] \delta^{IJ}
+2\,G\left[\partial^{(I}\Theta^{J)}-\frac 13\,\delta^{IJ}\,\partial \cdot \Theta\right]
\eqlabel{crystal}
\end{equation}
where $\calp$ is the equilibrium thermodynamic pressure,
$\Theta^J$ are the phononic degrees of freedom which can be identified with the Stueckelberg fields
of the translational symmetry breaking;
\begin{equation}
\partial_{(I}\Theta_{J)}\equiv u_{IJ}
\eqlabel{strina}
\end{equation}
is the usual strain tensor.
Notice that, contrarily to a perfect fluid, $T^{I}_{crystal,I}\neq 3\calp$, where $\calp=-\Omega$, 
related to the grand potential density. In a CFT the full stress-energy tensor vanishes ${T^\mu}_\mu=0$. Taking into account the definition of the energy density ${T^t}_t=\cale$ and the form of the spatial components of the stress tensor \eqref{crystal} we immediately obtain
\begin{equation}
\cale =3\,\calp+9\,\calk
\eqlabel{trace}
\end{equation}
which in absence of elastic properties is the usual conformal relation $\cale=(d-1)\calp$.\\
Using the Smarr relation $\cale+\calp=\cals T+\mu \calq$ and combining it with \eqref{trace} we obtain the final formula for the bulk modulus:
\begin{equation}
\calk=\frac{4\cale-3(\cals\, T+\mu \,\calq)}{9}\eqlabel{Kmod}
\end{equation}
At this point is important to notice that the bulk modulus defined in \eqref{Kmod} is just the ''solid'' contribution, which indeed vanishes in the pure RN ''fluid'' solution. On the contrary, the bulk modulus (intended as the total one) is nonzero even in the fluid phase and it gives rise to the common finite speed of longitudinal sound , \textit{i.e.} $c_L^2=\calk/(\epsilon+p)$,  which has already been analyzed in the literature. In other words, we can think of $\calk$ in \eqref{Kmod} as the additional contribution to the bulk modulus, and therefore the longitudinal sound speed, due to the solid nature of the media\footnote{See \cite{Andrade:2017cnc} for an explicit computation of such a correction due to the spontaneous symmetry breaking of translational invariance in a slightly different holographic model.}. \\
All the physical observables in the previous formulas can be obtained explicitly using holographic renormalization (see Appendix \ref{App4} for details).
The numerical results for the
reduced bulk elastic modulus $\tilde{\calk}=16\pi G_N \calk/k^4$ are collected in
fig.~\ref{figure9a}.  Left panel presents results as a function of $\frac kT$
for select values of $\frac{T}{\mu}=\{\frac {1}{12},\frac 16\}$, $\{$red,green$\}$ curves, at the
criticality; right panel demonstrates the critical behavior of the modulus at fixed $\frac kT=5$.
The critical exponent for the bulk elastic modulus $\a_\calk$ in our model is
\begin{equation}
\tilde{\calk}\ \propto\ \left(1-\frac{T}{T_{crit}}\right)^{\a_\calk}\,,\qquad \a_{\calk}=1 
\eqlabel{critbulk}
\end{equation}

\subsection{Viscoelastic transport with explicit $\zet_2$ symmetry breaking}\label{zet2explicit}

Another interesting regime of the model is the explicit breaking of the $\zet_2$ symmetry at fixed value of the
symmetry breaking parameter $\dd_\Delta$ relative to the equilibrium temperature $T$, but with varying
lattice spacing $\Delta  x$ (or equivalently $\frac{k}{T}$). To limit the
parameter space, we set the chemical potential $\mu=0$  and further restrict
\begin{equation}
\Delta=2\,,\qquad \{\g,\l_1,\l_2\}=\{1,0,1\}
\eqlabel{solidrestriction}
\end{equation}

This regime is represented holographically by the same  gravitational bulk equations \eqref{warpfg}-\eqref{fg4}
as in section \ref{fullg}, albeit with the vanishing bulk gauge potential
\begin{equation}
a(z)\equiv 0
\eqlabel{aiszero}
\end{equation}
The details of the setup and the numerical computations are presented in Appendix \ref{App4}.
The AdS boundary condition for the scalar which reflects a $\dd_2\ne 0$ source term of the explicit $\zet_2$ symmetry breaking are:
\begin{equation}
\phi=(p_2 +\a \ln z)z^2+\calo(z^6\ln^2 z)
\eqlabel{uvfgbs}
\end{equation}
where we introduced a dimensionless parametrization\footnote{Comparing with \cite{Buchel:2007vy}
$\dd_2\propto +m^2 >0$.} of
$\dd_2>0$  as
\begin{equation}
\dd_2=\frac{\a}{x_h^2}
\eqlabel{defdd2}
\end{equation}

\begin{figure}[t]
\begin{center}
  \includegraphics[width=2.75in]{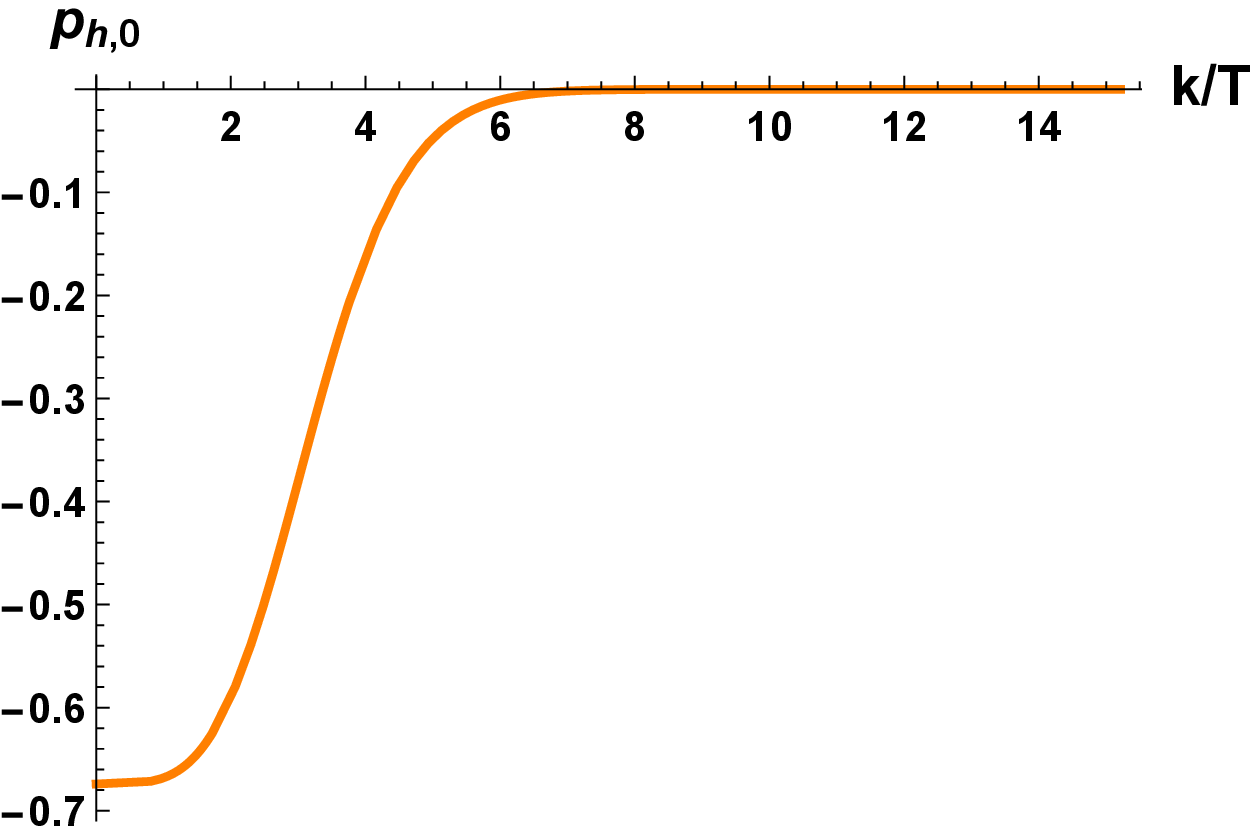}\quad
  \includegraphics[width=2.75in]{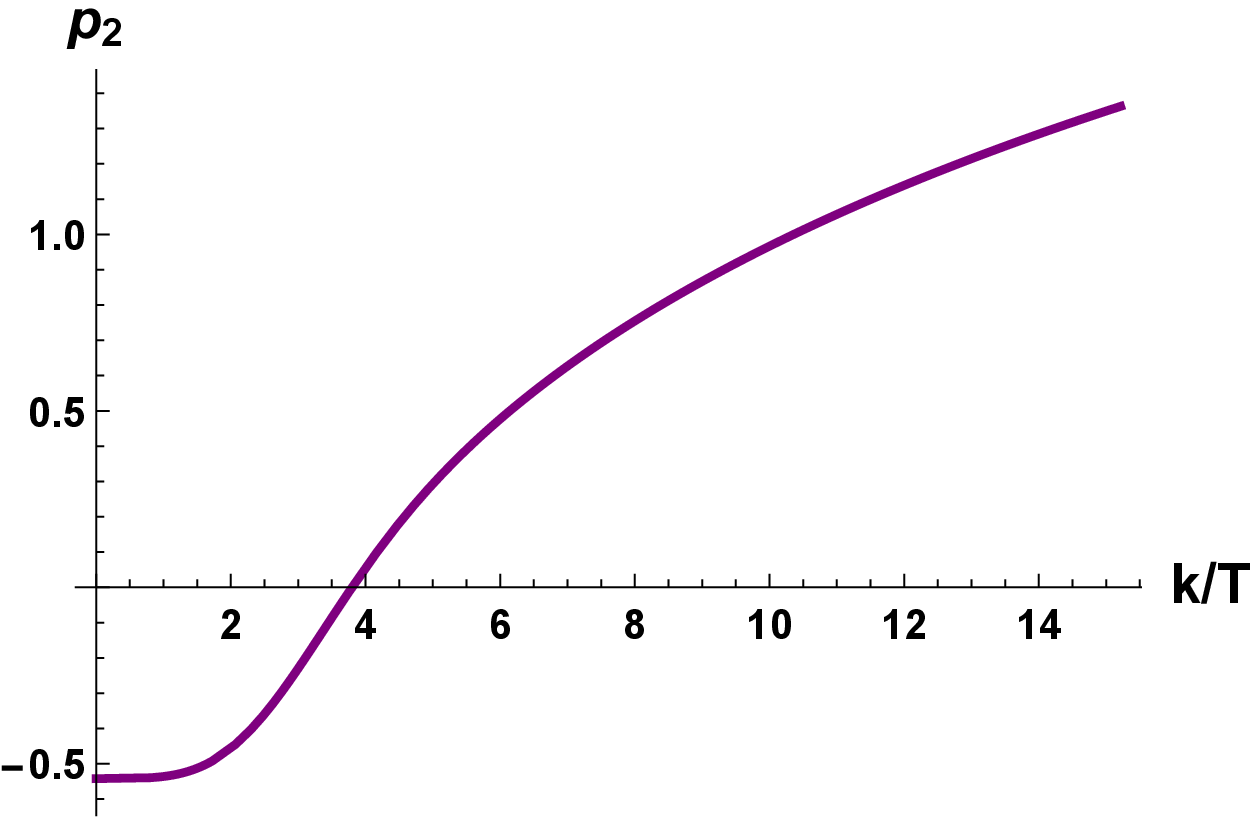}
\end{center}
 \caption{Bulk scalar behavior in the background geometry dual to a phase of the model
 with explicit $\zet_2$ symmetry breaking at fixed $\frac{\dd_2}{T^2}$.} \label{figure10} 
\end{figure}

\begin{figure}[t]
\begin{center}
  \includegraphics[width=2.75in]{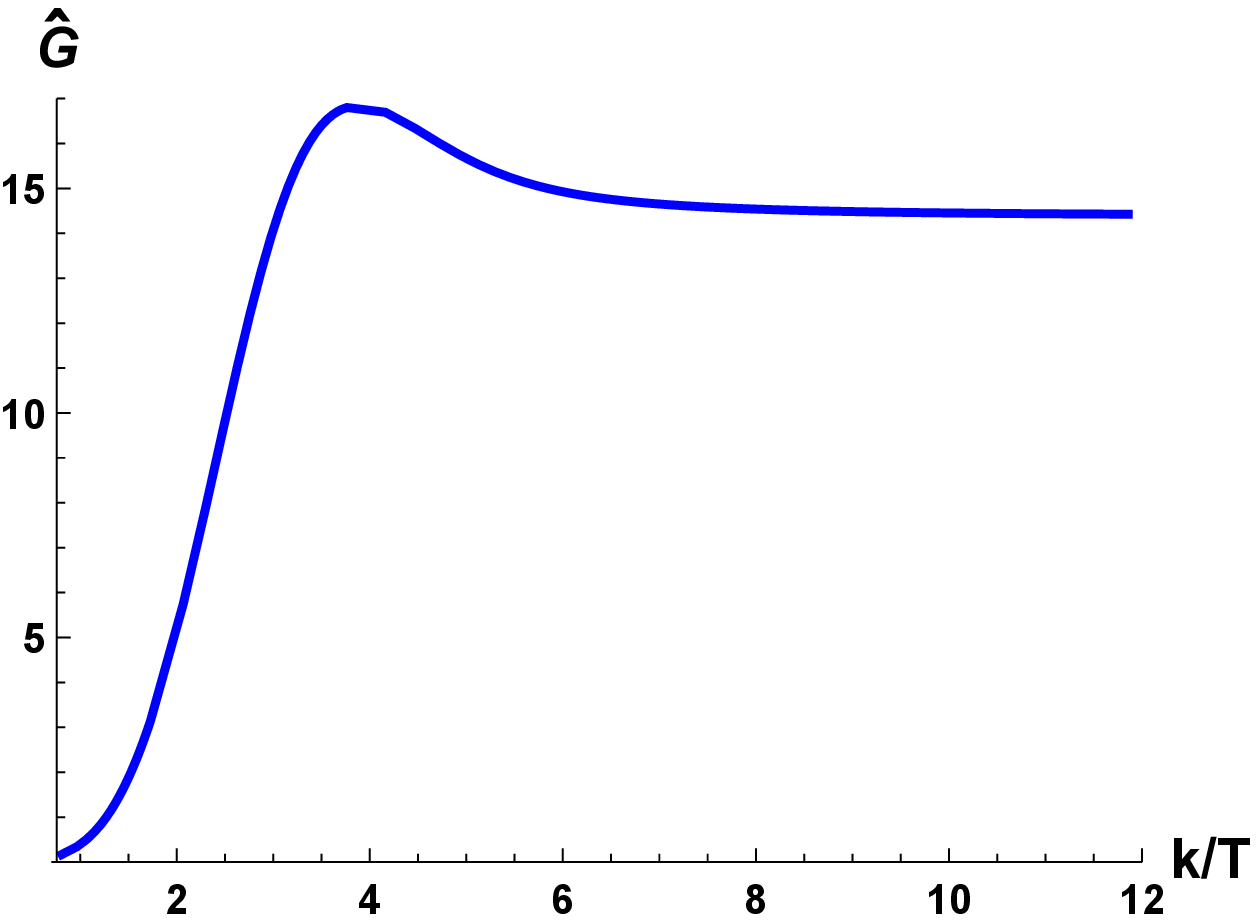}\quad
  \includegraphics[width=2.95in]{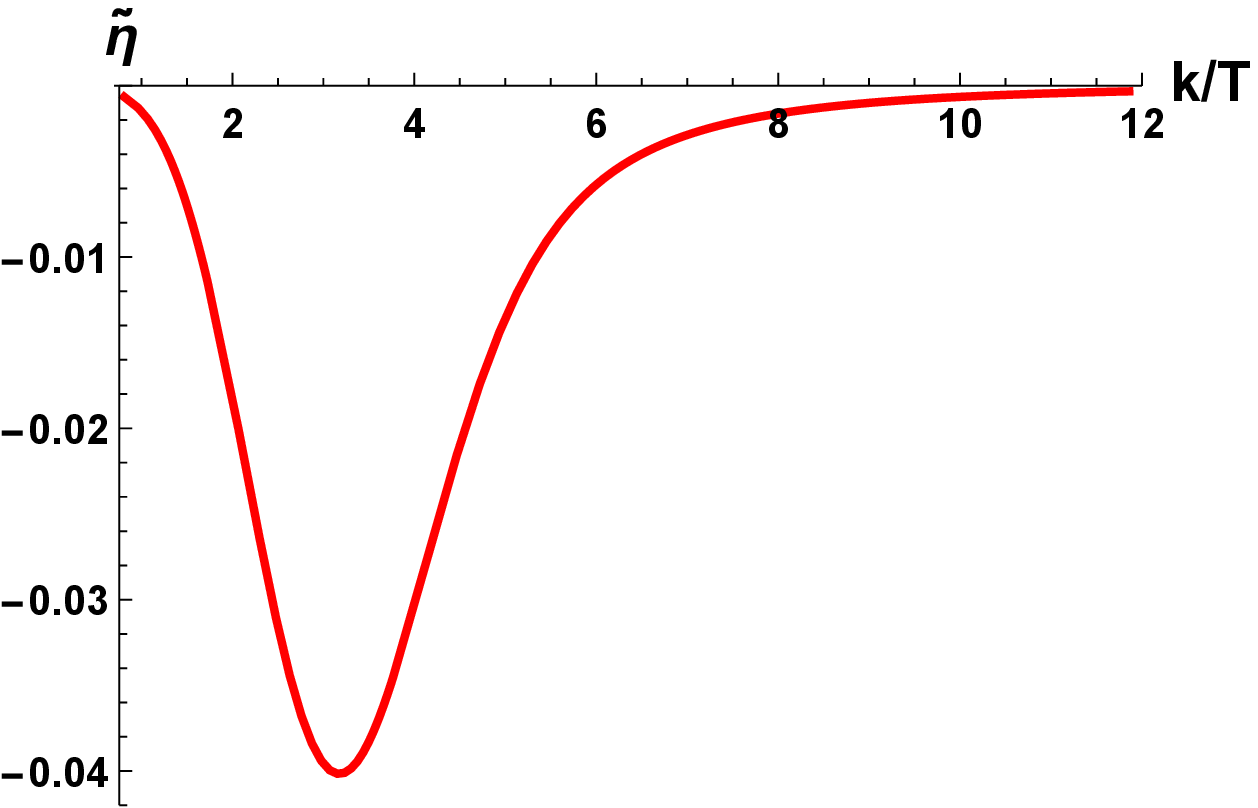}
\end{center}
 \caption{Elastic shear modulus $\hat{G}=16\pi G_N\ G/T^4$ and the shear viscosity
$\tilde{\eta}=4\pi \eta/\cals-1$ at fixed $\dd_2/T^2$.} \label{figure11} 
\end{figure}

Once again, the background has to be determined numerically.
In fig.~\ref{figure10} we present the behavior of the scalar field at the horizon $p_{h,0}$,
and $p_2$ (related to the expectation value of the dual $\calo_2$ operator) at fixed
$\frac{\dd_2}{T^2}$ as a function of $\frac kT$. Notice that in the limit $\frac kT\to \infty$
(vanishing lattice spacing) the scalar is exponentially suppressed at the horizon, while
the expectation value of the corresponding dual operator is finite. As a result, in this limit,
the horizon of the gravitation background is 'hairless' Schwarzschild,  leading to the universal
result for the shear viscosity \cite{Buchel:2003tz}. On the contrary, finiteness of $p_2$
as $\frac kT\to \infty$ suggests that the shear elastic modulus remains finite. These
expectations are indeed supported by direct computations. Fig.~\ref{figure11} collects the numerical results for
\begin{equation}
\hat{G}\equiv 16\pi G_N\ \frac{G}{T^4}\,,\qquad \tilde{\eta}=\frac{4\pi \eta}{\cals}-1
\eqlabel{defsolidtrans}
\end{equation}
Notice that while $\tilde{\eta}\to 0$ as $\frac kT\to \infty$, $\hat{G}$ remains finite.

\begin{figure}[t]
\begin{center}
  \includegraphics[width=2.7in]{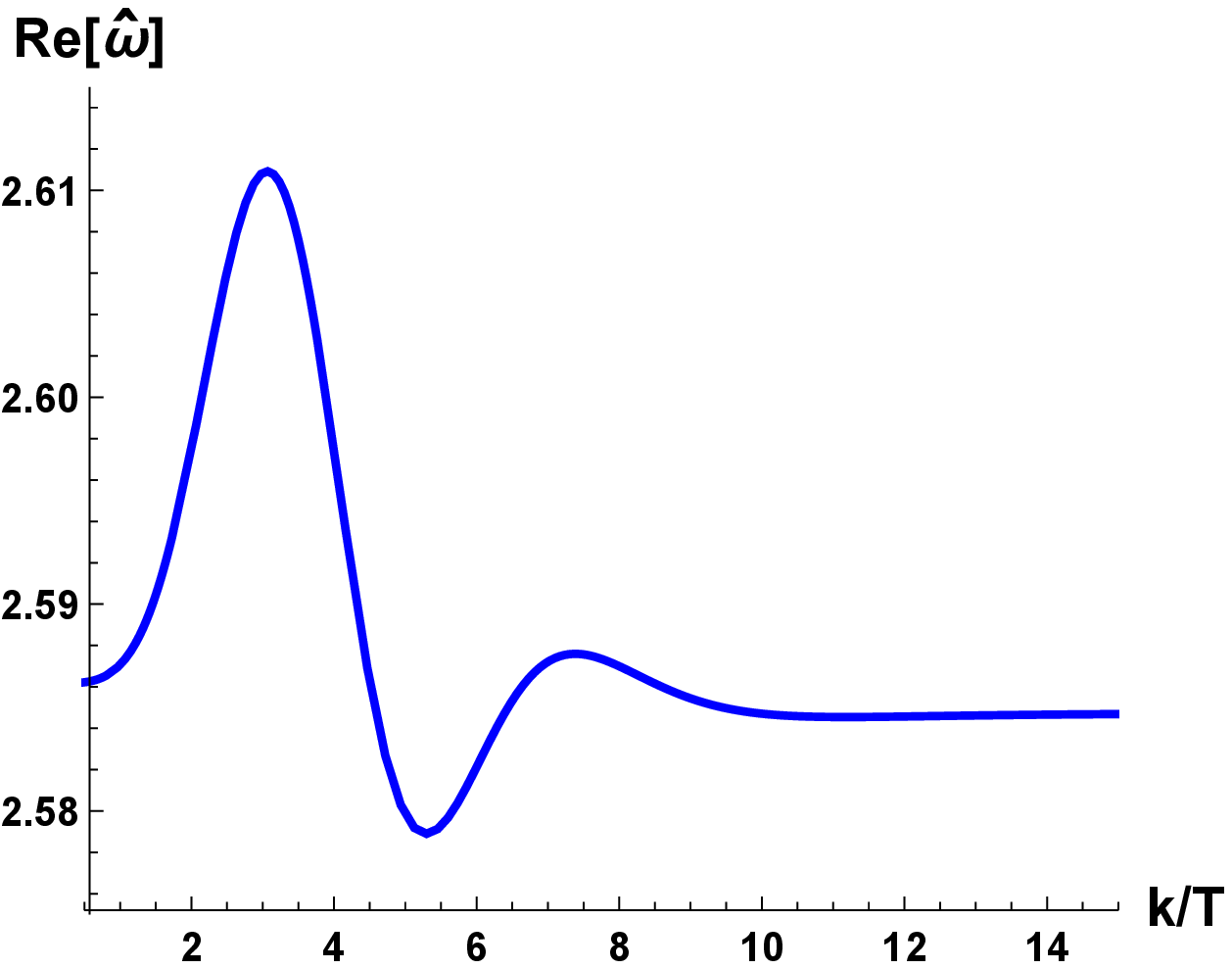}\quad
  \includegraphics[width=2.89in]{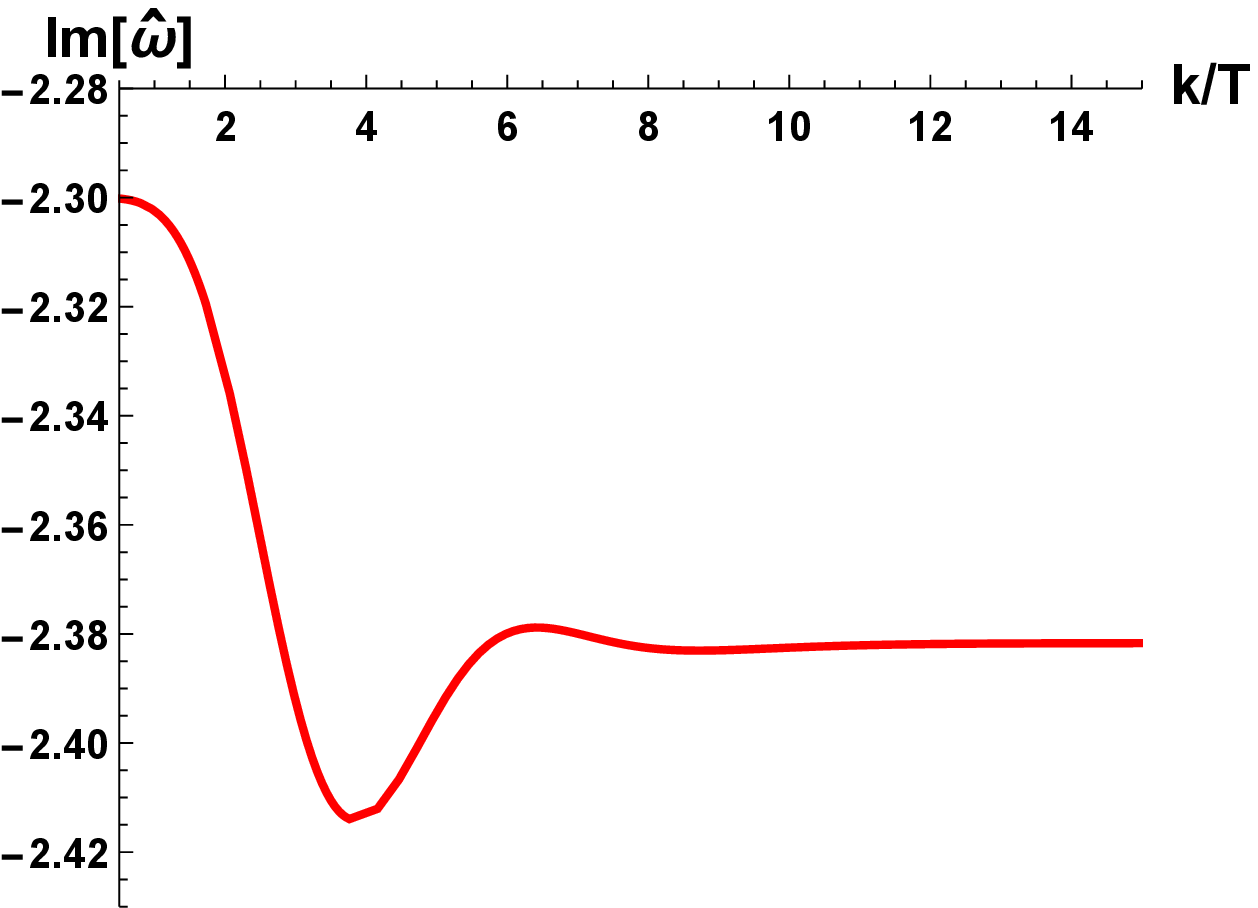}
\end{center}
 \caption{Lowest non-hydrodynamic mode at zero spatial momentum in the shear/scalar channel
 at fixed $\dd_2/T^2$. We define $\hat{\omega}=\omega/(2\pi T)$ the reduced frequency.} \label{figure12} 
\end{figure}

The fluid-gravity correspondence  \cite{Bhattacharyya:2008jc} identifies the effective theory of
small fluctuations of the holographic horizon
with the hydrodynamics of the boundary plasma. In the limit $\frac kT\to \infty$
the horizon fluctuations are sensitive to the translational symmetry breaking only
through the mediator --- the bulk scalar $\phi$, which is vanishingly small.
Thus, we expect that hydrodynamics (at least for the first few orders in the
gradient expansion) should not feel the spatial lattice of our viscoelastic media.
The non-hydrodynamic plasma excitations should know about the
translational symmetry breaking. Following \cite{Kovtun:2005ev} we study
the quasinormal modes in the shear/scalar  channels of the holographic
dual\footnote{Sound channel fluctuations are more involved and will not be discussed here.}.
Because of the spatial lattice, the momentum $\vec{q}$ of the modes must be quantized. All the modes
except those with $\vec{q}=0$ become infinitely heavy in the limit of vanishing lattice spacing,
and decouple. Thus, we focus on the $\vec{q}=0$ sector. In this sector there is no
distinction between the scalar channel and the shear channel QNMs. Numerical results for the lowest nonhydrodynamic QNM are collected in
fig.~\ref{figure12}. As expected, here
\begin{equation}
\lim_{k/T\to 0}\ \frac{\w}{T} \ \  \ne\ \ \lim_{k/T\to \infty}\ \frac{\w}{T}
\eqlabel{notthesame}
\end{equation}

\section{Hydrodynamics of homogeneous and isotropic flows}\label{hydro}

In this section we study the all-order hydrodynamics of the viscoelastic media undergoing the homogeneous and
isotropic expansion following \cite{Buchel:2016cbj,Buchel:2018ttd}. We focus on the explicit $\zet_2$
flavor symmetry breaking to leading nontrivial order in the source, \ie $\calo(\dd_\Delta^2)$. 
We omit technical details and refer the interested reader to the earlier work.

There are many probes of the all-order hydrodynamics: expectation values of the correlation functions,
entanglement entropy, etc. We consider the entropy production of the viscoelastic plasma
in the background metric \eqref{flrw}. To compute the entropy production \eqref{erate}
to order $\calo(\delta_\Delta^2)$ it is sufficient to study the linearized, order $\calo(\dd_\Delta)$, scalar field dynamics
in \eqref{rnflrw}:
\begin{equation}
\begin{split}
0=&\del_{\t z}^2\phi-\frac{(z^2-1)(q^2z^4-3z^2-3)}{6a}\ \del^2_{zz}\phi-
\frac{3q^2z^6-q^2z^4-3z^4-9}{6za}\ \del_z \phi\\
&-\frac {3}{2z}\ \del_\t \phi+\frac{m_{eff}^2}{2  z^2a}\ \phi\,,\\
m_{eff}^2\equiv &\Delta(\Delta-4)-z^2(4\gamma q^2 z^4-3\lambda_2 s^4 z^2-3 \lambda_1 s^2)\,,\qquad \phi=\phi(\t,z)\,,\qquad a=a(\t)
\end{split}
\eqlabel{phieoma}
\end{equation}
where we introduced 
\begin{equation}
\t\equiv\frac{t}{x_h}\,,\qquad z\equiv \frac{x}{ax_h}\in [0,1]\,,\qquad Q=\frac{q}{x_h^3}\,,\qquad k=\frac{s}{x_h}
\eqlabel{defzqsa}
\end{equation}
leading to the comoving entropy production rate
\begin{equation}
\frac{d}{dt}\ln\left(a^3\cals\right)=\frac{1}{4\pi T(t)}\  (d_+\phi)^2\bigg|_{x=x_h a}\,,\qquad T(t)\equiv \frac{6-q^2}{6\pi x_h a(t)} 
\eqlabel{entprod}
\end{equation}
$T(t)$ is a local temperature, red-shifting as $T(t)=\frac{T_0}{a(t)}$, $T_0\sim \frac{1}{x_h}$.
A general solution to \eqref{phieoma} can be written as a series expansion in the derivatives of
the scalar factor $a(t)$, $x_h \frac{da}{dt} =\frac{da}{d\t}\equiv \dot a \ll 1 $,
\begin{equation}
\begin{split}
\phi(\t,z)=\dd_\Delta a^{4-\Delta}x_h^{4-\Delta}\ \sum_{n=0}^\infty {\calt_{n,\Delta}[a(\t)]}\ F_{\Delta,n}(z)
\end{split}
\eqlabel{phiexpand}
\end{equation}
where the functional $\calt_{n,\Delta}[a]$ involves $n$ $\t$-derivatives of the scale factor, and 
\begin{equation}
\begin{split}
&\calt_{n,\Delta}=a \dot{\calt}_{n-1,\Delta}+(4-\Delta) \dot a\calt_{n-1}\,,\qquad \calt_{0,\Delta}=1\\
&0=F_{n,\Delta}''+\frac{3 q^2 z^6-q^2 z^4-3 z^4-9}{z (q^2 z^4-3 z^2-3) (z^2-1)} F_{n,\Delta}'
+\frac{3m_{eff}^2}{(q^2 z^4-3 z^2-3) (1-z^2) z^2}  F_{n,\Delta}
\\&+\frac{6}{(q^2 z^4-3 z^2-3) (1-z^2)} \left(F_{n-1,\Delta}'-\frac{3}{2 z} F_{n-1,\Delta}\right)\,,
\qquad n>0\\
&0=F_{0,\Delta}''+\frac{3 q^2 z^6-q^2 z^4-3 z^4-9}{z (q^2 z^4-3 z^2-3) (z^2-1)} F_{0,\Delta}'
+\frac{3m_{eff}^2}{(q^2 z^4-3 z^2-3) (1-z^2) z^2}  F_{0,\Delta}\\
&F_{0,\Delta}=z^{4-\Delta}\left(1+\calo(z)\right)\,,\qquad F_{n>0,\Delta}= z^{4-\Delta} \calo(z)\,,
\qquad F_{n,\Delta}=\calo((1-z)^0)
\end{split}
\eqlabel{eomseries}
\end{equation}
The all-order entropy production expression \eqref{entprod} then takes form
\begin{equation}
\frac{d}{dt}\ln\left(a^3\cals\right)=\frac{1}{4\pi T(t)}\  {(x_ha)^{6-2\Delta}\dd_\Delta^2}
 \ \left(\Omega_\Delta[a]\right)^2\,,\qquad \Omega_\Delta=\sum_{n=0}^\infty F_{n,\Delta}\bigg|_{z=1} \calt_{n+1}[a]
\eqlabel{entprod2}
\end{equation}
The recursive relation for $\calt_{n,\Delta}$ can be solved analytically for simple scale factors ---
\eg
choosing de Sitter expansion,  $a(t)=e^{H t}=\exp(Hx_h \t)$, we find 
\begin{equation}
\calt_{n,\Delta}=\frac{\Gamma(n+4-\Delta) (H x_h)^na^n}{\Gamma(4-\Delta)}
\eqlabel{solveT}
\end{equation}

\begin{figure}[t]
\begin{center}
  \includegraphics[width=2.85in]{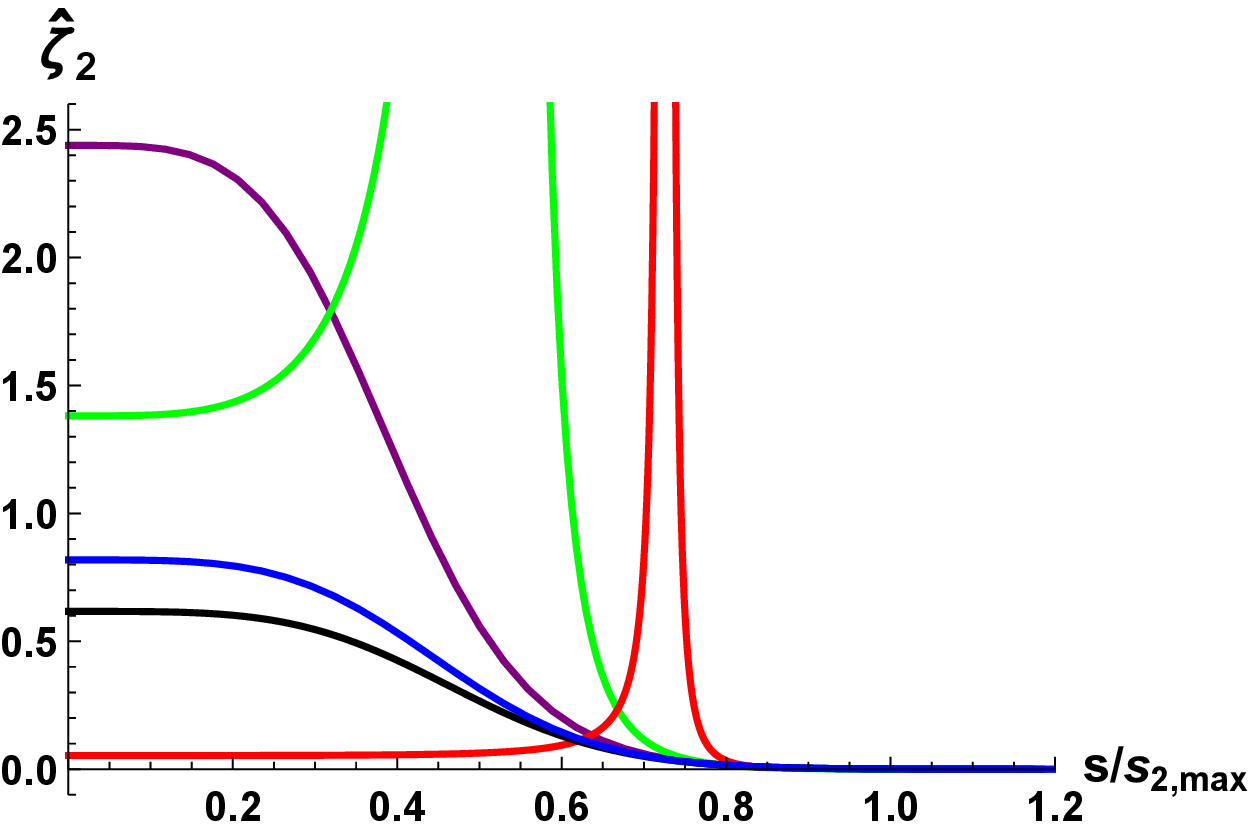}\quad
  \includegraphics[width=2.8in]{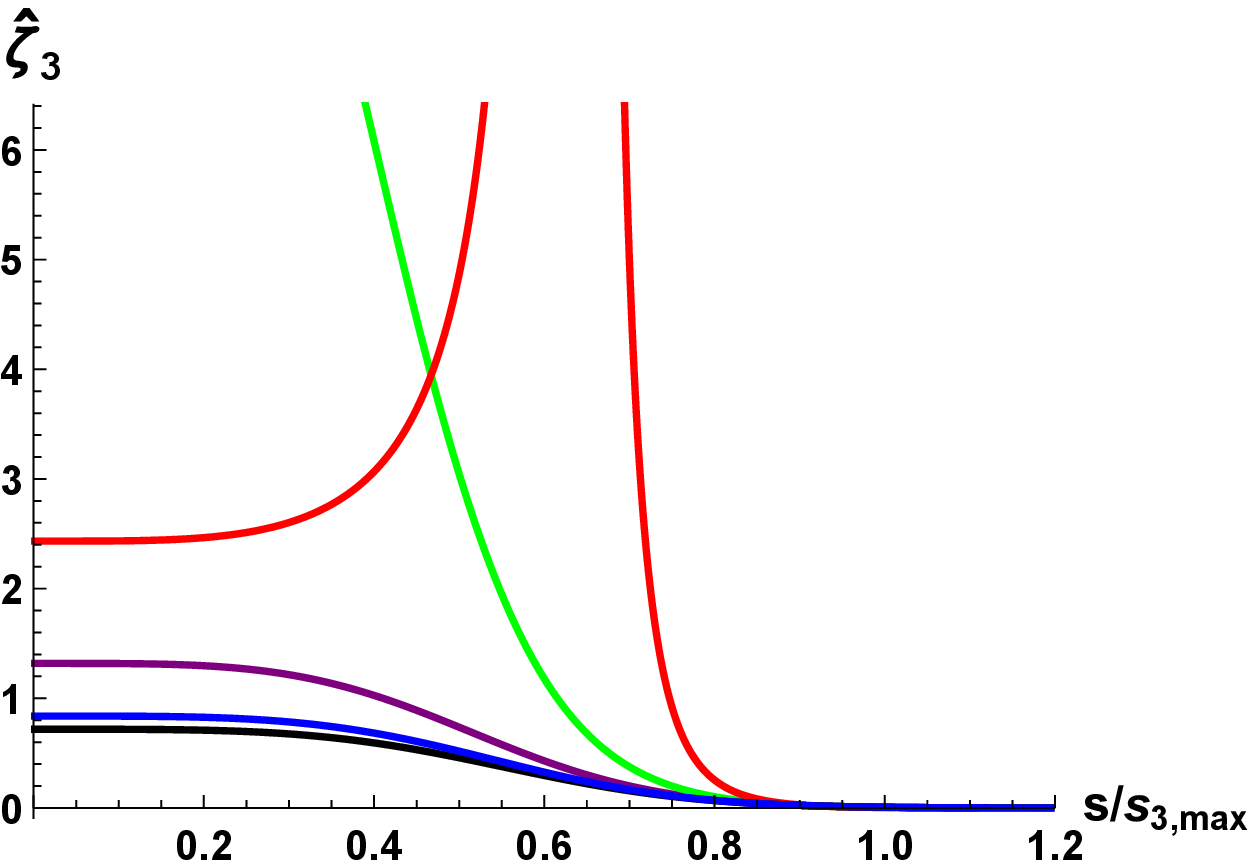}
\end{center}
 \caption{Reduced bulk viscosity $\hat{\zeta}_\Delta$ (left panel: $\Delta=2$; right panel: $\Delta=3$)
 of the holographic viscoelastic media to leading order in the explicit $\zet_2$ symmetry breaking
 (see \eqref{zetas} for parametrization). Color coding corresponds to
 different values of the ratio of temperature $T$ to the chemical potential $\mu$:
 $\frac T\mu=\frac {1}{12}$ (red),  $\frac T\mu=\frac {1}{6}$ (green),
  $\frac T\mu=\frac {1}{3}$ (magenta) and  $\frac T\mu=\frac {2}{3}$ (blue).  The former two values
  are below the critical values of ${T}/{\mu}|_{crit}$ at $s=0$, and the latter two are above for $\Delta=2$.
  For $\Delta=3$ model only the red curve corresponds to $\frac{T}{\mu}$ below the critical value.
  Black curves correspond to neutral viscoelastic media, \ie $\mu=0$.
 } \label{figure14} 
\end{figure}

Notice that to leading order in the derivative expansion,
\begin{equation}
\frac{d}{dt}\ln\left(a^3\cals\right)=\frac{1}{T}\ (3H)^2\ \biggl\{
\frac{(4-\Delta)^2}{36\pi}\ \frac{\dd_\Delta^2}{(\pi T)^{8-2\Delta}}\
(F_{0,\Delta}(1))^2 \left(1-\frac {q^2}{6}\right)^{8-2\Delta}\
\biggr\} +\calo(\frac{H^4}{T^3}) 
\eqlabel{extbv1}
\end{equation}
which when compared with the entropy production for the homogeneous and isotropic
flow in the hydrodynamic approximation\footnote{It was shown in \cite{Buchel:2016cbj} that
the bulk viscosity extracted from \eqref{bulkhydro} correctly reproduces the viscosity extracted from
Minkowski spacetime equilibrium two-point correlation function of the boundary stress-energy tensor.}
\begin{equation}
\frac{d}{dt}\ln\left(a^3\cals\right)\approx\frac{1}{T}\ \left(\nabla \cdot u\right)^2\ \frac\zeta\cals 
\eqlabel{bulkhydro}
\end{equation}
identifies the bulk viscosity of the viscoelastic medial to order $\calo(\dd_\Delta^2)$ as 
\begin{equation}
\frac{\zeta}{\cals}=\frac{(4-\Delta)^2}{36\pi}\ \frac{\dd_\Delta^2}{(\pi T)^{8-2\Delta}}\
\hat{\zeta}_\Delta\,,\qquad  \hat{\zeta}_\Delta \equiv
(F_{0,\Delta}(1))^2 \left(1-\frac {q^2}{6}\right)^{8-2\Delta}
\eqlabel{zetas}
\end{equation}
The reduced bulk viscosity $\hat{\zeta}_\Delta=\hat{\zeta}_\Delta(\g,\l_1,\l_2,s,q)$
is presented in fig.~\ref{figure14} for the viscoelastic models with $\Delta=\{2,3\}$ 
and $\{\g,\l_1,\l_2\}=\{1,0,1\}$ for select values of
\begin{equation}
\frac T\mu=\frac{6-q^2}{6\pi q}
\eqlabel{bulkTmu}
\end{equation}
Notice that the bulk viscosity vanishes in the  large-$s$ (small lattice spacing) limit. 
Again this is a reflection of the fact that the hydrodynamic transport is determined by the near horizon regime
of the gravitational dual, and the bulk scalar field is exponentially suppressed at the horizon in this limit,
see fig.~\ref{figure10}. Much like the shear elastic modulus $\hat{G}$ and the correction to the shear viscosity
$\k$ (see fig.~\ref{figure6}), the bulk viscosity diverges in the phase with explicit $\zet_2$ symmetry breaking
close to criticality.

In the rest of this section we present results for all-order computation of $\Omega_\Delta$
for the viscoelastic model with $\Delta=2$. We consider $\{\g,\l_1,\l_2\}=\{1,0,1\}$ neutral
viscoelastic medial in section \ref{neutral}, and charged plasma with spontaneous symmetry breaking in
section \ref{charged}.

\subsection{Neutral viscoelastic media}\label{neutral}

Setting $\l_1=q=0$ and $\g=\l_2=1$, we can solve equation \eqref{eomseries} for $F_{0,2}$ and $F_{1,2}$
analytically\footnote{Analytical solution is available for all $\Delta$.}:
\begin{equation}
\begin{split}
F_{0,2}=&-\frac{\pi z^2}{4\cosh(\frac{\pi\sqrt{3}}{4}s)}\ _2 F_1\left[\frac 12-\frac 14 i \sqrt{3} s\,,\,
\frac 12+\frac 14 i \sqrt{3} s\,;\, 1\,;\, 1-z^4\right]\\
F_{1,2}=&2 \left(\arctan(z) + \arctanh(z)\right)F_{0,2}(z)\\
&+\frac{\pi^2}{4\cosh^2(\frac{\pi\sqrt{3}}{4}s)}z^2\ _2 F_1\left[\frac 12-\frac 14 i \sqrt{3} s\,,\,
\frac 12+\frac 14 i \sqrt{3} s\,;\, 1\,;\, z^4\right]
\end{split}
\eqlabel{f02}
\end{equation}
We solved numerically \eqref{eomseries} for $n=2\cdots 300$, Borel transformed $\Omega_{\Delta=2}\to \Omega_{\Delta=2}^{(B)}$,
Pade approximated the result,
extracted the poles and compared them with the appropriate QNMs \cite{Buchel:2016cbj,Buchel:2018ttd}.

\begin{figure}[t]
\begin{center}
  \includegraphics[width=2.95in]{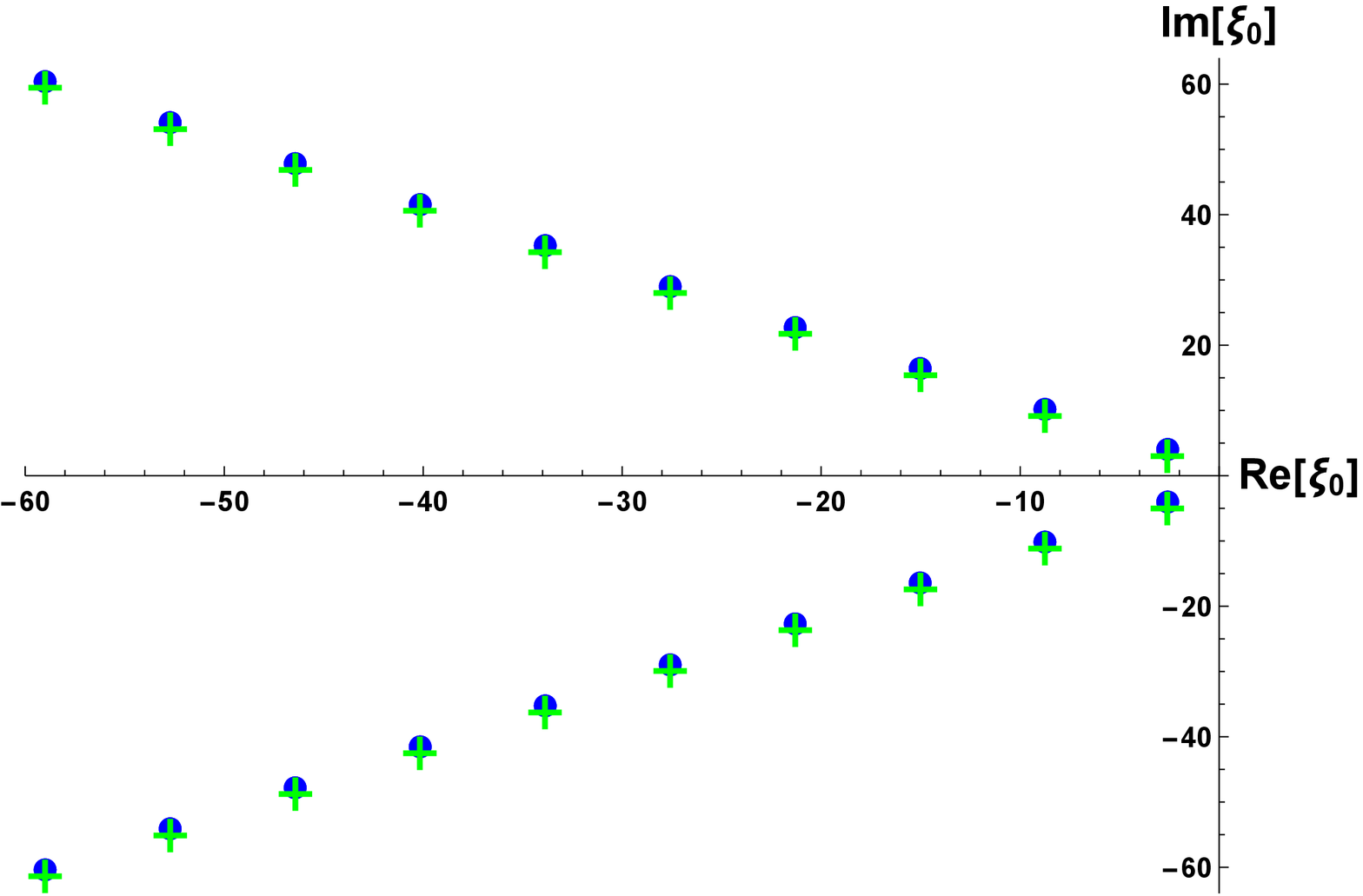}\quad 
  \includegraphics[width=2.95in]{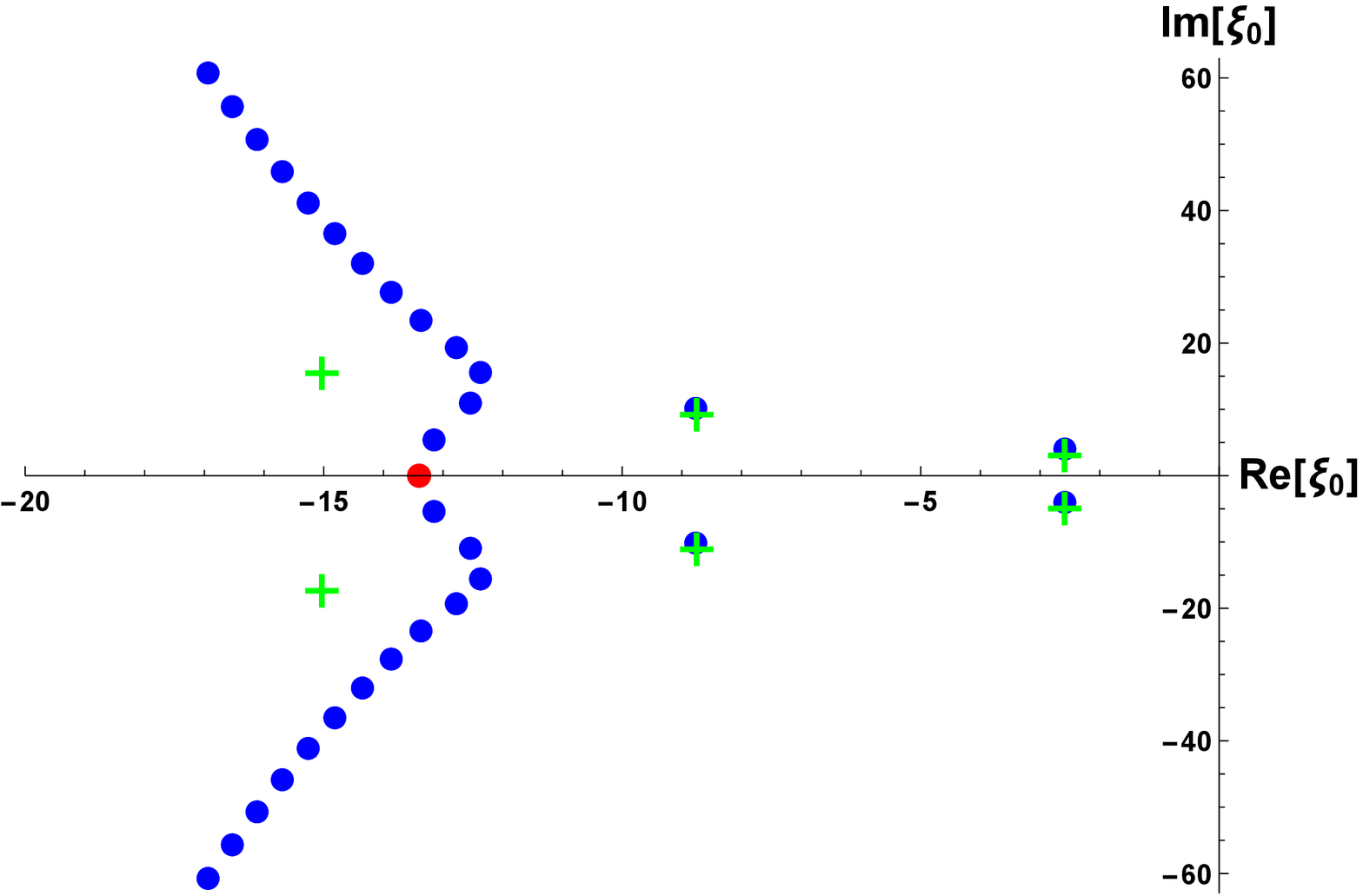}
\end{center}
 \caption{Positions on the Borel plane of leading singularities $\xi_0$ closest to the origin
 for $\Omega_{\Delta=2}^{(B)}$ are given by solid circles for $s=0$ (left panel) and $s=10^{-4}$ (right panel).
 Crosses correspond to QNM frequencies $\w_{QNM}(T)=\hat{\w}_{QNM} T$ and $\xi_0=-i \hat{w}_{QNM}$ for $\Delta=2$ bulk scalar
 for corresponding $s$. Red solid circle characterizes the location of the  wall-of-poles. 
 } \label{figure15} 
\end{figure}

\begin{figure}[t]
\begin{center}
  \includegraphics[width=5in]{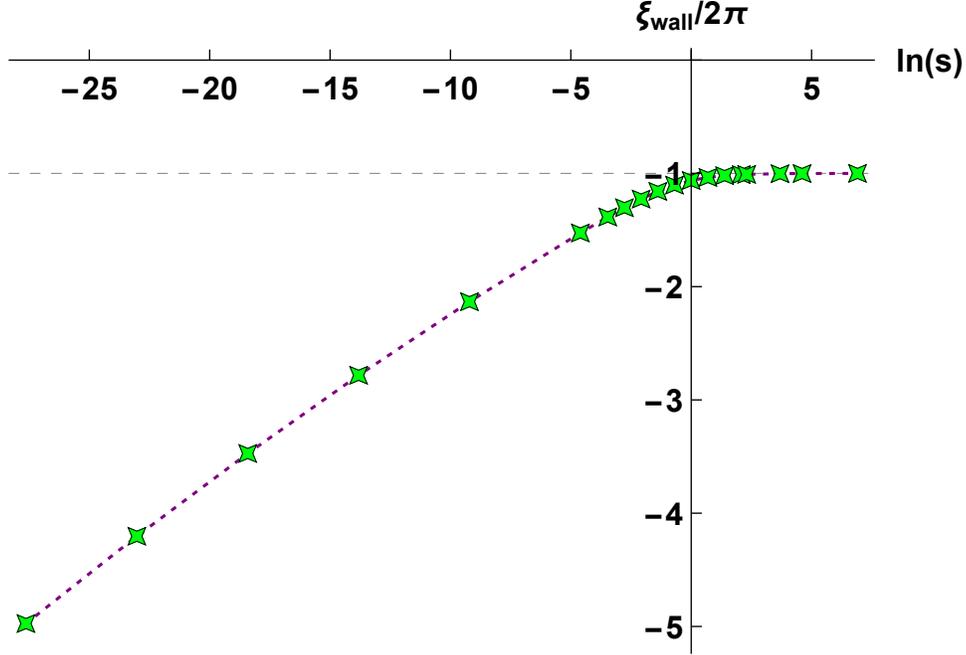}
\end{center}
 \caption{Location of the wall-of-poles as a function of $s$.  The dashed line indicates the asymptotic value $\xi_{wall}=-2\pi i$ for $s \rightarrow \infty$.
 } \label{figure16} 
\end{figure}

\begin{figure}[t]
\begin{center}
  \includegraphics[width=2.95in]{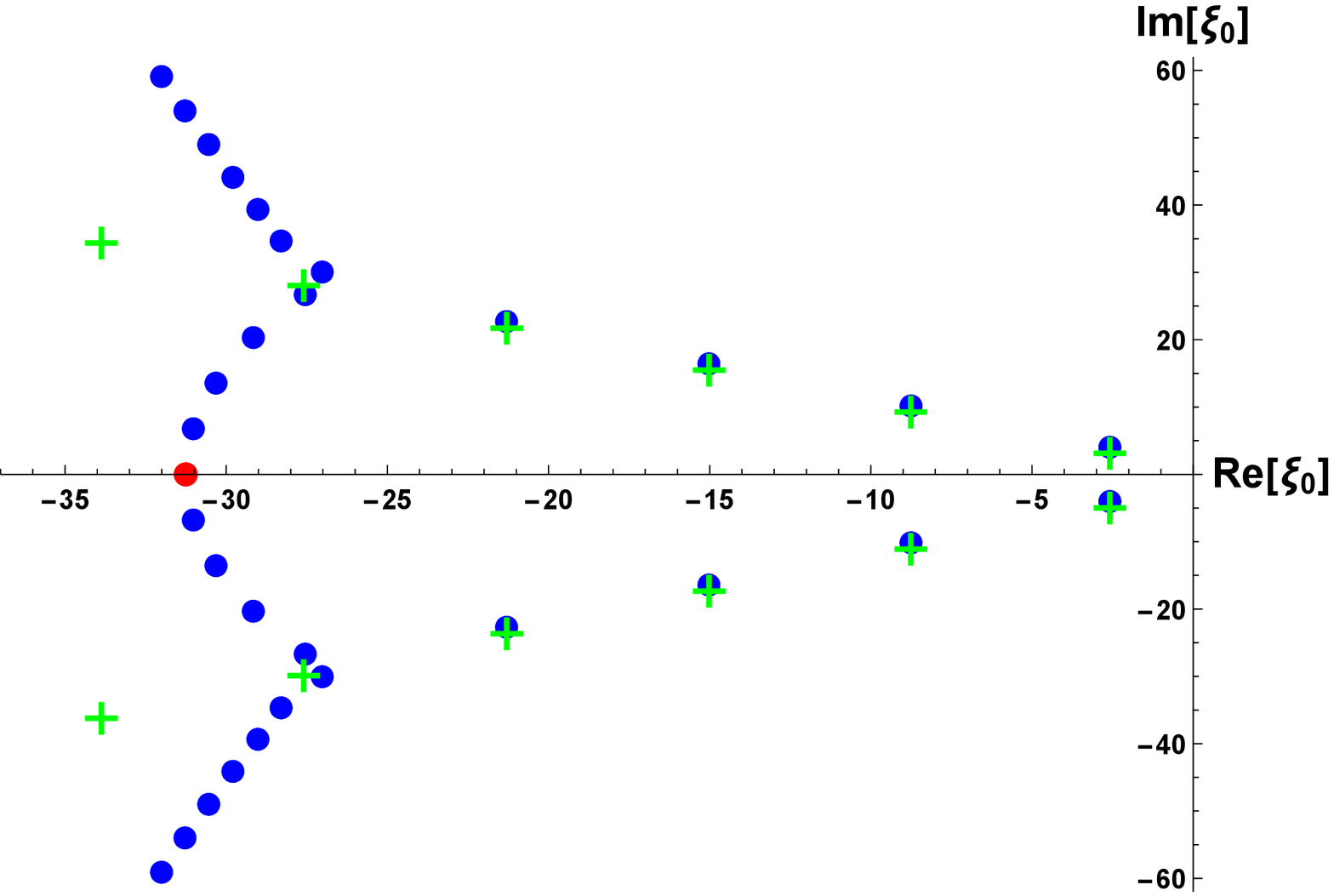}\quad
  \includegraphics[width=2.95in]{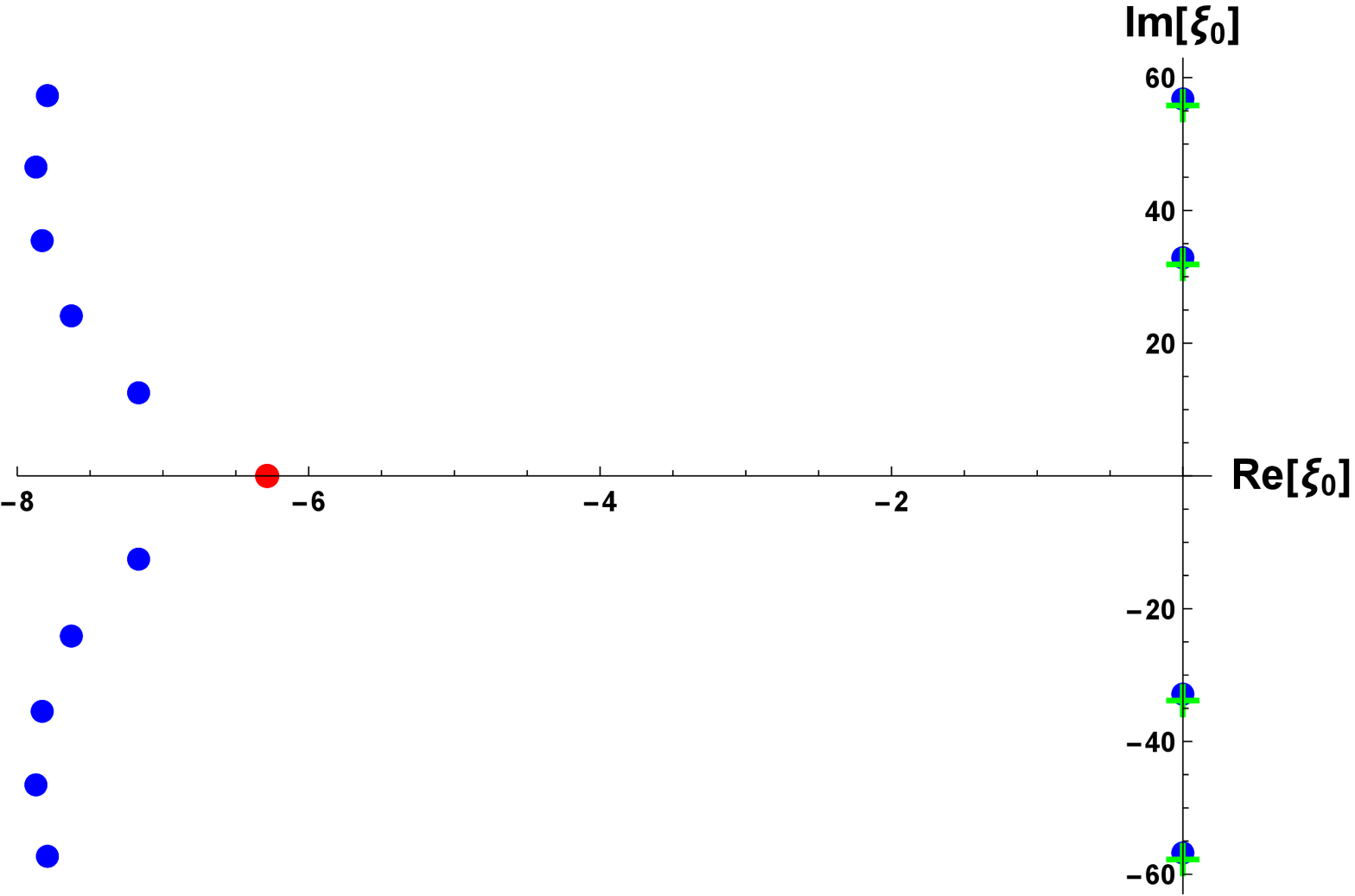}\qquad\qquad
\end{center}
 \caption{Borel plane singularities
 (solid circles) and the QNM frequencies (crosses) at $s=10^{-12}$ (left panel) and $s=1000$ (right panel).
 Red solid circles characterize  the location of the
 wall-of-poles: the wall is relevant for the singularity structure of the Pade approximate of
 $\Omega_{\Delta=2}^{(B)}$ for wide range of values of $s$.
 } \label{figure17} 
\end{figure}

Fig.~\ref{figure15} (left panel) reproduces the result reported in \cite{Buchel:2016cbj} for the agreement between
the QNMs
\begin{equation}
\hat{\w}_{QNM}=\frac{\w}{T}
\eqlabel{defhatw}
\end{equation}
and Borel plane leading singularities $\xi_0$,
\begin{equation}
\xi_0\,\, \Longleftrightarrow\,\, -i \hat{\w}_{QNM}
\eqlabel{fluid}
\end{equation}
in the plasma (fluid) case, \ie  $s=0$. For arbitrary small $s\ne 0$ a \testletter{W}-shaped wall-of-poles appears (right panel,
$s=10^{-4}$): the singularities of the Borel plane to the right of the wall are faithfully reproduced by the QNMs, while those
behind the wall do not. As $s$ further decreases, the wall moves to the left, revealing the agreement between the
singularities and the high-order QNMs. We can track the location of the wall $\xi_{wall}$ as a function of $s$ by its
purely real pole closest to the origin --- depicted by a red circle (right panel),
\begin{equation}
\Im\left[\xi_{wall}\right]=0
\eqlabel{locwall}
\end{equation}
Location of the wall-of-poles is shown in fig.~\ref{figure16}.
Note that the wall is removed very slowly as $s\to 0$; fig.~\ref{figure17} (left panel) shows the structure of the singularities at
$s=10^{-12}$ --- even for this tiny value of $s$ the singularity structure of the Pade approximate of $\Omega_{\Delta=2}^{(B)}$
is distinctly different from the one at $s=0$, see left panel fig.~\ref{figure15}.
The wall of poles persist for large values of $s$, see fig.~\ref{figure17} (right panel):
in the limit $s\gg 1$ its location approaches $\xi_{wall}\to -2\pi$. We want to stress that there is no QNM in the spectrum
with the frequency
\begin{equation}
\hat{\w}_{QNM,wall-solid}=\lim_{s\to \infty} i \xi_{wall} = -2\pi i
\eqlabel{solidwall}
\end{equation}
It is not clear to us whether the wall-of-poles is an artifact of the Pade approximation of $\Omega_{\Delta=2}^{(B)}$
or is a genuine reflection of the viscoelastic properties of the media. 

\begin{figure}[t]
\begin{center}
  \includegraphics[width=2.95in]{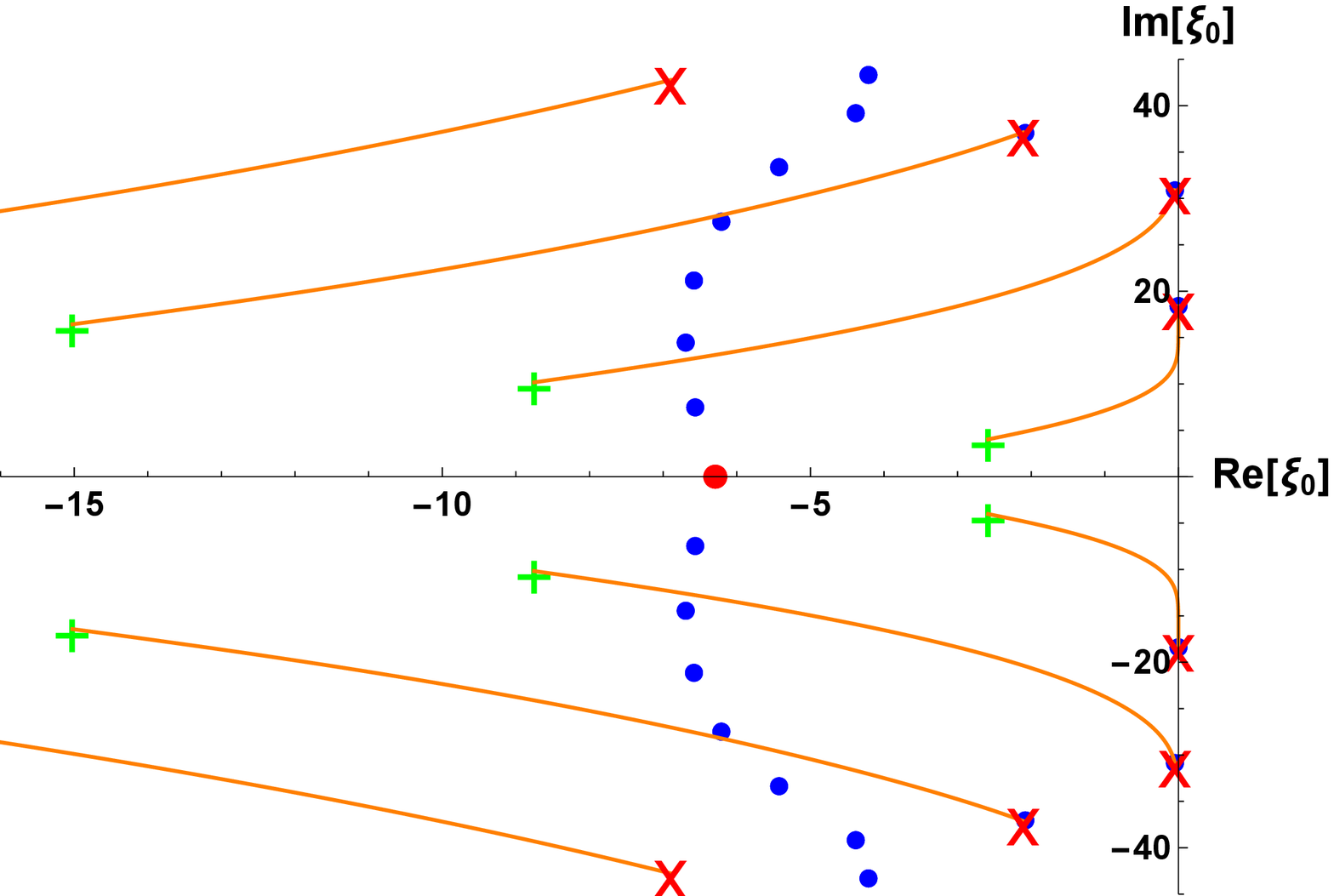}\quad
  \includegraphics[width=2.95in]{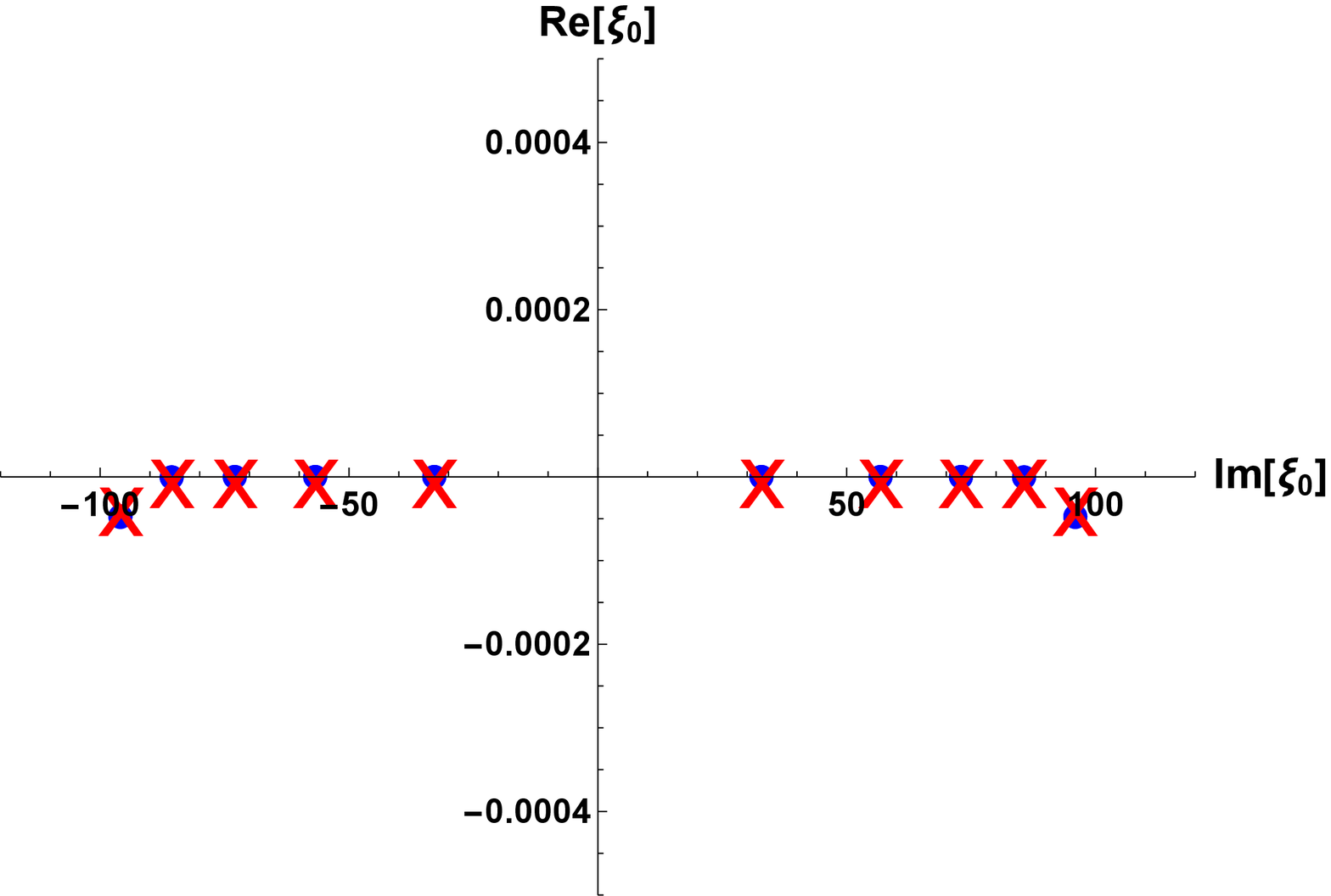}
\end{center}
 \caption{Borel plane singularities
 (solid circles) and the QNM frequencies (crosses) at $s=100$ (left panel) and $s=1000$ (right panel).
 Red solid circles characterize  the location of the
 wall-of-poles. Green crosses represent QNM frequencies at $s=0$, red crosses correspond QNM frequencies at $s=100$
 (left panel) and $s=1000$ (right panel). Orange lines represent the 'flow' of the QNM frequencies as $s\in[0,100]$.
 } \label{figure18} 
\end{figure}

Fig.~\ref{figure18} represents the Borel plane singularities and the QNM frequencies at $s=100$ (left panel) and $s=1000$.
In the left panel we indicated the 'flow' of the QNM frequencies as $s$ changes in the interval $[0,100]$. Again,
the singularities to the right of the wall-of-poles are in excellent agreement with the corresponding QNM frequencies.
Note that as $s$ increases, the Borel plane singularities to the right of the wall move closer to the imaginary axes
(the corresponding $\hat{\w}_{QNM}$ have vanishingly small imaginary part). This is emphasized in the right panel,
where the ratio $\Re[\xi_0]/\Im[\xi_0]$ varies from $\sim 10^{-19}$ to $\sim 10^{-6}$ for the poles closest to the origin
to the one furthest away. In the limit $s\to \infty$ we expect the 'solid' properties of the viscoelastic media to
be enhanced: the QNMs on the branch corresponding to the gravitational bulk scalar fluctuations
approach 'normal modes' --- the nearly non-dissipative fluctuations characteristic of those of the
perfect lattice.

\subsection{Charge plasma with spontaneous symmetry breaking}\label{charged}

\begin{figure}[t]
\begin{center}
  \includegraphics[width=5in]{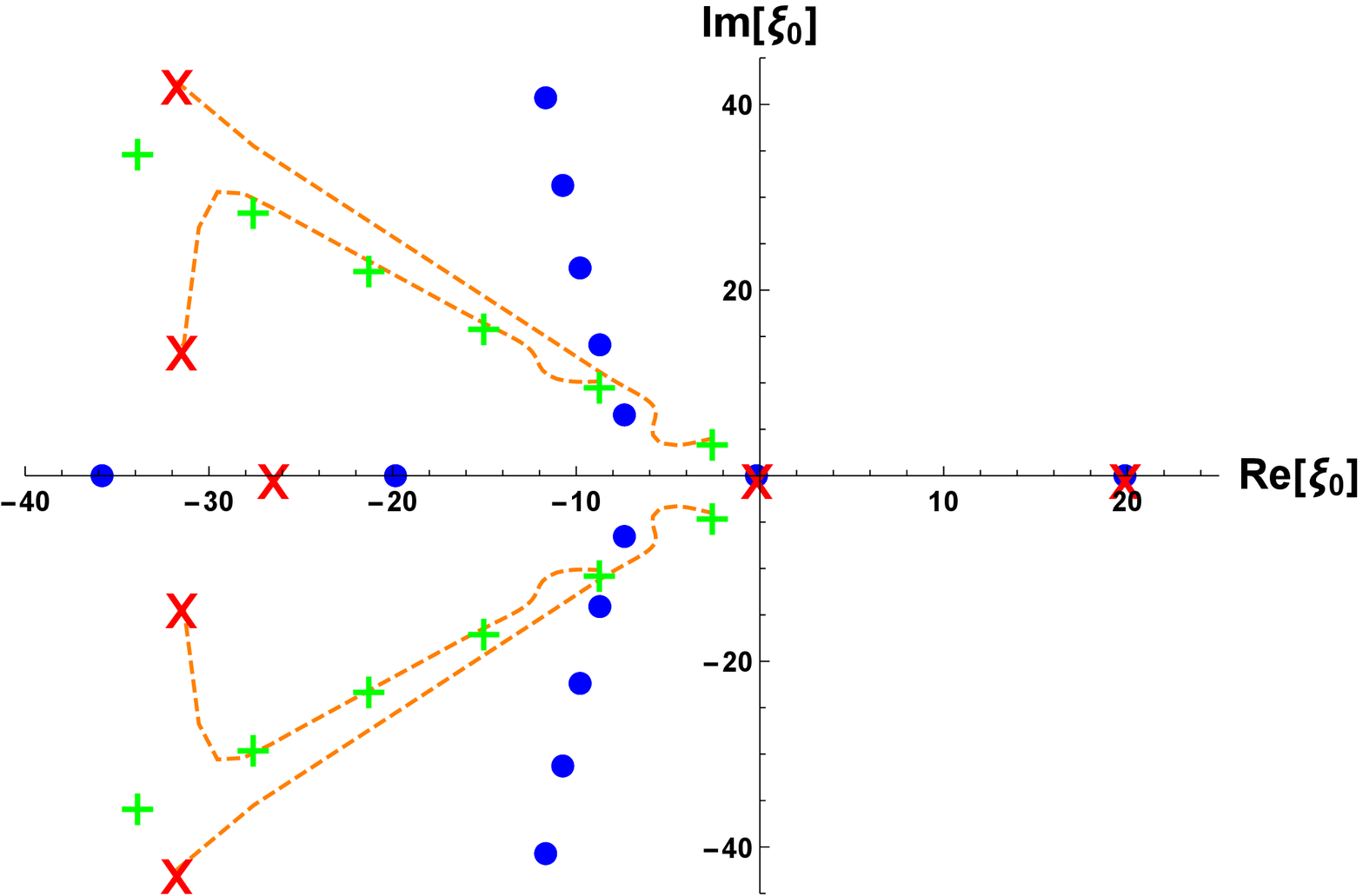}
\end{center}
 \caption{Borel plane singularities
 (solid circles) and the QNM frequencies (crosses) at $s=0$  with
 $\frac{T}{\mu}\approx \frac {1}{10} \frac{T}{\mu}\bigg|_{crit,s=0}$.  Green crosses represent QNM frequencies at
 $\mu=0$, red crosses represent  QNM frequencies at corresponding $\frac{T}{\mu}$. Orange lines
 represent the 'flow' of the QNM frequencies as $\mu$ changes from zero.
 } \label{figure19} 
\end{figure}

We now consider the 'charged fluid/plasma' regime of the viscoelastic media:
$\g=1$ and $s=0$ but with $q\ne 0$. Specifically, we choose $q=2.24$, corresponding to (see fig.~\ref{figure3})
\begin{equation}
\frac{T}{\mu}\approx \frac{1}{10}\ \frac{T}{\mu}\bigg|_{crit,s=0} 
\eqlabel{chargedtmu}
\end{equation}
Since we are below the critical temperature for the spontaneous $\zet_2$ symmetry breaking, the plasma is
unstable: there is a branch of the QNMs with $\Re[\hat{w}_{QNM}]=0$. Some of the modes on this
branch have $\Im[\hat{w}_{QNM}]>0$, signaling the perturbative instability. Of course,
there is also the branch of the QNMs with complex frequencies --- this is $\mu\ne 0$ deformation
of the neutral fluid/plasma QNM frequencies computed in \cite{Nunez:2003eq}. The QNM frequencies
at $q=2.24$ are represented by the red crosses, while those at $q=0$ ($\mu=0$) are represented
by the green crosses in fig.~\ref{figure19}. The orange lines represent the 'flow' as $q\in[0,2.24]$
of the complex QNM frequencies. As before, solid blue circles represent the Borel plane singularities.
There is a good agreement only with the first two QNMs on the $\Re[\hat{w}_{QNM}]=0$ branch.
One of these two modes is unstable --- we see that the Borel resummation
of the hydrodynamic derivative expansion captures the plasma instabilities. 
As in examples in section \ref{neutral}, there is a (distinct) wall-of-poles: there is no
agreement between the QNM frequencies and the Borel plane singularities close-to or to the left of the
wall. As we turn on $s\ne 0$ (at fixed $q\ne 0$), viscoelastic media instabilities
are suppressed, provided $s$ is sufficiently large (see fig.~\ref{figure3}) --- we observed
that in this case all the QNMs on $\Re[\hat{w}_{QNM}]=0$ branch are stable. While we did not study this
in details, we believe that the wall-of-poles  will move to $\Re[\xi_0]\to -\infty$
as $q\to 0$, revealing better agreement with the Borel plane singularities
(for small enough $q$ the QNM branch with  $\Re[\hat{w}_{QNM}]=0$ will disappear).
It would be interesting to better understand why there is worse agreement
between the Borel plane singularities and the QNM frequencies for the charged viscoelastic media.
This might be related to the present of additional
singularities\footnote{Similar phenomenon was observed in \cite{Buchel:2018ttd}.}
in \eqref{eomseries} at $q\ne 0$ (also for $\l_1\ne 0$), which renders Pade approximation
not reliable for the high-order poles of the Borel transform of $\Omega_\Delta$ ---
\ie there might be additional singularities (besides the poles) of the Borel transform.

\section{Conclusions and discussion}\label{conclude}

The goal of this project was understanding the large-order perturbative expansion
of a viscoelastic media with a control parameter smoothly interpolating between the
fluid and the solid. Previously, both the all-order fluid hydrodynamics
\cite{Heller:2013fn} and the all-order theory of elasticity \cite{1996PhRvL..77.1520B}
were argued to be asymptotic expansions (in velocity gradients and strain gradients correspondingly).
The nonperturbative effects responsible for the zero radius of convergence of the
perturbative expansions were argued to be  distinct in both cases:
these are the QNMs of the fluid and cracks in brittle solids. 
To analyze the problem in a controlled setting, we embedded it in the framework of
holographic gauge theory/gravity correspondence. Specifically, we focused on a class of viscoelastic
holographic models introduced in \cite{Donos:2013eha}:
\nxt the condensate of the holographic superconductor is coupled to the axion sector
with a flavor symmetry;
\nxt turning on a source term for the axions necessitates the introduction of the
spatial lattice, explicitly breaking the translational invariance;
\nxt once the axion flavor symmetry is spontaneously (or explicitly) broken,
the homogeneous and isotropic states of the viscoelastic media 'feel' the breaking of the
translational invariance, leading to the a nonzero shear elastic modulus --- an important
differentiating feature between fluids and solids;
\nxt controlling the scale of the source term for the axions
(inversely proportional to the spatial lattice spacing $\Delta x$) we can interpolate
between the fluid and the solid regimes of the viscoelastic media.

Physically, the intuitive picture of our holographic viscoelastic media is that of the
crystalline lattice of spacing $\Delta x$, immersed in a fluid: as $\Delta x\to \infty$
the lattice is removed and one is left with the fluid only.
We presented a rather detailed exploration of the  media \eqref{action}:
\begin{itemize}
\item We studied the equilibrium phase diagram of the model with both explicit and
spontaneous flavor symmetry breaking. We identified the critical behavior and the instabilities
in the microcanonical and grand canonical ensembles. As common in holographic models,
the critical behavior is mean-field.  
\item We showed that either with explicit or spontaneous flavor symmetry breaking
in the model there is a nonzero shear elastic modulus $G$ when the lattice spacing
${\Delta x}$ is finite.
Furthermore, the shear elastic modulus vanishes in the 'fluid limit':
\begin{equation}
\lim_{\Delta x\to \infty} G=0
\eqlabel{fluidlimit}
\end{equation}
\item In case of the spontaneous symmetry breaking, we utilized the underlying
conformal symmetry of the model to show that there is a nonzero bulk elastic modulus $\calk$ as well,
when $\Delta x$ is finite. Both the shear and the bulk elastic moduli vanish at the
criticality, $G\sim (T_{crit}-T)^{\a_G}$ and $\calk\sim (T_{crit}-T)^{\a_\calk}$,
with the same critical exponent $\a_G=\a_\calk=1$. It is clear that these critical exponents are not
universal, and instead depend on the coupling of the condensate to the axion sector of the model.
Specifically, modifying the coupling of the bulk scalar $\phi$ to the axions $\psi_I$ from
$\phi^2$ to $\phi^{2n}$ is expected to produce critical exponents $\a_G=\a_\calk=n$.
The reason for this is simple: close to critically $\phi \sim (T_{crit}-T)^{1/2}$, characteristic
of the mean-field behavior, and the elastic moduli respond proportional to the coupling strength,
\ie $\sim \phi^{2n}$.
\item In case of the explicit flavor symmetry breaking we found that the shear elastic modulus
$G\to {\rm const}$ as $\Delta x\to 0$, indicative of the 'solid limit'.  
\item We analyzed the hydrodynamic transport of the model: the shear viscosity
(extracted via a Kubo formula from the retarded two-point correlation function
of the stress-energy tensor) and the bulk viscosity. The shear viscosity  violates the
 bound $\frac{\eta}{s} \ge \frac{1}{4\pi}$
 both for the explicit and the spontaneous flavor symmetry breaking
 at finite $\Delta x$. This violation vanishes in the fluid limit, and at the criticality
 (in case of the spontaneous symmetry breaking). Somewhat unexpectedly,
 both the violation of the shear viscosity bound and the bulk viscosity
 vanish in the solid limit as well, \ie when $\Delta x\to 0$. The physical reason is unclear to us,
 but at a technical level this is related to the
 exponential suppression of the bulk scalar field $\phi$ in our model
 at the horizon of the gravitational dual in the limit $\Delta x\to 0$. Since the hydrodynamic
 transport is fully determined by the small fluctuations of the horizon, in the solid limit, the
 horizon is scalar-hair free, leading to the universal shear viscosity and the vanishing bulk viscosity.
 Note that the bulk scalar $\phi$ is not suppressed near the AdS boundary in the solid limit,
 which is likely responsible for the nonvanishing of the elastic shear modulus in this limit.
\item  Although the hydrodynamic transport (shear and bulk viscosities) are that of the conformal
fluid in the 'solid limit' of our viscoelastic media, the nonhydrodynamic excitations
feel the lattice. In the limit $\Delta x \to 0$ only the modes with vanishing spatial momentum remain
in the spectrum. Modes in the scalar/shear channels (according to classification of
\cite{Kovtun:2005ev}) experience a small but nonzero off-set, see \eqref{notthesame}. The nonhydrodynamic
modes of the condensate are profoundly affected: the complex frequencies on this branch have
exponentially suppressed imaginary part compare to the real part, see the right panel of
fig.~\ref{figure18}. They represent the nearly non-dissipative fluctuations characteristic of those of
the perfect lattice.
\item We studied the homogeneous and the isotropic expansion of the viscoelastic media
to all orders in the velocity gradients, but to leading order in the
explicit flavor symmetry breaking.
We focused on a particular observable: the non-equilibrium entropy production rate.
For neutral viscoelastic media we found a good agreement between the poles of the
Pade approximant of the Borel transform of the observable, and the corresponding QNM frequencies.
The Borel plane singularities closest to the
origin follow the flow of the QNM with $\Delta x$, correctly reproducing the solid limit
with nearly nondissipative nonhydrodynamic excitations of the
flavor symmetry breaking condensate, see  fig.~\ref{figure18}. We observed a \testletter{W}-shaped
wall-of-poles, which presence hampers the agreement between the Borel plane singularities and
the QNM frequencies. We observed the lingering effects of the wall in the fluid limit, \ie
$\Delta x\to \infty$, see fig.~\ref{figure16}. Unfortunately we could not establish whether the
presence of the wall indicates a new nonperturbative effect contributing to the asymptotic character
of the viscoelastic gradient expansion, or is simply an artifact of the Pade
approximation. We did establish that the presence of the wall is robust with respect
to the truncation order in series expansion \eqref{entprod2}. For charged viscoelastic media
there is also a wall-of-poles, albeit a different one --- in this case the agreement
between the Borel plane singularities and the QNM frequencies is limited to 1-2 modes, closest
to the origin. We established that the all-order derivative expansion for the charged fluid
correctly identifies nonhydrodynamic instabilities associated with the spontaneous flavor
symmetry breaking, see fig.~\ref{figure19}.
\end{itemize}

There are many open questions left for future analysis:
\nxt First and foremost, we did not identify the cracks present within the theory of elasticity for brittle
materials. We don't know whether our holographic
viscoelastic media is brittle. The wall-of-poles on the Borel plane hints to additional
singularity structure(s), beyond the poles of the Pade approximant.
\nxt We focused on one observable, \ie
the nonequilibrium entropy production rate. What are the large-order hydrodynamic expansions
of one- and higher-point correlation functions and of the entanglement entropy? 
\nxt To  leading order in the explicit symmetry breaking, the elastic moduli and the transport
coefficients diverge as one approaches the criticality. The observed divergences
(see figs.~\ref{figure6}-\ref{figure7} for example) are outside the regime of the validity of the probe
approximation and the full nonlinear analysis is necessary.
\nxt We are not aware of any discussion of the bulk elastic modulus in
holographic nonconformal models\footnote{With the exception of a formula given, but not studied in any direction, in \cite{Amoretti:2017frz}.}.
\nxt To facilitate the computations, we restricted the all-order hydrodynamic analysis to a particular
flow: the homogeneous and isotropic expansion. It would be interesting to analyze hydrodynamic
perturbation theory of the boost-invariant expansion. 
\nxt We studied nonhydrodynamic excitations in our viscoelastic media
in the scalar/shear channels, and for the condensate. The analysis of the QNMs in the sound channel is left for the future work.
\nxt It would be interesting to analyze the fate of the bulk viscosity bound \cite{Buchel:2007mf} in the presence of the elastic features.
\nxt Quenching the holographic media and study of its thermalization, as done for example in \cite{Buchel:2013lla,Buchel:2014gta,Buchel:2015saa}, could provide further insights about its viscoelastic features and in particular the possible presence of slow relaxation and aging typical of glassy systems (see \cite{Kachru:2009xf,Anninos:2013mfa} for some early attempts).
\nxt Lastly, it is important to explore different holographic viscoelastic models 	\cite{Alberte:2015isw,Grozdanov:2018ewh}
within the context of all-order hydrodynamic expansion. A particularly interesting
case is represented by those models where the shear viscosity vanishes in the 'solid limit' \cite{Hartnoll:2016tri,Alberte:2016xja,Burikham:2016roo}.

\section*{Acknowledgments}
We would like to thank Martin Ammon, Andrea Amoretti, Colin Denniston, Blaise Gouteraux, Elias Kiritsis, Napat Poovuttikul, Oriol Pujolas and Dam Thanh Son for useful discussions and comments about this work.\\
Research at Perimeter
Institute is supported by the Government of Canada through Industry
Canada and by the Province of Ontario through the Ministry of
Research \& Innovation. This work was further supported by
NSERC through the Discovery Grants program.
MB is supported in part by the Advanced ERC grant SM-grav, No 669288.\\
We would like to thank the organizers and participants of the
Holography and Supergravity 2018 (Chile) for the inspiring atmosphere
where this projects started.\\AB would further
like to thank the Hebrew University of Jerusalem, Aspen Center for Physics
and the Institute for Nuclear Theory at the University of Washington
for the hospitality and partial support during the completion of this work.\\
MB would like to thank the Hebrew University of Jerusalem, the University of Iceland, the Enartia Academy and Kritikos Fournos for the hospitality during the completion of this work. MB thanks Marianna Siouti for the unconditional support.

\appendix
\section{Details about the model and its EOMs}\label{App1}
We start with an effective Maxwell-Enstein-Hilbert action in asymptotically $AdS_5$
geometry with
three complex scalar fields $\Phi_I$ enjoying $SU(3)$ flavor symmetry:
\begin{equation}
\begin{split}
S=&\frac{1}{16\pi G_N}\int_{\calm_{5}}d^5 x \sqrt{-g} \biggl[R-\frac {\dd^{IJ}}{3}\del\Phi_I\del\bar{\Phi}_J-\frac 14 Z(\Phi_I\bar{\Phi}^I)\ F^2
-V(\Phi_I\bar{\Phi}^I)\biggr]\\
V=&-12+ \frac{m^2}{3}\ \Phi_I\bar{\Phi}^I\,,\qquad Z=1+\frac{2\g}{3}\ \Phi_I\bar{\Phi}^I\,,\qquad m^2=\Delta(\Delta-4)
\end{split}
\eqlabel{action1}
\end{equation}
where $\Delta$ is a scaling dimension of the dual boundary operator, and the bulk coupling constant obeys $\g>0$.  
Introducing
\begin{equation}
\Phi_I=\frac{\phi}{\sqrt{2}} e^{i\sqrt{3}\psi_I}\,,\qquad {\rm with\ identification}\qquad \psi_I\sim \psi_I+\frac{2\pi}{\sqrt{3}}\,,
\eqlabel{defpsi}
\end{equation}
we rewrite \eqref{action1} with manifest $U(1)^3$ flavor symmetry
\begin{equation}
\begin{split}
S=\frac{1}{16\pi G_N}\int_{\calm_{5}}d^5 x \sqrt{-g} \biggl[&R+12-\frac12 (\del\phi)^2-\frac 14 (1+\g \phi^2)\
F^2+\frac{\Delta(4-\Delta)}{2}\ \phi^2\\
&-\frac {1}{2}\phi^2\ \sum_{I=1}^3 \biggl\{(\del\psi_I)^2\biggr\}\biggr]
\end{split}
\eqlabel{action2}
\end{equation}
Notice that a consistent truncation of \eqref{action2} with $\psi_I\equiv 0$
leads to an effective action $S_{sc}$ of a neutral holographic superconductor:
\begin{equation}
\begin{split}
S_{sc}=\frac{1}{16\pi G_N}\int_{\calm_{5}}d^5 x \sqrt{-g} \biggl[&R+12-\frac12 (\del\phi)^2-\frac 14 (1+\g \phi^2)\
F^2+\frac{\Delta(4-\Delta)}{2}\ \phi^2\biggr]
\end{split}
\eqlabel{holsup}
\end{equation}
where the operator $\calo_\phi$, dual to the bulk mode $\phi$, has a conformal dimension $\Delta$.\\
The effective action we use throughout the paper is a simple generalization of \eqref{action2}.\\
As in \cite{Hartnoll:2008kx}, in the absence of the source term for
a real bulk scalar $\phi$, spatially homogeneous and isotropic thermal charged states of \eqref{holsup}  
will become unstable with respect to a $\phi$-condensate (below some critical temperature or the energy density);
the role of the bulk coupling constant $\g>0$ is to stimulate this condensation process, \ie increase the
critical temperature/energy density. \\[0.1cm]
The equations of motion which follows from the action \eqref{action} presented in the main text are,
$d_+\equiv \del_t-x^2 A\del_x $ and $'\equiv \del_x$, $\cdot \equiv \del_t$,
\begin{equation}
\begin{split}
0=&(d_+\phi)'+\frac 32(\ln\Sigma)' d_+\phi+\frac 32\phi' d_+(\ln\Sigma)-\frac \g2 x^2(a_0')^2\phi+\frac{\phi}{2x^2}
\biggl(\Delta(\Delta-4)\\
&+\frac{3\l_1k^2}{\Sigma^2}+\frac{3\l_2k^4}{\Sigma^4}\biggr)
\end{split}
\eqlabel{eom1}
\end{equation}
\begin{equation}
\begin{split}
&0=(d_+\Sigma)'+2(\ln\Sigma)'d_+\Sigma-\frac{x^2}{12}\Sigma(a_0')^2(1+\g\phi^2)+\frac{2}{x^2}\Sigma
-\frac{\phi^2\Sigma}{12x^2}\biggl(\Delta(\Delta-4)\\
&+\frac{3\l_1k^2}{\Sigma^2}+\frac{3\l_2k^4}{\Sigma^4}\biggr)
\end{split}
\eqlabel{eom2}
\end{equation}
\begin{equation}
\begin{split}
&0=A''+\frac 2x A'+\frac{6}{x^2}(\ln\Sigma)'d_+(\ln\Sigma)-\frac{\phi'}{2x^2}d_+\phi-\frac{7}{12}
(a_0')^2(1+\g\phi^2)+\frac{2}{x^4}\\&-\frac{\phi^2}{12x^4}\biggl(\Delta(\Delta-4)
+\frac{9\l_1k^2}{\Sigma^2}+\frac{15\l_2k^4}{\Sigma^4}\biggr)
\end{split}
\eqlabel{eom3}
\end{equation}
\begin{equation}
\begin{split}
&0=a_0''+a_0'\ \left(\ln(x^2\Sigma^3)(1+\g\phi^2)\right)'
\end{split}
\eqlabel{eom4}
\end{equation}
Additionally, there are 'constraints':
\begin{equation}
\begin{split}
0=&\Sigma''+\frac 2x\Sigma'+\frac 16\Sigma (\phi')^2
\end{split}
\eqlabel{const1}
\end{equation}
\begin{equation}
\begin{split}
0=&d_+^2\Sigma+A' x^2d_+\Sigma+\frac 16\Sigma (d_+\phi)^2
\end{split}
\eqlabel{const2}
\end{equation}
\begin{equation}
\begin{split}
0=&\del_t\left(a_0'\Sigma^3 (1+\g\phi^2)\right)
\end{split}
\eqlabel{const3}
\end{equation}
Once the equation \eqref{const1} is satisfied at some time $t=0$, by virtue of \eqref{eom1}-\eqref{eom4} it is satisfied
at any time $t>0$. Similarly, once equations \eqref{const2}-\eqref{const3} are satisfied at any radial location,
\eg the $AdS_5$ boundary at $x=0$, for all times, they are satisfied at any other radial location. 
These equations represent the conservation of the energy density (eq.~\eqref{const2}) and the charge density
(eq.~\eqref{const3}). \\[0.1cm]

\textbf{EOMs in Fefferman-Graham coordinates}\\[0.1cm]
The EOMs in the FG coordinates \eqref{fgmetric} are given by 
\begin{equation}
\begin{split}
0=&C''+\frac{C (\g \phi^2+1)}{6A} (a_0')^2-4 C B+\frac C6 (\phi')^2-\frac{C' B'}{2B}
\\&+ \frac{B \phi^2}{6} \left(C \Delta (\Delta-4)+\frac{3 k^4 \l_2}{C}+3 k^2 \l_1\right)
\end{split}
\eqlabel{fg10}
\end{equation}
\begin{equation}
\begin{split}
0=&A''-\frac{(A')^2}{2A}+\left(\frac{C'}{C}-\frac{B'}{2B}\right) A'-\frac{A}{2C^2} (C')^2
+\frac A6 (\phi')^2-\frac 56 (1+\g \phi^2) (a_0')^2-4 A B
\\&+\frac{\phi^2 A B}{6} \left(\Delta (\Delta-4)-\frac{9 k^4 \l_2}{C^2}-\frac{3 k^2 \l_1}{C}\right)
\end{split}
\eqlabel{fg20}
\end{equation}
\begin{equation}
\begin{split}
0=&\phi''+\left(\frac{A'}{2A}+\frac{3C'}{2C}
-\frac{B'}{2B}\right) \phi'+\phi \left(\frac \g A (a_0')^2-B \left(
\Delta (\Delta-4)+\frac{3 k^4 \l_2}{C^2}+\frac{3 k^2 \l_1}{C}\right)
\right)
\end{split}
\eqlabel{fg30}
\end{equation}
\begin{equation}
\begin{split}
0=&(\g \phi^2+1) a_0''+a_0' \left( (\phi^2)'  \g-\left(
\frac{A'}{2A}+ \frac{B'}{2B}
- \frac{3C'}{2C}
\right) (1+\g \phi^2)\right)
\end{split}
\eqlabel{fg40}
\end{equation}
and the first-order constraint, related to reparametrization of the radial coordinate, 
\begin{equation}
\begin{split}
0=&(a_0')^2 (1+\g \phi^2)-A (\phi')^2+\frac{3 A (C')^2}{C^2}
+\frac{3 C' A'}{C}-24 A B
\\&+A B \phi^2\left(\Delta (\Delta-4)+\frac{3 k^4 \l_2}{C^2}+\frac{3 k^2 \l_1}{C}\right) 
\end{split}
\eqlabel{fgc}
\end{equation}
\section{Shear elastic modulus and shear viscosity at leading order}\label{App3}
Notice that to leading order in $\calo(\phi^2)$,
both the metric warp factors $\{A,B,C\}$ and the bulk gauge potential
$a_0$ are corrected. However, as we show, the viscoelastic transport is completely determined
by the corresponding quantities without the bulk scalar backreaction, \ie the
$\zet_2$ symmetric or 
the Reissner-Nordström black brane solution:
\begin{equation}
\begin{split}
&A_{\zet_2}=\frac{f}{x^2}\,,\qquad B_{\zet_2}=\frac{1}{x^2 f}\,,\qquad C_{\zet_2}=\frac {1}{x^2}\\
&f=1-\frac{x^4}{x_h^4}-\frac{\mu^2 x^4}{3x_h^2}\left(1-\frac{x^2}{x_h^2}\right)\,,\qquad a_{0,\zet_2}=\mu\left(1
-\frac{x^2}{x_h^2}\right)
\end{split}
\eqlabel{rnfg}
\end{equation}
where $x_h$ is the location of the horizon, related to temperature $T$,
\begin{equation}
T=\frac{1}{\pi x_h}-\frac{\mu^2 x_h}{6\pi}
\eqlabel{tfg}
\end{equation}
and  $\mu$ is the chemical potential.

To extract $\{G,\eta\}$ we need to solve \eqref{heom} to order $\calo(\w)$, subject to the boundary
condition \eqref{expaH} and an incoming wave boundary condition at the horizon.
It is convenient to represent
\begin{equation}
H=H_0(x;\w)+H_1(x;\w)
\eqlabel{hsplit}
\end{equation}
where the two functions satisfy, correspondingly\footnote{The term $\calo(\w^2)$ is important to properly
set up the incoming wave boundary conditions, but can be neglected otherwise.}
\begin{equation}
\begin{split}
0&=H_0''+\frac 12 H_0' \left( \ln \frac{AC^3}{B}\right)'\\
0&=H_1''+\frac 12 H_1' \left( \ln \frac{AC^3}{B}\right)'
+H_0 \left(-\frac{B \phi^2 k^2}{C^2} \left(2 k^2 \lambda_2+C \lambda_1\right)\right)
\end{split}
\eqlabel{hspliteom}
\end{equation}
where $H_0$ and $H_1$ has to be solved to order $\calo(\phi^2)$ --- the latter guarantees that
\eqref{hsplit} solves \eqref{heom} to order $\calo(\phi^2)$.
The first equation in \eqref{hspliteom} can be solved, to all orders in $\phi$, following
\cite{Benincasa:2006fu}:
\nxt The general solution takes form 
\begin{equation}
H_0=h_{0,0}+h_{0,1}\ \int_0^x dy\ \frac{B(y)^{1/2}}{A(y)^{1/2} C(y)^{3/2}}
\eqlabel{h0}
\end{equation}
where the integration constants $h_{0,i}$ are fixed from the boundary conditions as
\begin{equation}
h_{0,0}=1\,,\qquad h_{0,1}=\frac{i \w}{x_h^3}
\eqlabel{bc0}
\end{equation}
Expanding \eqref{h0} near the boundary, we find
\begin{equation}
H_0=h_{0,0}+\frac{h_{0,1}}{4}\ x^4+\calo(x^8\ln^2 x)= 1+\frac{i\w}{4x_h^3}\ x^4+\calo(x^8\ln^2 x)
\eqlabel{expansion}
\end{equation}
leading to (see \eqref{expaH}-\eqref{viscotransport})
\begin{equation}
\begin{split}
&h_{4,0}^b=\frac{i\w}{4 x_h^3}\qquad \Longrightarrow\qquad \calg^{R}_{x_1x_2,x_1x_2}=-\frac{i\w}{4\pi}\ \frac{1}{4 G_N x_h^3}
=-\frac{i\w}{4\pi}\ \cals\\
&\Longrightarrow\qquad 
G\bigg|_0=0\,,\qquad \frac{\eta}{\cals}\bigg|_{0}=\frac{1}{4\pi}
\end{split}
\eqlabel{order0}
\end{equation}
\ie to this order, elastic shear modulus vanishes, and the shear viscosity is universal, as first established
at this level of generality in \cite{Benincasa:2006fu}.
\nxt Having $H_0$ \eqref{h0}, we can analytically
solve\footnote{The "$-$'' sign is due to the fact that the inner integral is $\int_y[\cdots]$.}
the second equation in \eqref{hspliteom}:
\begin{equation}
\begin{split}
H_1(x)=&-\int_0^x dy\ \frac{B(y)^{1/2}}{A(y)^{1/2}C(y)^{3/2}}\\
&\times \biggl\{\int_{y}^{x_h}
dz \frac{H_0(z)A(z)^{1/2}B(z)^{1/2}\phi(z)^2}{C(z)^{1/2}}(\l_1 k^2 C(z)+2k^4\l_2)\biggr\}
\end{split}
\eqlabel{solveH1}
\end{equation}
Notice that $H_1(x\to 0)=0$, and so the asymptotic normalization of $H$ in \eqref{expaH} is preserved;
additionally, because the inner integral $\int_y^{x_h}[\cdots] \to 0$ as $y\to x_h$ $H_1(x)$ does not
have a log-singularity as $x\to x_h$ implying that the incoming wave boundary conditions set by $H_0(x)$
is not modified. From \eqref{solveH1} we extract
\begin{equation}
h_{4,1}^b=-\frac 14\ \int_{0}^{x_h}
dz \frac{H_0(z)A(z)^{1/2}B(z)^{1/2}\phi(z)^2}{C(z)^{1/2}}(\l_1 k^2 C(z)+2k^4\l_2)
\eqlabel{order1}
\end{equation}
Recalling \eqref{h0}-\eqref{bc0}
we conclude
\begin{equation}
\begin{split}
\Re[h_{4,1}^b]=&-\frac 14\ \int_{0}^{x_h}
dz \frac{A(z)^{1/2}B(z)^{1/2}\phi(z)^2}{C(z)^{1/2}}(\l_1 k^2 C(z)+2k^4\l_2)\\
\Im[h_{4,1}^b]=&-\frac {\w x_h^3}4\ \int_{0}^{x_h}
dz \frac{A(z)^{1/2}B(z)^{1/2}\phi(z)^2}{C(z)^{1/2}}(\l_1 k^2 C(z)+2k^4\l_2)\\
&\times \biggl\{\int_0^z  dt\ \frac{B(t)^{1/2}}{A(t)^{1/2} C(t)^{3/2}}\biggr\} 
\end{split}
\eqlabel{h41}
\end{equation}
resulting in
\begin{equation}
\begin{split}
&16\pi G_N\  G\bigg|_1=\int_{0}^{x_h}
dz \frac{A(z)^{1/2}B(z)^{1/2}\phi(z)^2}{C(z)^{1/2}}(\l_1 k^2 C(z)+2k^4\l_2)\\
&\frac{\eta}{\cals}=-\frac{1}{4\pi}  \int_{0}^{x_h}
dz \frac{A(z)^{1/2}B(z)^{1/2}\phi(z)^2}{C(z)^{1/2}}(\l_1 k^2 C(z)+2k^4\l_2)
\biggl\{\int_0^z  dt\ \frac{B(t)^{1/2}}{A(t)^{1/2} C(t)^{3/2}}\biggr\} 
\end{split}
\eqlabel{geta1}
\end{equation}
To $\calo(\dd_\Delta)$, using \eqref{rnfg}, we find
\begin{equation}
\begin{split}
0=&p''+\frac{(3 z^2-1) \b^2 z^4-3 z^4-9}{z (z^2-1) (\b^2 z^4-3 z^2-3)}  p'
-\frac{3 (3 s^2 z^2 (\l_2 s^2 z^2+ \l_1) -4 \b^2 \g z^6+\Delta(\Delta-4))}{(\b^2 z^4-3 z^2-3) (z^2-1) z^2} p\,,
\end{split}
\eqlabel{pfg}
\end{equation}
where $p\equiv \frac{\phi}{\dd_\Delta x_h^{4-\Delta}}$, and similar to \eqref{phieom} we introduced
\begin{equation}
z=\frac{x}{x_h}\in [0,1]\,,\qquad k=\frac{s}{x_h}\,,\qquad \mu=\frac{\b}{x_h}\ \Longrightarrow  \frac \mu T= \frac{6\b\pi}{6-\b^2}
\eqlabel{notationsfg}
\end{equation}
with the extremality reached as $\b\to \sqrt{6}_-$. Eq.~\eqref{pfg} has to be solved subject to the following
boundary conditions ($\Delta=2$):
\begin{equation}
p=-z^2\ln z+\calo(z^2)\,,\qquad p=\calo(1) 
\eqlabel{bcfg}
\end{equation}
Once the solution for the $p$ is determined (numerically), and parameterizing
\begin{equation}
\frac{16\pi G_N\ G }{\dd_2^2}=\hat{G}(z=1)\,,\qquad \frac \eta \cals= \frac{1}{4\pi}\left(1-\frac{\dd_2^2}{T^4}\ 
\k(z=1)\right)
\eqlabel{getaphi2n}
\end{equation}
we find from \eqref{getalead} the differential equations for ${\hat G}(z)$ and $\k(z)$:
\begin{equation}
\begin{split}
0=&\hat{G}'-\frac{2s^4p^2}{z}\,,\qquad \hat{G}(z=0)=0\\
0=&\k'-\frac{(\b^2-6)^3 s^4 p^2}{96z (12 \b^2+9)^{1/2} \pi^4} \biggl(
\ln\frac{\sqrt{12 \b^2+9}+3}{\sqrt{12 \b^2+9}-3}+\ln\frac{\sqrt{12 \b^2+9}+2 \b^2 z^2-3}
{\sqrt{12 \b^2+9}-2 \b^2 z^2+3}\\
&+\frac{\sqrt{12\b^2+9}}{9} \ln\frac{3 (1-z^2)}{3(1+z^2)-\b^2 z^4} \biggr)\,,\qquad \k(z=0)=0
\end{split}
\eqlabel{getares}
\end{equation}
We show some example of numerical results in the main text.
\section{The spontaneously/explicitly broken phases}\label{App4}
\textbf{The spontaneously broken phase}\\[0.1cm]
We collect here the details about the results presented in section \ref{fullg} dealing with the spontaneously broken phase and its transport properties.\\
As a first step, we construct $\zet_2$ symmetry breaking phase in FG coordinate system.
Assuming $x\in[0,x_h]$, we introduce
\begin{equation}
\begin{split}
&A=\frac{1}{x^2}\ f_1\left(\frac{x}{x_h}\right)\,,\qquad B=\frac{1}{x^2}\ f_2\left(\frac{x}{x_h}\right)\,,
\qquad C=\frac{1}{x^2}\,,\\
&a_0'=-\frac{2\b}{x_h^2}\ a\left(\frac{x}{x_h}\right))\,, 
\qquad \phi=\phi\left(\frac{x}{x_h}\right)\,,\qquad k=\frac{s}{x_h}
\end{split}
\eqlabel{warpfg}
\end{equation}
and using the radial coordinate $z= \frac{x}{x_h}$ we find the following
equations:
\begin{equation}
\begin{split}
0=&f_1'+\frac16 z (\phi')^2 f_1-\frac23 z^3 a^2 (\phi^2+1) \b^2-\frac{4 f_1 (f_2-1)}{z f_2}
-\frac{\phi^2 f_1 (3 s^4 z^4-4)}{6z f_2}
\end{split}
\eqlabel{fg1}
\end{equation}
\begin{equation}
\begin{split}
0=&f_2'-\frac16 z (\phi')^2 f_2-\frac{2z^3 a^2 f_2 (\phi^2+1) \b^2}{3f_1}
-\frac{\phi^2 (3 s^4 z^4-4)}{6z}+\frac{4(1-f_2)}{z}
\end{split}
\eqlabel{fg2}
\end{equation}
\begin{equation}
\begin{split}
0=&a'+\frac{a}{6z (\phi^2+1)} \left(z^2 (\phi')^2 (\phi^2+1)+12 \phi \phi' z-6 \phi^2-6\right)
\end{split}
\eqlabel{fg3}
\end{equation}
\begin{equation}
\begin{split}
0=&\phi''+\frac{2z^2 a^2\b^2}{3f_1} \left(\phi' z (\phi^2+1)+6 \phi\right)
+\frac{\phi^2 (3 s^4 z^4-4)+6 f_2-24}{6z f_2} \phi'- \frac{3 s^4 z^4-4}{z^2 f_2}\phi
\end{split}
\eqlabel{fg4}
\end{equation}
Eqs.~\eqref{fg1}-\eqref{fg4} have to be solved subject to the following boundary conditions:
\begin{equation}
\begin{split}
&f_1=1+f_{1,2} z^4+\calo(z^6)\,,\qquad f_2=1+\left(\frac 13 p_2^2+f_{1,2}\right) z^4+\calo(z^6)\\
&a=z-\frac{7}{6}p_2^2 z^5+\calo(z^9)\,,\qquad \phi=p_2 z^2+\calo(z^6)
\end{split}
\eqlabel{uvfgbn}
\end{equation}
and
\begin{equation}
\begin{split}
&f_1=f_{1,h,1} (1-z)+\calo((1-z)^2)\,,\qquad a=a_{h,0}+\calo((1-z))\,,\qquad \phi=p_{h,0}+\calo((1-z))\\
&f_2=\frac{f_{1,h,1,}(24+(4-3 s^4)p_{h,0}^2)}{2(2 a_{h,0}^2\beta^2 (p_{h,0}^2+1)+3f_{1,h,1})}(1-z)+\calo((1-z)^2)
\end{split}
\eqlabel{irfgb}
\end{equation}
Fixing $\{\b,s\}$ (the charge density/chemical potential and the lattice spacing) uniquely determine the solution, specified by
$\{p_2,f_{1,2},f_{1,h,1},a_{h,0},p_{h,0}\}$. The equilibrium temperature is given by:
\begin{equation}
T=\frac{f_{1,h,1}}{4\sqrt{2}\pi x_h}\ \sqrt{\frac{24-p_{h,0}^2(3 s^4-4)}{2\b^2 a_{h,0}^2(1+p_{h,0}^2)+3f_{1,h,1}}}
\eqlabel{tfg1}
\end{equation}
There is a simple solution to \eqref{fg1}-\eqref{irfgb}
\begin{equation}
f_1=f_2=1-z^4-\frac{\b^2 z^4}{3} (1-z^2)\,,\qquad a=z\,,\qquad  \phi=0
\eqlabel{reprn}
\end{equation}
which is the $\zet_2$ symmetric solution \eqref{rnfg}.

Once the $\zet_2$ symmetry breaking background is determined, we need to solve \eqref{heom}:
\begin{equation}
0=H''+\frac{4 \b^2 a^2 z^4 f_2 (\phi^2+1)+\phi^2 f_1 (3 s^4 z^4-4)+6 f_1 (f_2-4)}{6z f_1 f_2} H'
+\frac{x_h^2  \w^2-2  z^2 \phi^2 s^4 f_1}{f_1 f_2}H
\eqlabel{fgheom}
\end{equation}
We find it convenient to introduce $\{H_0,H_1\}$, which are independent of $\w$,  as
\begin{equation}
H=(1-z)^{-i\w/(4\pi T)}\ \left(1+z^4 H_0(z)+\frac{i\w}{T}\left(z^4 H_1(z)-\frac{z^3}{12\pi} -\frac{z^2}{8\pi}
-\frac{z}{4\pi}\right)+\calo(\frac{\w^2}{T^2})\right)
\eqlabel{defh0h1}
\end{equation}
Notice that $H$ as defined above automatically satisfied an incoming wave boundary condition
at the horizon, provided $\{H_0, H_1\}$ are regular at the horizon. Near the AdS boundary we have:
\begin{equation}
\begin{split}
&H_0=h_{4,0}+\left(-\frac{1}{12}h_{4,0}p_2^2-\frac12 f_{1,2} h_{4,0}+\frac{1}{16}p_2^2 s^4\right) z^4+\calo(z^6)\\
&H_1=h_{4,1}-\frac{1+5 h_{4,0}}{20\pi}\ z+\calo(z^2)
\end{split}
\eqlabel{uvfgh}
\end{equation}
Following \eqref{expaH} we find
\begin{equation}
\begin{split}
h_4^b(\w)=\frac{1}{x_h^4}\left(h_{4,0}+\frac{i\w}{T}\left(\frac{1}{16\pi}+h_{4,1}\right)+\calo(\frac{\w^2}{T^2})\right)
\end{split}
\eqlabel{defh4b}
\end{equation}
leading to (see \eqref{viscotransport})
\begin{equation}
16\pi G_N\ \frac{G}{k^4}= -\frac{4 h_{4,0}}{s^4}\,,\qquad \frac{\eta}{\cals}=\frac{1}{\pi T x_h}\left(\frac{1}{16\pi}+h_{4,1}\right)
\eqlabel{resviscospon}
\end{equation}
In the $\zet_2$ symmetric phase $\{H_0, H_1\}$ can be determined analytically, producing
\begin{equation}
\{h_{4,0},h_{4,1}\}=\left\{0,\frac{3}{16\pi}-\frac{\b^2}{24\pi}\right\}\qquad \Longrightarrow\qquad  \left\{\frac{G}{k^4}=0\,,\ \frac{\eta}{\cals}=\frac{1}{4\pi}\right\}
\eqlabel{rncaseh}
\end{equation}
From the holographic renormalization,
\begin{equation}
\begin{split}
&16\pi G_N\ \calq=\frac{2\b}{x_h^3}\,,\qquad 16\pi G_N\ \cals=\frac{4\pi}{x_h^3}\\
&16\pi G_N\ \cale=-\frac{3f_{1,2}}{x_h^4}\,,\qquad \calp=\cals T+\mu\calq -\cale
\end{split}
\eqlabel{thermo2}
\end{equation}
where the temperature $T$ is given by \eqref{tfg1}.\\
The numerical results are shown in section \ref{fullg}.\\[0.1cm]
\textbf{The explicitly broken phase}\\[0.1cm]
In this section we provide more details about the results presented in section \ref{zet2explicit} regarding the explicitly broken $\zet_2$ phase.\\
The AdS boundary conditions which reflect a finite $\dd_2\ne 0$ source term of the explicit $\zet_2$ symmetry breaking are:
\begin{equation}
\begin{split}
&f_1=1+f_{1,2} z^4+\calo(z^8\ln^2 z)\,,\qquad \phi=(p_2 +\a \ln z)z^2+\calo(z^6\ln^2 z)
\\
&f_2=1+\left(\frac 13 p_2^2+f_{1,2}+\frac{\a^2}{24}+\frac{\a p_2}{6}
+\left(\frac{\a^2}{6}+\frac{2\a p_2}{3}\right)\ln z+\frac 13\ln^2 z
\right) z^4+\calo(z^8\ln^4 z) 
\end{split}
\eqlabel{uvfgbsn}
\end{equation}
where we introduced a dimensionless parametrization\footnote{Comparing with \cite{Buchel:2007vy}
$\dd_2\propto +m^2 >0$.} of
$\dd_2>0$  as
\begin{equation}
\dd_2=\frac{\a}{x_h^2}
\eqlabel{defdd2n}
\end{equation}
The near horizon asymptotes are as in \eqref{irfgb} with $a_{h,0}=0$.

Following section \ref{fullg},
we need  to solve \eqref{fgheom} (again $a(z)\equiv 0$) with parametrization
\eqref{defh0h1} and modified asymptotically AdS boundary conditions:
\begin{equation}
\begin{split}
&H_0=h_{4,0}+\biggl(
\frac{1}{16} p_2^2 s^4-\frac{1}{128} \a^2 h_{4,0}-\frac{1}{48} \a h_{4,0} p_2
-\frac{1}{12} h_{4,0} p_2^2-\frac12 f_{1,2} h_{4,0}-\frac{3}{64} \a p_2 s^4
\\&+\frac{7}{512} \a^2 s^4
+\left(\frac 18 \a p_2 s^4-\frac{1}{48} \a^2 h_{4,0}-\frac16 \a h_{4,0} p_2
-\frac{3}{64} \a^2 s^4\right) \ln z\\
&+\left(-\frac{1}{12} \a^2 h_{4,0}+\frac{1}{16} \a^2 s^4\right) \ln^2 z
\biggr) z^4+\calo(z^8\ln^4 z)\\
&H_1=h_{4,1}-\frac{1+5 h_{4,0}}{20\pi}\ z+\calo(z^2)
\end{split}
\eqlabel{uvfghn}
\end{equation}
The shear elastic modulus and the shear viscosity are extracted from \eqref{defh4b}
and \eqref{resviscospon}.\\
The numerical results are presented in the main text.

\newpage

\bibliographystyle{JHEP}
\bibliography{viscoelastic}

\end{document}